\documentclass[aps,prb,twocolumn,a4paper,10pt,superscriptaddress]{revtex4-1}

\usepackage{graphicx}
\usepackage{graphics}
\usepackage{amsmath}
\usepackage{amssymb}
\usepackage{amsfonts}
\usepackage{dsfont}
\usepackage{braket}
\usepackage{color}
\usepackage{braket,slashed}
\usepackage[mathscr]{euscript}
\usepackage{hyperref}
\usepackage{subfigure}
\usepackage{xfrac}
\usepackage{bm}
\usepackage{pgfplots}

\allowdisplaybreaks[1]

\newcommand{\ZZ}{\ensuremath{\mathbb{Z}}}
\newcommand{\CC}{\ensuremath{\mathbb{C}}}

\DeclareMathOperator{\tr}{tr}

\newcommand{\one}{\mathds{1}}

\renewcommand{\dag}{^\dagger}
\renewcommand{\d}{\ensuremath{\mathrm{d}}}
\renewcommand{\ol}[1]{\overline{#1}}
\newcommand{\e}{\ensuremath{\mathrm{e}}}

\newcommand{\ham}{\ensuremath{\hat{H}}}

\newcommand{\vect}[1]{\ensuremath{\boldsymbol{#1}}}
\newcommand{\lv}{\ensuremath{\vect{v_L^\dagger}}}
\newcommand{\rv}{\ensuremath{\vect{v_R}}}
\newcommand{\spst}{\ensuremath{\ket{\{s\}}}}
\newcommand{\E}[2]{\ensuremath{E^{#1}_{#2}}}
\renewcommand{\O}[2]{\ensuremath{O^{#1}_{#2}}}
\renewcommand{\H}[4]{\ensuremath{H^{#1#2}_{#3#4}}}

\newcommand{\HH}[6]{\ensuremath{H^{#1#2#3}_{#4#5#6}}}
\newcommand{\J}[4]{\ensuremath{J^{#1#2}_{#3#4}}}
\newcommand{\tA}{\ensuremath{A}}

\newcommand{\gammap}{\ensuremath{{\gamma'}}}
\newcommand{\perm}{\ensuremath{{\mathcal{P}}}}
\newcommand{\vv}{\ensuremath{\vect{v}}}

\usepackage{soul}

\date{\today}

\begin{document}

\title{Scattering particles in quantum spin chains}

\author{Laurens Vanderstraeten}
\affiliation{Ghent University, Department of Physics and Astronomy, Krijgslaan 281-S9, B-9000 Gent, Belgium}
\author{Frank Verstraete}
\affiliation{Ghent University, Department of Physics and Astronomy, Krijgslaan 281-S9, B-9000 Gent, Belgium}
\affiliation{Vienna Center for Quantum Science, Universit\"at Wien, Boltzmanngasse 5, A-1090 Wien, Austria}
\author{Jutho Haegeman}
\affiliation{Ghent University, Department of Physics and Astronomy, Krijgslaan 281-S9, B-9000 Gent, Belgium}

\begin{abstract}
A variational approach for constructing an effective particle description of the low-energy physics of one-dimensional quantum spin chains is presented. Based on the matrix product state formalism, we compute the one- and two-particle excitations as eigenstates of the full microscopic Hamiltonian. We interpret the excitations as particles on a strongly-correlated background with non-trivial dispersion relations, spectral weights and two-particle S matrices. Based on this information, we show how to describe a finite density of excitations as an interacting gas of bosons, using its approximate integrability at low densities. We apply our framework to the Heisenberg antiferromagnetic ladder: we compute the elementary excitation spectrum and the magnon-magnon S matrix, study the formation of bound states and determine both static and dynamic properties of the magnetized ladder.
\end{abstract}

\maketitle

\tableofcontents

\newlength\figureheight 
\newlength\figurewidth 
\setlength\figureheight{4cm}
\setlength\figurewidth{0.85\columnwidth} 

\section{Introduction}

Finding the ground state of strongly-correlated quantum many-body systems poses one of the main challenges of contemporary condensed matter physics. The physical properties of these systems, however, are not determined by the ground state but rather by the low-lying, elementary excitations relative to this ground state. In contrast to the strongly-correlated ground state, these elementary excitations are of a particularly simple character: in most cases they can be treated as a collection of independent, weakly-interacting particles living on non-trivial vacuum state \cite{Landau1941, Anderson1963}.
\par In condensed matter theory these ``quasi-particles'' are typically defined starting from some non-interacting limit. In Fermi liquid theory for interacting electron systems  \cite{Landau1956, *Landau1957, Nozieres1964, *Pines1966} -- the most prominent example of this approach -- the quasi-particles are defined in the free-electron system, but remain well-defined modes when turning on the interactions. The effect of a finite lifetime and quasi-particle interactions can be treated in perturbation theory. In strongly-correlated lattice systems, however, there is typically no obvious way to start from a non-interacting theory to define the quasi-particles that determine the system's properties (notable counter-examples in one dimension include integrable systems \cite{Essler2004a} and continuous unitary transformations \cite{Knetter2003a}). The variational approach, which we will advocate in this paper, is orthogonal to the perturbative approach by starting from the strongly-correlated ground state and finding the low-lying excitations of the interacting system variationally. As exact eigenstates, these excitations have an infinite lifetime, but a priori it is not clear that they should have a local, particle-like nature.
\par
\begin{figure}
\includegraphics[width=\columnwidth]{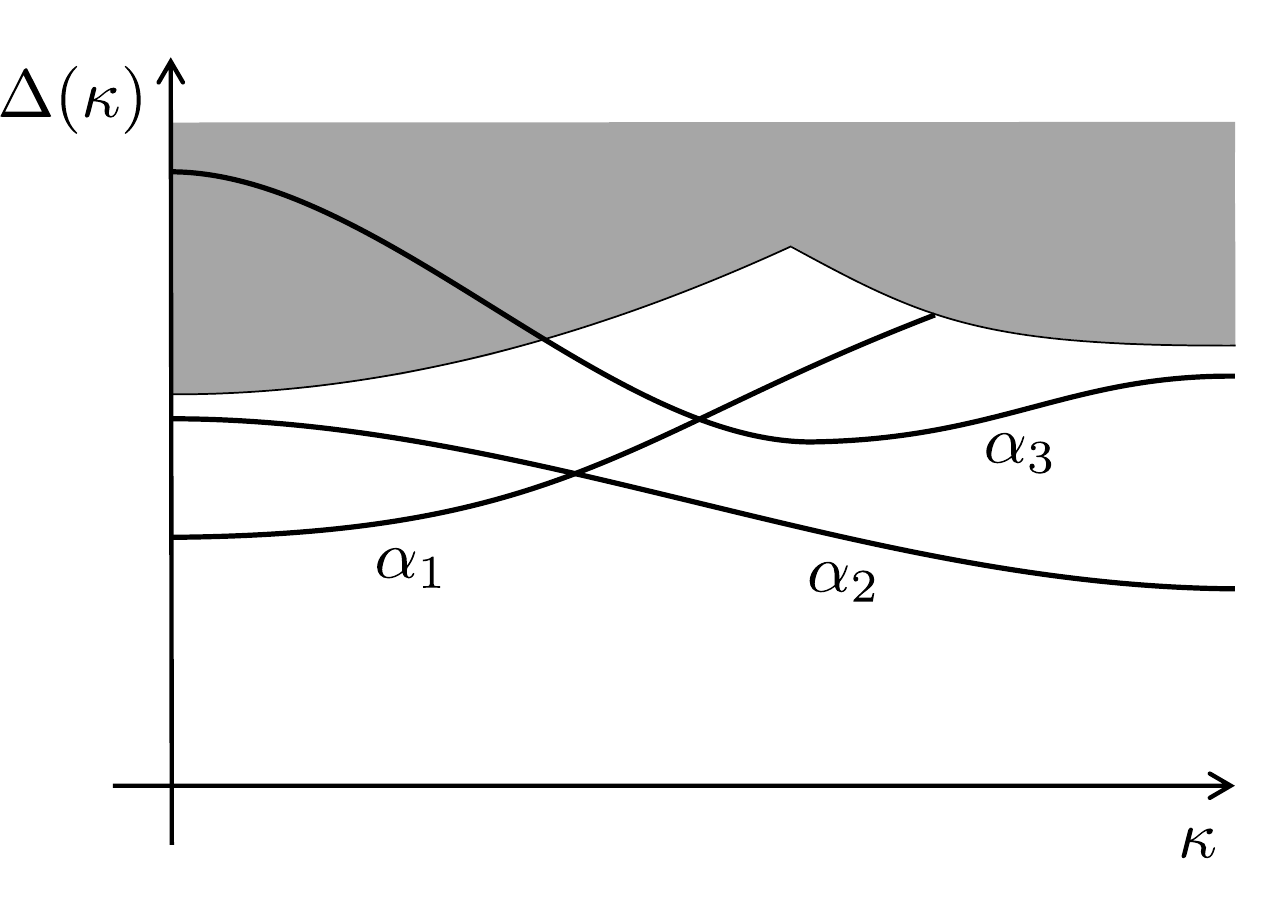}
\caption{A typical momentum-energy excitation spectrum of a one-dimensional lattice system. We have depicted three elementary (one-particle) excitations (full lines) and the many-particle continuum (grey). Both $\alpha_2$ and $\alpha_3$ are stable in the whole Brillouin zone; the latter remains stable even inside the continuum, possibly because it cannot decay in a two-particle state through symmetry constraints. Particle $\alpha_1$ becomes unstable upon entering the continuum, so that it ceases to be a one-particle excitation (cannot be created by a local operator).}
\label{fig:typicalspectrum}
\end{figure}

 In relativistic quantum field theory, a picture of localized elementary excitations on top of a strongly-correlated vacuum has been formulated in a rigorous fashion. Haag-Ruelle scattering theory \cite{Haag1958, *Ruelle1962, *Haag1996} does indeed construct a many-particle Fock space by acting with local operators on the vacuum and even defines an S matrix describing the interactions between these particles. Because this formalism depends heavily on Lorentz invariance, there is a priori no straightforward translation to lattice systems. Indeed, on the lattice there are fewer restrictions on the spectrum: different elementary excitation branches typically have different characteristic velocities and are not bound to be stable in the whole Brillouin zone -- a typical spectrum is shown in Fig.~\ref{fig:typicalspectrum}.
\par Recently though, it was realized that by using Lieb-Robinson bounds \cite{Lieb1972} as the soft lattice analog of strict causality in relativistic QFT, the locality of elementary excitations can be established in a rigorous way. More specifically, it was shown in Ref.~\onlinecite{Haegeman2013a} that an elementary excitation that lives on an isolated branch in the energy-momentum spectrum and has a finite overlap with an arbitrary local operator, can be created out of the ground state by the action of a momentum superposition of a local operator (to an exponential precision in the size of the support of this operator). In Ref.~\onlinecite{Bachmann2014} the scattering problem of these particle excitations was formulated by translating the Haag-Ruelle formalism to the lattice setting.
\par These theoretical developments provide a motivation for the variational approach towards a particle picture of the low-energy physics of lattice systems. Indeed, by making use of the fact that gapped excitations should be local, it might prove possible to describe them with only a small number of variational parameters. This is the idea of the single-mode approximation, or Feynman-Bijl ansatz, pioneered by Feynman in his study on liquid helium \cite{Feynman1953a, *Feynman1954, *Feynman1956, Girvin1986} and later successfully applied to quantum spin systems \cite{Arovas1988, Takahashi1988, Sorensen1994}. Although providing qualitative insight into the nature of the elementary excitations, the single-mode approximation is often too crude as a variational ansatz to obtain quantitative results on the low-lying spectrum of generic spin chains.
\par Indeed, constructing a variational ansatz for excitations with quantitative accuracy requires both an accurate parametrization of the ground state and a systematic way to change this ground state locally. For one-dimensional systems, the framework of matrix product states \cite{Schollwock2011a, Verstraete2008a} (MPS) has proven to meet both requirements. The ground state of one-dimensional quantum spin systems can indeed be efficiently parametrized by the class of MPS \cite{Verstraete2006, Hastings2007}; the success of the density matrix renormalization group \cite{White1992} is based on MPS serving as the class of states over which it optimizes \cite{Ostlund1995a, *Rommer1997a, Dukelsky1997}. The defining characteristic of MPS -- or tensor network states in general -- is the presence of a ``virtual'' level that carries the (quantum) correlations in the many-body wave function. By acting both on the physical and  the virtual level, a variational ansatz for elementary excitations on an MPS ground state was introduced in Refs.~\onlinecite{Haegeman2012a} and \onlinecite{Pirvu2012}. The ansatz was used for calculating dispersion relations and dynamical correlation functions of quantum spin chains \cite{Haegeman2013b, Zauner2015, Keim2015}, quantum field theories \cite{Draxler2013, Milsted2013} and gauge theories \cite{Buyens2014}.
\par A more general understanding of these efforts is obtained by realizing that the low-energy dynamics correspond to small variations around the variational ground state and are therefore not necessarily contained within the variational class itself. Indeed, for the smooth manifold of MPS \cite{Haegeman2014} it is the tangent space constructed around the MPS ground state that provides a natural parametrization of the low-energy dynamics. For example, the best approximation to time evolution within the MPS manifold can be obtained by projecting the Schr\"{o}dinger equation into the MPS tangent bundle, according to the time-dependent variational principle (TDVP) \cite{Haegeman2011d,Haegeman2014a}. Similarly, the ansatz for an elementary excitation corresponds exactly to a vector in the tangent space around the MPS ground state. These ideas can be grouped under the concept of \emph{post-MPS} methods \cite{Haegeman2013b} as an alternative for the standard MPS algorithms for tackling the low-energy dynamics around an MPS ground state. 
\par The crucial next step in this approach -- after the construction of single-particle excitations -- consists of studying the interactions between these particles and, more specifically, computing the two-particle S matrix \cite{Vanderstraeten2014}. This information can then be used as the input for the ``approximate Bethe ansatz'' \cite{Krauth1991, Kiwata1994a, Okunishi1999a} (i.e. neglecting all three-particle scattering processes) in order to provide a first-quantized description of a finite density of excitations on top of the strongly-correlated vacuum.
\par These developments should eventually lead to the ab initio construction of an effective second-quantized Hamiltonian, acting in a Fock space of interacting particles. In contrast to standard effective field theory constructions, the variational approach would automatically incorporate all symmetries and correlations of the vacuum state on which these particles live without relying on phenomenological considerations.
\par In this paper we further elaborate on the framework that we introduced in Ref.~\onlinecite{Vanderstraeten2014}. In Sec.~\ref{sec:section2} we show how to construct one- and two-particle excitations on an MPS vacuum state. We formulate a definition of the S matrix based on the form of the two-particle wave function and prove that it is unitary. Finally, we construct the projector on the one- and two-particle subspace which shows up in the spectral representation of dynamical correlation functions. In Sec.~\ref{sec:section3} we take a step back and show that the S matrix as defined in Sec.~\ref{sec:section2} corresponds to the one that shows up in standard dynamical scattering theory. Next we elaborate on the approximate Bethe ansatz as a way of dealing with a finite density of excitations in a first-quantized many-particle formalism. In Sec.~\ref{sec:section4} we apply our variational method to study the Heisenberg antiferromagnetic two-leg ladder. We compute the elementary excitation spectrum, the two-particle S matrix and one- and two-particle contributions to dynamical correlation functions. Afterwards, we apply the approximate Bethe ansatz to the magnetization process, at zero and finite temperature, and compute both thermodynamic properties and correlation functions of the magnetized ladder. In the last section, we provide an overview of some interesting extensions of our framework and give an outlook towards the construction of effective field theories in second quantization.

\section{Constructing scattering states}
\label{sec:section2}

In this section we construct variational one- and two-particle states on an MPS background \footnote{Note that our approach was inspired by the works of Kohn \cite{Kohn1948} and Feynman \cite{Feynman1972}}. Based on the form of these wave functions, we define the S matrix and introduce the projectors on the one- and two-particle subspaces (i.e. the low-energy subspace) which can be used to compute the low-energy part of dynamical correlation functions. Note that, while the complete framework is presented in the main body, technical details and long equations are taken up in the appendix. A short remark on notation is also in order. Vectors of any length will be denoted in boldface, whereas matrices will use a sans serif font. Vector entries will be referred to using a superscript (in which case the boldface will be dropped), whereas subscripts of a boldface vector typically refer to a label of a set of vectors, such as a basis. The only exception to these rules is that physical states are denoted using Dirac's bra-ket notation and the matrices appearing in the definition of the matrix product state (which can also be interpreted as rank three tensors) are typeset using the normal serif type (italic).

\subsection{Ground state}

Consider a one-dimensional quantum spin system with local dimension $d$ in the thermodynamic limit, described by a local and translation invariant Hamiltonian. While in no way crucial, we restrict to nearest-neighbour Hamiltonians, i.e. $\ham=\sum_{n\in\ZZ}\hat{h}_{n,n+1}$, for reasons of simplicity. We furthermore assume that the translation invariant ground state of this system (we restrict to a unique ground state, see Sec.~\ref{sec:section5} for extensions) can be accurately described by an injective uniform matrix product state (uMPS) \cite{Fannes1992,Vidal2007b,Haegeman2011d}
\begin{equation} \label{eq:mps}
    \ket{\Psi[A]} = \sum_{\{s\}=1}^d \lv \left[ \prod_{m} A^{s_m} \right] \rv \spst,
\end{equation}
where $A^s$ is a set of $D\times D$ matrices for $s=1,\dots,d$, or, equivalently, $A$ can be interpreted as a $D\times d \times D$ tensor; $\lv$ and $\rv$ are $D$-dimensional boundary vectors. In the thermodynamic limit, all physical observables are independent of these boundary vectors \cite{Haegeman2014}, so that the tensor $A$ provides a complete description of the ground state $\ket{\Psi[A]}$.
\begin{figure}
\subfigure{\includegraphics[width=0.4\columnwidth]{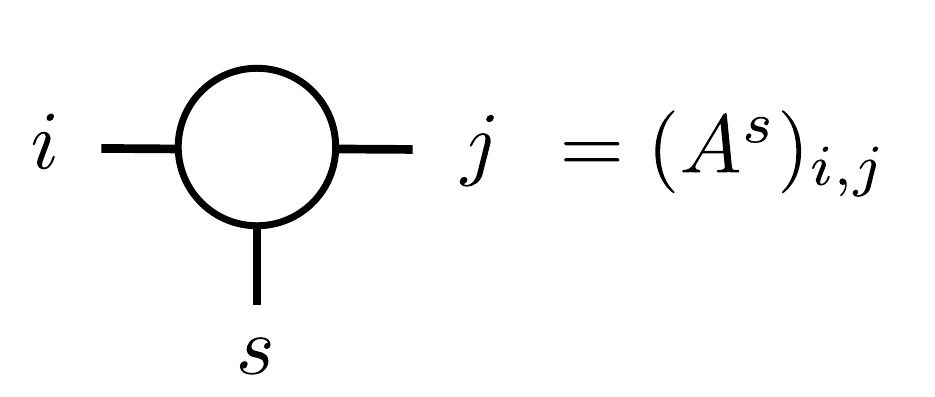}}\\
\subfigure{\includegraphics[width=0.85\columnwidth]{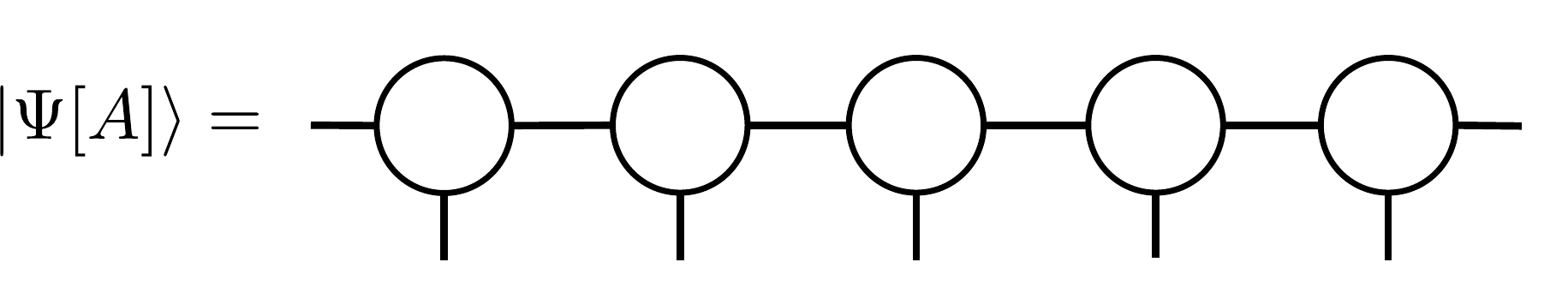}}
\caption{Graphical representation of an MPS ground state. The circles represent the ($D\times d \times D$)-dimensional tensor $A$: every outgoing leg corresponds to a tensor index. Whenever two legs are connected, this corresponds to a contraction of the two indices. In the MPS all virtual indices are contracted, while the physical indices correspond to the physical degrees of freedom in the MPS wave function \eqref{eq:mps}. The matrix product structure contains the (quantum) correlations of this ground state.}
\end{figure}
\par The set of injective MPS of a certain bond dimension constitute a complex manifold \cite{Haegeman2014}. Finding the best approximation of the ground state of a certain Hamiltonian within this manifold can be achieved using different algorithms \cite{Schollwock2011a} -- in our simulations we will always use the TDVP algorithm\cite{Haegeman2011d,Haegeman2014a}.

\subsection{One-particle excitations}

This ground state serves as our vacuum, on top of which we will build localized, particle-like excitations. A first guess for the wave function of an elementary excitation with momentum $\kappa$ is the single-mode approximation
\begin{equation} \label{eq:sma}
\ket{\Phi_\text{SMA}(\kappa)} = \sum_n \e^{i\kappa n} \hat{O}_n \ket{\Psi[A]},
\end{equation}
where $\hat{O}_n$ is an operator acting at site $n$. The choice of operator can be inspired by physical intuition \cite{Feynman1954,Girvin1986,Arovas1988,Takahashi1988, Sorensen1994, Talstra1997} or determined by numerical optimization \cite{Chung2010a}. Though providing some qualitative insight into elementary excitation spectra, this ansatz is typically not a good quantitative approximation for the true wave function of the excitation. Systematically improving on this would ask for the introduction of bigger local operators $\hat{O}_n$. It was indeed proven \cite{Haegeman2013a} that, in the case of an isolated excitation branch, the exact wave function can be arbitrary well approximated in this way. More specifically, it was shown that the localized nature of an excitation depends on the gap to the nearest eigenvalue of the Hamiltonian in the same momentum sector.
\begin{figure}
\includegraphics[width=\columnwidth]{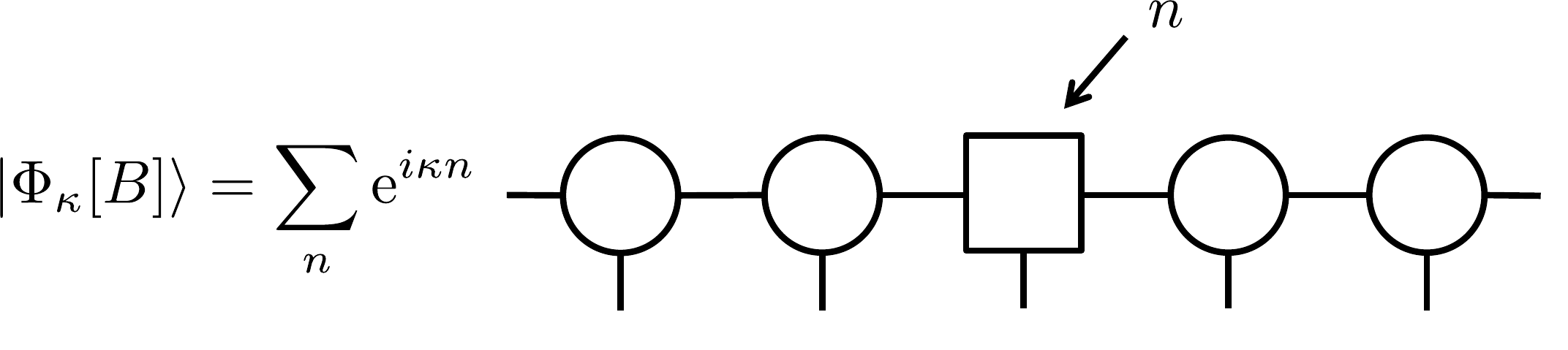}
\caption{Graphical representation of the one-particle excitation ansatz. The ground state tensor $A$ is changed at site $n$ into a new tensor $B$ (square) and a momentum superposition is taken. The matrix product structure allows that the tensor $B$ can change the ground state over a finite distance.}
\label{fig:ansatz}
\end{figure}
\par Within the framework of matrix product states, it is possible to construct a variational ansatz that is able to capture the localized nature of the excitation by directly modifying the local tensors. Indeed, instead of only operating on the physical level, we can change one ground state tensor $A^s$ with a new tensor $B^s$ and take a momentum superposition \cite{Ostlund1995a, *Rommer1997a, Haegeman2012a,Haegeman2013b}
\begin{multline} \label{oneparticle}
    \ket{\Phi_\kappa[B]} = \sum_n \e^{i\kappa n} \sum_{\{s\}} \lv \left[ \prod_{m<n} A^{s_m} \right] \\
    \times B^{s_n} \left[ \prod_{m>n} A^{s_m} \right] \rv \spst.
\end{multline}
Through the virtual level of the MPS, this ansatz is able to perturb the ground state over a finite length determined by the bond dimension $D$.  All variational freedom of this ansatz is contained within the tensor $B^s$. As the parametrization of the state \eqref{oneparticle} is linear in the elements of $B^s$, variationally optimizing amounts to solving the Rayleigh-Ritz problem
\begin{equation} \label{eig1p}
 \mathsf{H}_\text{eff,1p}(\kappa) \vect{u} = \lambda \vect{u}
\end{equation}
with $\mathsf{H}_\text{eff,1p}(\kappa)$ the momentum dependent effective one-particle Hamiltonian and vector $\vect{u}$ containing the coefficients $u^i$ to expand tensor $B$ in the state \eqref{oneparticle} with respect to a suitably chosen basis $\{B_{(i)}, i=1,\ldots,(d-1)D^2\}$. We refer to Appendix \ref{sec:1pA} for details on how to calculate $\mathsf{H}_\text{eff,1p}$.
\par Upon solving the eigenvalue problem in Eq.~\eqref{eig1p}, we obtain a set of $(d-1)D^2$ eigenvalues for every momentum $\kappa$. Some of those eigenvalues $\Delta_{\alpha}(\kappa)$ offer a good approximation to the exact dispersion relations of the elementary excitations, i.e.\ the isolated branches in the spectrum of the Hamiltonian. Moreover, from the corresponding eigenvectors $\vect{u}_{\alpha}(\kappa)$ we obtain an explicit expression for the wave function of the elementary excitations by inserting the tensors $B_{\alpha}(\kappa)=\sum_i u_{\alpha}^i(\kappa) B_{(i)}$ in Eq.~\eqref{oneparticle}. This expression can be used to calculate the spectral weights of the excitations and, consequently, the one-particle contribution to dynamical correlation functions.
\par Other eigenvalues obtained from Eq.~\eqref{eig1p} will fall in the continuous part of the spectrum of the Hamiltonian, i.e.\ in the set of scattering states. Scattering states cannot be described by a single local perturbation, so we expect the ansatz \eqref{oneparticle} to fail. In fact, instead of a scattering state, the variational optimization will create a localized wave packet of two-particle states within some energy range. Obviously, the variational eigenstates of the form in Eq.~\eqref{oneparticle} will not provide a good approximation to the exact scattering eigenstates of the full Hamiltonian. A more appropriate variational ansatz for two-particle scattering states is discussed in the remainder of this section.
\par Remark that we have so far not discussed the case of bound states. When defining (quasi-) particles along a path of Hamiltonians using e.g. perturbation theory or continuous unitary transformations \cite{Knetter2003a}, bound states can be identified with isolated eigenstates emerging from a multi-particle continuum along the path. In our variational framework, we consider one particular Hamiltonian which is not necessarily related to a one-parameter family. All isolated branches in the spectrum are equally elementary (see Ref.~\onlinecite{Zimmermann1958} for the analogous result in QFT). While there might be quantum numbers that indicate the ``history'' of an elementary excitation along a path of Hamiltonians, there is typically no particle number symmetry to make the interpretation of bound states unambiguous. On a related note, elementary excitations are by this definition exact eigenstates of the Hamiltonian and therefore have an infinite life time. We cannot and do not target resonances within the continuous part of the spectrum. Therefore, the Hamiltonian does not contain interactions that link the one-particle sector with higher particle states.
\par As mentioned previously, the spectrum of general quantum spin chains can be very complex. Within certain regions of the Brillouin zone, the energy of elementary excitations can fall within the continuum (this typically requires a quantum number that protects them against decay) or there might be no elementary excitations at all (e.g. around momentum zero in the spin-1 Heisenberg antiferromagnet). We therefore need a way to determine which variational eigenvalues of Eq.~\eqref{eig1p} correspond to elementary excitations and therefore offer a good approximation to actual eigenstates of the Hamiltonian. Upon enlarging the variational one-particle space, e.g. by increasing the bond dimension or the spatial support of the local perturbation, eigenvalues that correspond to elementary excitations will converge quickly (related to the gap to the nearest eigenvalue) and remain at a fixed position. Eigenvalues in the continuous part of the exact spectrum, on the other hand, will not really converge and several new eigenvalues will appear in those regions. A more quantitative way to assess how well an exact eigenstate is approximated consists of calculating the variance of the Hamiltonian \cite{Davison1968}, i.e.\ $\bra{\Phi_\kappa[B]}(\ham-\Delta(\kappa))^2 \ket{\Phi_\kappa[B]}$. For elementary excitations, these variances should be small (see Sec.~\ref{sec:ladder1p} for numerical values). For the other solutions of the one-particle problem \eqref{eig1p}, which correspond to scattering states, the variance should be larger. For a typical gapped system, the difference will be some orders of magnitude. Consequently, this quantity allows for the identification of one-particle states, even within higher-particle bands and without exploiting symmetries.
\par Note finally that, without Galilean invariance on the lattice, the tensor $B_{\alpha}(\kappa)$, which describes the particle $\alpha$ on a dispersion branch $\Delta_\alpha(\kappa)$, is momentum dependent. On the other hand, we expect that for a well-defined particle in a certain momentum range this momentum dependence is not too strong. Indeed, it turns out that by a suitable choice of the basis tensors $\{B_{(i)}, i=1,\ldots,(d-1)D^2\}$, we can fully capture $B_{\alpha}(\kappa)$ for all elementary excitations $\alpha$ and for all momenta $\kappa$ in the span of just a small number $\ell\ll (d-1)D^2$ basis vectors $\{B_{(i)},i=1,\ldots,\ell\}$. Although more sophisticated optimization strategies should be possible, we construct this reduced basis from a number of $B$'s at different momenta. This reduced basis will be important for solving the scattering problem in the next sections.

\subsection{Variational ansatz for two-particle states}

In the previous section it became clear that we need another ansatz to capture the delocalized nature of a two-particle state. We will start from a one-particle spectrum consisting of a number of different types of particles, labelled by $\alpha$, with dispersion relations $\Delta_\alpha(\kappa)$. In the thermodynamic limit, constructing the two-particle spectrum is trivial: the momentum and energy are the sum of the individual momenta and energies of the two particles \cite{Anderson1963}. The two-particle wave function, however, depends on the particle interaction. The interactions, which depend on both the Hamiltonian and the ground state correlations, are reflected in the wave function in two ways: (i) the asymptotic wave function has different terms, with the S matrix elements as the relative coefficients, and (ii) the local part of the wave function.
\par In order to capture both we introduce the following ansatz for describing states with two localized, particle-like excitations with total momentum $K$ 
\begin{align} \label{eq:ansatz}
  \ket{\Upsilon(K)} = \sum_{n=0}^{+\infty} \sum_{j=1}^{L_n} c^j(n) \ket{\chi_{K,j}(n)}
\end{align}
where the basis states are
\begin{widetext}
\begin{align}
   & \ket{\chi_{K,j}(n=0)} = \sum_{n_1=-\infty}^{+\infty} \e^{i K n_1} \sum_{\{s\}=1}^d \lv \left[ \prod_{m<n_1} A^{s_m} \right] B_{(j)}^{s_{n_1}} \left[ \prod_{m>n_1} A^{s_m} \right] \rv \spst \label{local} \\
  & \ket{\chi_{K,(j_1,j_2)}(n>0)} = \sum_{n_1=-\infty}^{+\infty} \e^{i K n_1} \sum_{\{s\}=1}^d \lv \left[ \prod_{m<n_1} A^{s_m} \right] B_{(j_1)}^{s_{n_1}} \left[ \prod_{n_1<m<n_1+n} A^{s_m} \right]  B_{(j_2)}^{s_{n_1+n}} \left[ \prod_{m>n_1+n} A^{s_m} \right] \rv \spst. \label{nonLocal}
\end{align}
\end{widetext}
\begin{figure*}
\centering
\includegraphics[width=0.7\columnwidth]{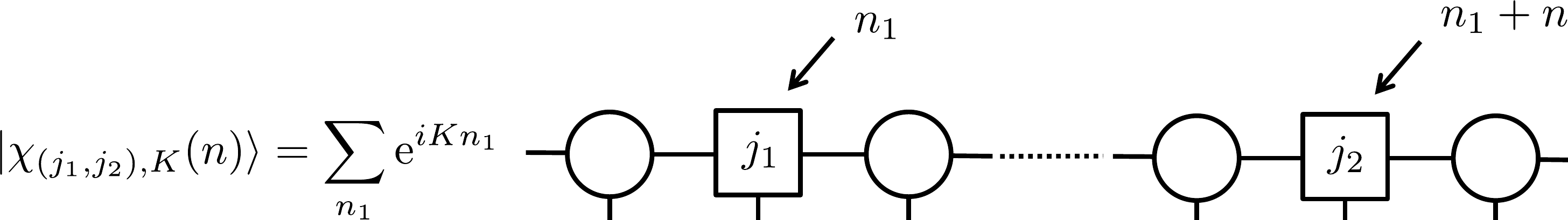}
\caption{Graphical representation of the basis states \eqref{nonLocal}. The ground state is changed at two sites at a distance of $n$ sites and a momentum superposition is taken (with the total momentum $K$).}
\end{figure*}
We collect the variational coefficients either in one half-infinite vector $\vect{C}$ with $C^{j,n} = c^j(n)$ or using the finite vectors $\vect{c}(n)$ with entries $\{c^j(n),j=1,\ldots,L_n\}$ for every $n=0,1,\ldots$. Here, we have $L_0 = (d-1) D^2$ and $L_{n>0} = [(d-1)D^2]^2$. Note that the sum in Eq.\eqref{eq:ansatz} only runs over values $n\geq0$, because a sum over all integers would result in an overcomplete basis.
\par Already at this point, we will reduce the number of variational parameters to keep the problem tractable. The terms with $n=0$ (corresponding to the basis vectors in Eq.~\eqref{local}) are designed to capture the situation where the two particles are close together. No information on how this part should look like is a priori available, so we keep all variational parameters $c^j(0)$, $j=1,\dots,L_0=D^2(d-1)$. The terms with $n>0$ corresponding to the basis vectors in Eq.~\eqref{nonLocal} represent the situation where the particles are separated. We know that, as $n\rightarrow\infty$, the particles decouple and we should obtain a combination of one-particle solutions. With this in mind, we restrict the range of $j_1$ and $j_2$ to the first $\ell$ basis tensors $\{B_{(i)}, i=1,\ldots,\ell\}$, which were chosen so as to capture the momentum dependent solutions of the one-particle problem. Consequently, the number of basis states of Eq.~\eqref{nonLocal} for $n>0$ satisfies $L_n=\ell^2$, which we will henceforth denote as just $L$.
\par This might seem like a big approximation for $n$ small: when the two particles approach the wave functions might begin to deform, so that the $B$ tensors of the one-particle problem no longer apply. Note, however, that the local ($n=0$) and non-local ($n>0$) part are not orthogonal, so that the local part is able to correct for the part of the non-local wave function where the one-particle description is no longer valid.
\par As the state \eqref{eq:ansatz} is linear in its variational parameters $\vect{C}$, optimizing the energy amounts to solving a generalized eigenvalue problem
\begin{equation} \label{eig}
\mathsf{H}_\text{eff} \vect{C} = \omega \mathsf{N}_{\text{eff}} \vect{C}
\end{equation}
with $\omega$ the total energy of the state and
\begin{align}
& (\mathsf{H}_\text{eff})_{n'j',nj} = \bra{\chi_{j',K}(n')} \hat{H} \ket{\chi_{j,K}(n)} \label{Heff} \\
& (\mathsf{N}_\text{eff})_{n'j',nj} = \braket{\chi_{j',K}(n') | \chi_{j,K}(n)} \label{Neff}
\end{align}
two half-infinite matrices. They have a block matrix structure, where the submatrices are labelled by $(n',n)$ and are of size $L_{n'} \times L_n$. The computation of the matrix elements is quite involved and technical, so we refer to the appendix for the explicit formulas.
\par Since the eigenvalue problem is still infinite, it cannot be diagonalized straightforwardly. Since we actually know the possible energies $\omega$ for a scattering state with total momentum $K$, we can also interpret Eq.~\eqref{eig} as an overdetermined system of linear equations for the coefficients $C^{j,n}=c^j(n)$. In the next two sections we will show how to reduce this problem to a finite linear equation.

\subsection{Asymptotic regime}

First we solve the problem in the asymptotic regime, where the two particles are completely decoupled. This regime corresponds to the limit $n',n\to\infty$, where the effective norm and Hamiltonian matrices, consisting of blocks of size $L\times L$, take on a simple form. Indeed, if we properly normalize the basis states, the asymptotic form of the effective norm matrix reduces to the identity, while the effective Hamiltonian matrix is a repeating row of block matrices centred around the diagonal
\begin{equation} \label{Am}
(\mathsf{H}_\text{eff})_{n',n}\rightarrow \mathsf{A}_{n-n'}, \qquad n,n' \rightarrow \infty.
\end{equation}
The blocks decrease exponentially as we go further from the diagonal, so we can, in order to solve the problem, consider them to be zero if $|n-n'|>M$ for some suitably chosen integer $M$. In this approximation, the coefficients $\vect{c}(n)$ obey
\begin{equation} \label{recurrence}
   \sum_{m=-M}^M \mathsf{A}_m \vect{c}(n+m) = \omega \vect{c}(n), \qquad n\rightarrow\infty.
\end{equation}
We can reformulate this as a recurrence relation for the $\vect{c}(n)$ vectors and therefore look for elementary solutions of the form $\vect{c}(n) = \mu^n \vect{v}$. For fixed $\omega$, the solutions $\mu$ and $\vect{v}$ are now determined by the polynomial eigenvalue equation
\begin{equation} \label{polynomial}
   \sum_{m=-M}^M \mathsf{A}_m \mu^m \vect{v} = \omega \vect{v} .
\end{equation}
From the special structure of the blocks $\mathsf{A}_m$ (see Appendix \ref{sec:asymptotic}) and their relation to the one-particle effective Hamiltonian $\mathsf{H}_\text{eff,1p}$, we already know a number of solutions to Eq.~\eqref{polynomial}. Indeed, if we can find $\Gamma$ combinations of two types of particles $(\alpha,\beta)$ with individual momenta $(\kappa_1,\kappa_2)$ such that $K=\kappa_1+\kappa_2$ and $\omega=\Delta_\alpha(\kappa_1)+\Delta_\beta(\kappa_2)$, then the polynomial eigenvalue problem will have $2\Gamma$ solutions $\mu$ on the unit circle. These solutions take the form $\mu=\e^{i\kappa_2}$ and the corresponding eigenvector is given by
\begin{equation} \label{asModes}
 \vect{v} = \vect{u}_{\alpha}(\kappa_1) \otimes \vect{u}_{\beta}(\kappa_2)
\end{equation}
(in the case of degenerate eigenvalues we can take linear combinations of these eigenvectors that no longer have this product structure). Every combination is counted twice, because we can have particle $\alpha$ on the left and particle $\beta$ on the right, and vice versa. 
\par Moreover, since $\mathsf{A}_m\dag=\mathsf{A}_{-m}$, the number of eigenvalues within and outside the unit circle should be equal. This allows for a classification of the eigenvalues $\mu$ as
\begin{align*}
  & \left|\mu_i\right| <1 \qquad \text{for} \qquad  i=1,\dots,LM-\Gamma\\
  & \left|\mu_i\right| =1 \qquad \text{for} \qquad i= LM-\Gamma+1,\dots,LM+\Gamma\\
  & \left|\mu_i\right| >1 \qquad \text{for} \qquad i=LM+\Gamma+1,\dots,2LM.
\end{align*}
The last eigenvalues with modulus bigger than one are not physical (because the corresponding $\vect{c}(n)\sim \mu_i^n \vect{v}_i$ yiels a non-normalizable state) and should be discarded. The $2\Gamma$ eigenvalues with modulus 1 are the oscillating modes discussed above; we will henceforth label them with $\gamma=1,\ldots,2\Gamma$ such that $\mu = \e^{i \kappa_{\gamma}}$ ($\kappa_\gamma$ being the momentum of the particle of the right) and the corresponding eigenvector is given by
\begin{equation*}
 \vect{v}_{\gamma} = \vect{u}_{\alpha_\gamma}(K-\kappa_\gamma) \otimes \vect{u}_{\beta_\gamma}(\kappa_\gamma).
\end{equation*}
Finally, the first eigenvalues are exponentially decreasing and represent corrections when the excitations are close to each other. We henceforth denote them as $e^{-\lambda_i}$ with $\Re(\lambda_i)>0$ for $i=1,\ldots,LM-\Gamma$ and denote the corresponding eigenvectors as $\vect{w}_{i}$.
\par With these solutions, we can represent the general asymptotic solution as
\begin{equation} \label{asymptotic}
   \vect{c}(n) \rightarrow \sum_{i=1}^{LM-\Gamma} p^i \e^{-\lambda_i n} \vect{w}_i + \sum_{\gamma=1}^{2\Gamma} q^\gamma \e^{i\kappa_\gamma n} \vect{v}_\gamma.
\end{equation}
Of course, we still have to determine the coefficients $\{p^i,q^\gamma\}$ by solving the local problem.

\subsection{Solving the full eigenvalue equation}

Since the energy $\omega$ was fixed when constructing the asymptotic solution, the generalized eigenvalue equation is reduced to the linear equation
\begin{equation*}
  (\mathsf{H}_\text{eff}-\omega \mathsf{N}_\text{eff}) \vect{C} = 0.
\end{equation*}
We know that in the asymptotic regime this equation is fulfilled if and only if $\vect{c}(n)$ is of the form of Eq.~\eqref{asymptotic}. We will introduce the approximation that the elements for the effective Hamiltonian matrix [Eq.~\eqref{Heff}] and norm matrix [Eq.~\eqref{Neff}] have reached their asymptotic values when either $n>M+N$ or $n'>M+N$, where $N$ is a finite value and can be chosen sufficiently large. This implies that we can safely insert the asymptotic form for $n>N$ in the wave function, which we can implement by rewriting the wave function as
\begin{equation} \label{Qx}
\vect{C}=\mathsf{Z}\times \vect{x}
\end{equation}
where
\begin{align*}
  \mathsf{Z} = \begin{pmatrix} \one_\text{local}  &  & \\ & \{ \e^{-\lambda_i n}\vect{w}_i \} & \{ \e^{-i\kappa_\gamma n} \vect{v}_{\gamma} \} \end{pmatrix}.
\end{align*}
The $\{ \e^{-\lambda_i n}\vect{w}_i \}$ and  $\{ \e^{-i\kappa_\gamma n} \vect{v}_{\gamma} \}$ are the vectors corresponding to the damped, resp. oscillating modes, while the identity matrix is inserted to leave open all parameters in $\vect{c}(n)$ for $n\leq N$. The number of parameters in $x$ is reduced to the finite value of $D^2(d-1)+NL+LM+\Gamma$.
\par Since the equation is automatically fulfilled after $M+N$ rows, we can reduce $\mathsf{H}_\text{eff}$ and $\mathsf{N}_\text{eff}$ to the first rows, so we end up with the following linear equation
\begin{equation} \label{scatEq}
[\mathsf{H}-\omega \mathsf{N}]_\text{red}\times \mathsf{Z} \times \vect{x} = 0
\end{equation}
with
\begin{equation*}
  \left[\mathsf{H}-\omega \mathsf{N}\right]_\text{red} =
  \left(\begin{array}{c|cccc}
  & 0 & 0 & \dots & 0 \\
  & \vdots & \vdots & \ddots & \vdots \\
  & 0 & 0 & \dots & 0 \\
  (\mathsf{H}-\omega \mathsf{N})_\text{ex} & \mathsf{A}_M & 0 & \dots & 0 \\
   & \mathsf{A}_{M-1} & \mathsf{A}_M & \dots & 0 \\
  & \vdots & \vdots & \ddots & \vdots \\
   & \mathsf{A}_1 & \mathsf{A}_2 & \dots & \mathsf{A}_M  \end{array} \right).
\end{equation*}
This ``effective scattering matrix'' consists of the first $(M+N)\times(M+N)$ blocks of the exact effective Hamiltonian and norm matrix and the $\mathsf{A}$ matrices of the asymptotic part [Eq.~\eqref{Am}] to make sure that these matrices remain the truncated versions of a hermitian problem. This matrix has $D^2(d-1)+(N+M)L$ rows, which implies that the linear equation \eqref{scatEq} has $\Gamma$ exact solutions, which is precisely the number of scattering states we expect to find. Every solution consists of a local part ($D^2(d-1) + NL$ elements), the $LM-\Gamma$ coefficients $\vect{p}$ of the decaying modes and the $2\Gamma$ coefficients $\vect{q}$ of the asymptotic modes.

\subsection{S matrix and normalization}
\label{sec:norm}

After having shown how to find the solutions of the scattering problem, we can now elaborate on the structure of the asymptotic wave function and define the S matrix. 
\par We start from $\Gamma$ linearly independent scattering eigenstates $\ket{\Upsilon_i(K,\omega)}$ ($i=1,\ldots,\Gamma$) at total momentum $K$ and energy $\omega$ with asymptotic coefficients $\vect{q}_i(K,\omega)$. The asymptotic form of these eigenstates is thus a linear combination of all possible non-decaying solutions of the asymptotic problem:
\begin{multline} \label{asForm}
\ket{\Upsilon_i(K,\omega)} = \sum_{\gamma=1}^{2\Gamma} q_i^\gamma(K,\omega) \\ \times \sum_{n>N}\sum_{j} \e^{i\kappa_\gamma n}v_\gamma^j(\kappa_\gamma) \ket{\chi_{j,K}(n)}
\end{multline}
where the coefficients are obtained from solving the local problem. The number of eigenstates equals half the number of oscillating modes that appear in the linear combination. With every oscillating mode $\gamma$ we can associate a function $\omega_\gamma(\kappa)$ giving the energy of this mode as a function of the momentum $\kappa_\gamma$ of the second particle at a fixed total momentum $K$. If $\gamma$ corresponds to the two-particle mode with particles $\alpha_\gamma$ and $\beta_\gamma$, this function is given by $\omega_{\gamma}(\kappa) = \Delta_{\alpha_\gamma}(K-\kappa) + \Delta_{\beta_\gamma}(\kappa)$. The derivative of this function, which will prove of crucial importance, is $\omega'_{\gamma}(\kappa) = \Delta'_{\beta_\gamma}(\kappa) - \Delta'_{\alpha_\gamma}(K-\kappa)$. It can be interpreted as the difference in group velocity between the two particles, i.e. the relative group velocity in the center of mass frame.
\par Much like the proof of conservation of particle current in one-particle quantum mechanics, it can be shown that (see Appendix \ref{proof}), if \eqref{asForm} is to be the asymptotic form of an eigenstate, the coefficients $q_i^\gamma(K,\omega)$ should obey
\begin{equation} \label{conservation}
\sum_{\gamma} \left|q^\gamma_i(K,\omega) \right|^2 \left(\frac{\d\omega_\gamma}{\d\kappa}(\kappa_\gamma)\right) = 0.
\end{equation}
This equation can indeed be read as a form of conservation of particle current, with $\omega_\gamma'(\kappa_\gamma)$ playing the role of the (relative) group velocity of the asymptotic mode $\gamma$. As any linear combination of eigenstates with the same energy $\omega$ is again an eigenstate, this relation can be extended to
\begin{equation*}
\sum_{\gamma}  \overline{q^\gamma_j(K,\omega)} q^\gamma_i(K,\omega) \left(\frac{\d\omega_\gamma}{\d\kappa}(\kappa_\gamma)\right) = 0.
\end{equation*}
With this equation satisfied, we can define the two-particle S matrix $S(K,\omega)$. Firstly, the different modes are classified according to the sign of the derivative: the incoming modes have $\frac{\d\omega}{\d\kappa}>0$ (two particles moving towards each other), the outgoing modes have $\frac{\d\omega}{\d\kappa}<0$ (two particles moving away from each other), so that we have
\begin{multline*}
\sum_{\gamma\in\Gamma_{\text{in}}}  \overline{q_j^\gamma(K,\omega)} q_i^\gamma(K,\omega) \left|\frac{\d\omega_\gamma}{\d\kappa}(\kappa_\gamma)\right| \\ = \sum_{\gamma\in\Gamma_{\text{out}}}  \overline{q_j^\gamma(K,\omega)} q_i^\gamma(K,\omega) \left|\frac{\d\omega_\gamma}{\d\kappa}(\kappa_\gamma)\right|.
\end{multline*}
If we group the coefficients of all solutions in (square) matrices $Q_\text{in}(K,\omega)$ and $Q_\text{out}(K,\omega)$, so that the $i$'th column is a vector with the coefficients $q^\gamma_i$ for the in- and outgoing modes of the $i$'th solution, we can rewrite this equation as
\begin{multline*}
Q_\text{in}(K,\omega)\dag V^2_\text{in}(K,\omega) Q_\text{in}(K,\omega) \\= Q_\text{out}(K,\omega)\dag V^2_\text{out}(K,\omega) Q_\text{out}(K,\omega),
\end{multline*}
with $V_\text{in,out}(K,\omega)_{ij}= \delta_{ij} \left|\frac{\d\omega_\gamma}{\d\kappa}(\kappa_\gamma)\right|^{1/2}$ a diagonal matrix. As $Q_\text{in}(K,\omega)$ and $Q_\text{out}(K,\omega)$ should be connected linearly, we can define a unitary matrix $S(K,\omega)$ as
\begin{multline*}
V_\text{out}(K,\omega) Q_\text{out}(K,\omega) = S(K,\omega) V_\text{in}(K,\omega) Q_\text{in}(K,\omega).
\end{multline*}
In Sec.~\ref{sec:moller} we will show that this definition corresponds to the standard S matrix. Note, however, that $S(K,\omega)$ is only defined up to a set of phases. Indeed, since the vectors $\vect{v}_\gamma$ can only be determined up to a phase, the coefficient matrices $C_\text{in}$ and $C_\text{out}$ are only defined up to a diagonal matrix of phase factors. These arbitrary phase factors show up in the S matrix as well. We will show how to fix them in the case of the elastic scattering of two identical particles (Sec. \ref{sec:oneType}); in the case where we have different outgoing channels only the square of the magnitude of the S matrix elements is physically well-defined (see Sec. \ref{sec:moller}).
\par This formalism allows to calculate the norm of the scattering states in an easy way. Indeed, the general overlap between two scattering states is given by
\begin{widetext}
\begin{align*}
\braket{\Upsilon_{i'}(K',\omega')|\Upsilon_i(K,\omega)} &= 2\pi\delta(K-K') \left( \sum_{\gamma,\gamma'}  \ol{q_{i'}^\gammap(K',\omega')} q_i^\gamma(K,\omega) \vect{v}_{\gamma'}\dag \vect{v}_\gamma \sum_{n,n'>N} \e^{i(\kappa_\gamma-\kappa'_{\gamma'} ) n} + \text{finite} \right) \\
&= 2\pi\delta(K-K') \left( \sum_{\gamma,\gamma'}  \ol{q_{i'}^{\gamma'}(K',\omega')} q_i^\gamma(K,\omega)\vect{v}_{\gamma'}\dag \vect{v}_\gamma \pi\delta(\kappa_\gamma(\omega)-\kappa'_{\gamma'}(\omega')) + \text{finite} \right).
\end{align*}
The $\delta$ factor for the momenta $\kappa_\gamma$ is obviously only satisfied if $\omega=\omega'$, so we can transform this to a $\delta(\omega-\omega')$. Moreover, if $\kappa_\gamma(\omega) = \kappa'_{\gamma'}(\omega')$ for $\gamma\neq\gamma'$, then necessarily $\vect{v}_{\gamma'}\dag \vect{v}_\gamma=0$, so we can reduce the double sum in $\gamma,\gamma'$ to a single one. If we omit all finite parts, we have
\begin{align*}
\braket{\Upsilon_{i'}(K',\omega')|\Upsilon_i(K,\omega)} &= 2\pi\delta(K-K') \pi \delta(\omega-\omega') \sum_{\gamma} \ol{q_{i'}^{\gamma}(K',\omega')} q_i^\gamma(K,\omega) \left|\frac{\d\omega_\gamma}{\d\kappa}(\kappa_\gamma)\right|.
\end{align*}
With the $Q_\text{in/out}$ as defined above the overlap reduces to
\begin{align*}
\braket{\Upsilon_{i'}(K',\omega')|\Upsilon_i(K,\omega)} &= 2\pi\delta(K-K') 2\pi \delta(\omega-\omega') \left[Q_\text{in}(K,\omega)\right]_{i'} \dag V^2_\text{in}(K,\omega) \left[Q_\text{in}(K,\omega)\right]_i \\
&= 2\pi\delta(K-K') 2\pi \delta(\omega-\omega') \left[Q_\text{out}(K,\omega)\right]_{i'} \dag V^2_\text{out}(K,\omega) \left[Q_\text{out}(K,\omega)\right]_i.
\end{align*}
\end{widetext}

\subsection{One type of particle}
\label{sec:oneType}

Let us make things more concrete by working out the case where the one-particle spectrum consists of just one type of particle with dispersion relation $\Delta(\kappa)$. Suppose we have only one combination of momenta $\kappa_1$ and $\kappa_2$ such that they add up to total momentum $K=\kappa_1+\kappa_2$ and total energy $\omega=\Delta(\kappa_1)+\Delta(\kappa_2)$. There are two asymptotic modes -- one mode with $\kappa_1$ on the left and $\kappa_2$ on the right, and one mode with the momenta interchanged -- that combine into one scattering state with the asymptotic form
\begin{equation*}
 \vect{c}(n) \rightarrow q^1 \e^{i\kappa_2 n} \vect{v_1} + q^2 \e^{i\kappa_1n} \vect{v_2}.
\end{equation*}
The conservation equation that was derived in the previous section takes on the simple form
\begin{equation*}
\left|q^1\right|^2 = \left|q^2\right|^2
\end{equation*}
because $\omega'(\kappa_1) = -\omega'(\kappa_2)$. As we mentioned above in the general case, the relative phase of the two vectors $\vect{v_1}$ and $\vect{v_2}$ can be chosen arbitrarily. However, since the two modes correspond to the interchanging of two identical particles, it makes sense to fix the phase such that $\vect{v_2}\dag\vect{v_1}>0$. Due to the momentum dependence of the one-particle solutions, this overlap will be slightly smaller than one.
\par The S matrix reduces to a phase factor and is defined as
\begin{equation*}
S(\kappa_1,\kappa_2) = S(K,\omega) = \frac{q^2}{q^1}.
\end{equation*}
The asymptotic wave function takes the form
\begin{multline} \label{oneType}
\ket{\Upsilon(\kappa_1,\kappa_2)} \rightarrow \sum_{n_1<_2} \e^{i(\kappa_1n_1+\kappa_2n_2)} \left[ \text{$B_{\kappa_1}$ at $n_1$,$B_{\kappa_2}$ at $n_2$} \right] \\ + S(\kappa_1,\kappa_2)  \e^{i(\kappa_2n_1+\kappa_1n_2)} \left[ \text{$B_{\kappa_2}$ at $n_1$,$B_{\kappa_1}$ at $n_2$} \right].
\end{multline}
From simple arguments \cite{Sachdev2011} one can argue that in one dimension the S matrix should have the universal limiting value for low-energy scattering \cite{Sachdev1997,Damle1997}
\begin{equation*}
S(\kappa_1,\kappa_2) \rightarrow -1 \quad \text{as} \quad |\kappa_1-\kappa_2|\rightarrow 0.
\end{equation*}
We define the scattering phase $\theta$ as the phase shift of the S matrix relative to its universal low-energy value $S(\kappa_1,\kappa_2)=-\e^{i\theta(\kappa_1,\kappa_2)}$.

\subsection{Spectral functions}

With the variational wave functions of one- and two-particle states, we can now calculate the low-energy part of spectral functions at zero temperature. We consider the following function
\begin{equation*}
S(\kappa,\omega) = \sum_n \int \d t \, \e^{i(\omega t - \kappa n)} \bra{\Psi_0} O_n\dag(t) O_0(0) \ket{\Psi_0}
\end{equation*}
with $O_n(t)$ an operator at site $n$ in the Heisenberg picture. In order to approximate the low-energy part, we insert a projector on the one- and two-particle subspaces
\begin{multline*}
P_\text{1p,2p} = \int \frac{\d\kappa}{2\pi} \sum_{\alpha\in\Gamma_1(\kappa)} \ket{\Phi_\alpha(\kappa)}\bra{\Phi_\alpha(\kappa)} \\
+ \int \frac{\d K}{2\pi} \int \frac{\d\omega}{2\pi} \sum_{\gamma\in\Gamma_2(K,\omega)}\ket{\Upsilon_\gamma(K,\omega)}\bra{\Upsilon_\gamma(K,\omega)}
\end{multline*}
where $\Gamma_{1}$ ($\Gamma_{2}$) is the set of all types of one-particle (two-particle) states at that momentum (momentum-energy). The states are orthonormalized as
\begin{align*}
 & \braket{\Phi_{\gamma'}(\kappa')|\Phi_\gamma(\kappa)} = 2\pi\delta(\kappa'-\kappa) \delta_{\gamma\gamma'} \\
 & \braket{\Upsilon_{\gamma'}(K',\omega')|\Upsilon_\gamma(K,\omega)} = 4\pi^2\delta(K'-K) \delta(\omega'-\omega) \delta_{\gamma\gamma'}
\end{align*}
so that we obtain the Lehmann representation \cite{Lehmann1954} for the spectral function up to two-particle contributions
\begin{align*}
 S(\kappa,\omega) &= \sum_{\alpha\in\Gamma_1(\kappa)} 2\pi\delta(\Delta_\alpha(\kappa)-\omega) \left|\bra{\Phi_\alpha(\kappa)} \hat{O}_0 \ket{\Psi_0} \right|^2 \\
 & \qquad + \sum_{\gamma\in\Gamma_2(\kappa,\omega)} \left| \bra{\Upsilon_\gamma(\kappa,\omega)}\hat{O}_0 \ket{\Psi_0} \right|^2 \\
 & + ...
\end{align*}
In gapped systems, the one- and two-particle contributions saturate the full spectral function below the three-particle threshold, while contributions from higher-particle excitations might become important for higher energies. Yet, it appears that typically the one- and two-particle sectors already contain the largest portion of the spectral function, see e.g. Ref.~\onlinecite{Caux2008}. The one- and two-particle form factors appearing in the above expression are calculated explicitly in Appendix \ref{sec:form1p}.
\par To get a quantitative estimate of how well the spectral function is approximated, we look at the zeroth and first frequency moment at a certain momentum, which are defined as
\begin{equation*}
s_0(\kappa) = \int \frac{\d \omega}{2\pi} S(\kappa,\omega) \quad \text{and} \quad s_1(\kappa) = \int \frac{\d \omega}{2\pi} \omega S(\kappa,\omega).
\end{equation*}
These quantities follow the sum rules \cite{Hohenberg1974}
\begin{equation*} \begin{split}
s_0(\kappa) &= \int \frac{\d \omega}{2\pi} \bra{\Psi_0} O_{-\kappa}\dag 2\pi\delta(\omega-\hat{H}) O_0(0) \ket{\Psi_0} \\
&= \bra{\Psi_0} O_{-\kappa}\dag O_0(0) \ket{\Psi_0}
\end{split} \end{equation*}
and
\begin{align*}
s_1(\kappa) &= \int \frac{\d \omega}{2\pi} \omega \bra{\Psi_0} O_{-\kappa}\dag 2\pi\delta(\omega-\hat{H}) O_0(0) \ket{\Psi_0} \\
&=  \bra{\Psi_0} O_{-\kappa}\dag \hat{H} O_0(0) \ket{\Psi_0}.
\end{align*}
If the ground state is taken to be an MPS, these quantities can be calculated exactly. Note that $s_0$ is just the static correlation function and that the ratio of the two is equal to the single mode approximation for the dispersion relation \cite{Arovas1992}
\begin{align*}
\Delta_\text{SMA}(\kappa) = \frac{s_1(\kappa)}{s_0(\kappa)} = \frac{\bra{\Psi_0} O_{-\kappa}\dag \hat{H} O_0(0) \ket{\Psi_0}}{\bra{\Psi_0} O_{-\kappa}\dag O_0(0) \ket{\Psi_0}}.
\end{align*}
By comparing the one- and two-particle contributions for $s_0$ and $s_1$ to the exact values, we can get an idea of how well these eigenstates capture the effect of the operators working on the ground state and, consequently, how well the spectral function is approximated by only looking at these contributions.

\section{Two-particle S matrix and approximate Bethe ansatz}
\label{sec:section3}

We now discuss how the variational formulation of scattering theory using matrix product states, as developed in the previous section, relates to standard scattering theory. We then discuss how we can use the information provided by the scattering matrix to build an effective description of the low-energy behaviour of the spin chain using the approximate Bethe ansatz.

\subsection{Stationary scattering states and the S matrix in one dimension}
\label{sec:moller}

In standard scattering theory the S matrix is typically defined from a dynamical point of view: its elements are the overlaps of asymptotically free in and out states with respect to the full time-evolution operator. Although it is a priori not clear that this definition corresponds to the one that was presented in the previous sections, we can show that this is indeed the case.
\par Appendix \ref{sec:mollerA} provides a summary of the standard scattering formalism of single-particle quantum mechanics \cite{Taylor1972}, which we have adapted to the one-dimensional setting with general Hamiltonians (e.g. potentials which are not diagonal in real space) and arbitrary dispersion relations (non-quadratic eigenvalue spectrum of the ``free'' Hamiltonian $\ham_0$). More specifically we have shown how the S matrix elements $f(q_\beta\leftarrow p_\alpha)$ show up in the asymptotic form of the scattering eigenstates $\ket{p_\alpha\pm}$ of the full Hamiltonian $\ham$.
\par To make the connection to the variational scattering states of Sec.~\ref{sec:section2}, we have to make a few modifications. First of all, we can reformulate the two-particle scattering problem as a one-particle problem by factoring out the conservation of total momentum and only focus on the matrix elements between different relative momenta. At every value of the total momentum, we can define relative momentum states $\ket{p_\gamma}$ with dispersions $\omega(p_\gamma)$, which are solutions of the free Hamiltonian $\hat{H}_0$. This free Hamiltonian corresponds to the effective two-particle Hamiltonian matrix in the asymptotic regime \eqref{Am} and the states $\ket{p_\gamma}$ are the asymptotic modes \eqref{asModes}.
\par Secondly, our ``one-particle'' Hilbert space is only defined on a half-infinite line, because the particles are essentially bosonic. The way around this consists of artificially assigning particle labels and distinguishing the situation where particle 1 (2) is on the left (right), and the opposite situation; the relative coordinate $n=n_2-n_1$ now ranges over the positive and negative integers. Alternatively, one could add to the free Hamiltonian $\ham_0$ a potential $\hat{V}$ which is infinite everywhere on the negative real line, making this a forbidden region. Scattering theory would still work (existence of the M\"{o}ller operators etc.), provided that we restrict the ``in'' states to momenta for which $\frac{\d \omega}{\d p}(p) <0$ and the out states to momenta for which $\frac{\d \omega}{\d p}(p)>0$. This corresponds exactly to how we defined the incoming and outgoing modes in Sec.~\ref{sec:norm}.
\par Translating the expression for the asymptotic wave function of the scattering states $\ket{p_\alpha+}$ to the framework of Sec.~\ref{sec:norm} amounts to the following form for the wave function $\vect{c}_\alpha(n)$
\begin{multline*}
\vect{c}_\alpha(n) \rightarrow \left| \frac{\d\omega}{\d\kappa}(\kappa_\alpha) \right|^{-1/2} \vect{v}_\alpha e^{i p_\alpha n} \\
+ \sum_{\gamma\in A^{+}(\kappa_\alpha)} f(\kappa_\gamma\leftarrow\kappa_\alpha) \left| \frac{\d\omega}{\d\kappa}(\kappa_\gamma) \right|^{-1/2} \vect{v}_\gamma e^{i\kappa_\gamma n}
\end{multline*}
for every incoming mode $\alpha=1,\dots,\Gamma$. In this representation, we choose one incoming mode $\alpha$ that couples only to all outgoing modes $\{\gamma\in A^{+}(\kappa_\alpha)\}$. The coefficient matrix for the incoming modes that was defined earlier takes on the form
\begin{equation*}
(Q_\text{in})_{\gamma,\alpha} = \delta_{\gamma\alpha} \left| \frac{\d\omega}{\d\kappa}(\kappa_\alpha) \right|^{-1/2}
\end{equation*}
while the coefficients for the outgoing modes are given by
\begin{equation*}
 (Q_\text{out})_{\gamma,\alpha} = \left| \frac{\d\omega}{\d\kappa}(\kappa_\gamma) \right|^{-1/2} f(\kappa_\gamma\leftarrow\kappa_\alpha).
\end{equation*}
The S matrix $S(K,\omega)$ that was defined takes on the simple form (as $V_\text{in}Q_\text{in}=\one$ in this representation)
\begin{align*}
S(K,\omega) &= V_\text{out}Q_\text{out} \\
&= f(\kappa_\gamma\leftarrow \kappa_\alpha).
\end{align*}
Through this identification the unitariness of the S matrix $S(K,\omega)$ that was proven in the previous section is indeed equivalent to the unitary S matrix defined through the M{\o}ller operators as $S=\Omega_-\dag\Omega_+$.

\subsection{Scattering length and bound states}
\label{sec:scatBound}

Suppose we have the scattering process of two identical particles in the limit of vanishing relative momentum. We expect that the equation for the relative wave function $\psi(x)$ should obey the zero energy and zero potential Schr\"odinger equation
\begin{equation*}
 \frac{\d^2\psi(x)}{\d x^2} = 0
\end{equation*}
in the region $x>x_0$ where $x_0$ is the length of the interaction. The solutions are of the form $\psi(x) \propto x - a$ for large $x$ which matches the asymptotic form of Sec.~\ref{sec:oneType} if the phase of the S matrix reduces to
\begin{equation} \label{linearTheta}
\theta(\kappa_1,\kappa_2) \approx - a (\kappa_1-\kappa_2) 
\end{equation}
in the limit for $\kappa_1-\kappa_2\rightarrow0$. The slope $a$ will be called the scattering length and still depends on the total momentum $\kappa_1+\kappa_2$.
\par Suppose now the existence of a bound state with very low binding energy $-\epsilon$. The wave function of this bound state should look like $\psi(x)=\e^{-\kappa x}\approx 1-\kappa x$ with $\omega(i\kappa)= -\epsilon \rightarrow0$. If we want the formation of this bound state to follow smoothly from a scattering state with vanishing energy, the scattering length should diverge. This means that the formation of a bound state out of a scattering continuum at a certain momentum should be accompanied by a diverging scattering length.

\subsection{Approximate Bethe Ansatz}
\label{sec:aba}

In this section we will develop a method to describe a finite density of excitations based on the coordinate Bethe ansatz. For simplicity, we will for the remainder of this section restrict to the case of one type of particle -- making the consistency conditions for factorized scattering (Yang-Baxter equation) trivial -- but the framework can be extended to multicomponent situations \cite{Sutherland2004}. We will interpret the strongly correlated MPS ground state as a vacuum state on which we can build $N$-particle states, described by a $N$-particle wave function $\Psi(x_1,\dots,x_N)$.  Although in general we have no particle conservation in the system, we will argue that the first-quantized approach gives a good approximation at low densities. Indeed, particle-number violating processes involve three or more particles and can be neglected at low densities. In Sec.~\ref{sec:section5} we will discuss how to develop a second-quantization approach.
\par We start with one particle. We can link the one-particle excitation $\ket{\Phi_\kappa[B]}$ with dispersion $\Delta(\kappa)$ in an obvious way with a one-particle wave function $\Psi_1(x)$ in first quantization as 
\begin{equation*}
\Psi_1(x) = \e^{i\kappa x}.
\end{equation*}
Adding a second particle can be done by only taking account of the asymptotic part of the two-particle wave function [Eq.~\eqref{oneType}] ($x_1<x_2$)
\begin{align*}
\Psi_2(x_1,x_2) = \e^{i(\kappa_1x_1 + \kappa_2x_2)} + S(\kappa_1,\kappa_2) \e^{i(\kappa_2x_1 + \kappa_1x_2)}.
\end{align*}
As we are working with identical particles, the wave function in the other sector ($x_1>x_2$) has to be determined from the statistics of the particles. On the level of the spin system, the addition of a particle is a local operation, so we will work with bosonic many-particle wave functions.
\par The addition of a third particle can only be done approximately. Indeed, a three-particle wave function has the general form \cite{Sutherland2004}
\begin{align} 
& \Psi_3(x_1,x_2,x_3) = \e^{i(\kappa_1x_1 + \kappa_2x_2 + \kappa_3x_3)} \nonumber\\
& \qquad + S(\kappa_1,\kappa_2) \e^{i(\kappa_2x_1 + \kappa_1x_2+\kappa_3x_3)} \nonumber \\
& \qquad + \dots \nonumber \\
& \qquad + \iiint \d\kappa_1'\d\kappa_2'\d\kappa_3' \; S(\kappa_1'\kappa_2'\kappa_3')\; \e^{i(\kappa_1'x_1 + \kappa_2'x_2 + \kappa_3'x_3)} \nonumber \\
& \qquad + \text{other particle numbers} .\label{three} 
\end{align}
The first terms represent a sum over all six permutations of the three momenta, with the S matrices for all possible two-particle scattering processes as prefactors. The next term is the diffractive part, which accounts for the three-particle scattering. For these scattering processes, the two conservation laws are not enough to preserve individual momenta and we can generate a whole continuum of other momenta. The last term accounts for the non-particle preserving scattering processes, which can generate two- or four-particle states as well. As a result, it is no longer possible to assign a set of individual momenta $\{\kappa_1,\kappa_2,\kappa_3\}$ (or even a particle number) to this wave function, because they are completely mixed up with all other possible sets of momenta that are compatible with conservation of total energy and momentum.
\par The crucial approximation of our approach is that we neglect the two last terms in Eq.~\eqref{three}: every many-particle scattering event can be decomposed into two-particle scatterings that preserve particle number and individual momenta. This implies that three-particle eigenstates can be labeled by three individual momenta and that the three-particle wave function is given by the permutation terms only. The absence of diffractive scattering is the hallmark of integrability \cite{Sutherland2004}, so we are essentially assuming that our many-particle system is integrable \cite{Krauth1991, Kiwata1994a, Okunishi1999a}.
\par If this approximation proves to be valid, we can apply the Bethe ansatz machinery\cite{Bethe1931,Sutherland2004,Korepin1997}. The first-quantized wave function of an integrable $N$-particle system, unambiguously defined by a set of momenta $\{\lambda_1,\dots,\lambda_M\}$, is a sum of plane waves with all possible permutations of the momenta
\begin{equation} \label{bethe}
\Psi(x_1,\dots,x_N) = \sum_{\perm} A(\perm) \e^{i(\lambda_{\perm 1}x_1 + \dots+\lambda_{\perm N}x_N)} 
\end{equation}
where $A(\perm)/A(\perm') = S(\lambda_i,\lambda_j)$ if the permutations $\perm$ and $\perm'$ differ by the interchange of the momenta $\lambda_i$ and $\lambda_j$. 
\par By imposing periodic boundary conditions on the Bethe wave function in the thermodynamic limit, we arrive at a description of the ground state as a Fermi sea of ``pseudo-momenta'' filled up to a certain Fermi level $q$. In contrast to the free-fermion case, the density of occupied modes is not constant but given by the function $\rho(\lambda)$ such that $\rho(\lambda)=0$ for $|\lambda|>q$. The energy of the modes $\epsilon(\lambda)$ can be determined from the integral equation
\begin{equation} \label{liebEq1}
\epsilon(\lambda) -\frac{1}{2\pi} \int_{-q}^{q} K(\lambda,\mu) \epsilon(\mu) \d\mu = \epsilon_0(\lambda)
\end{equation}
where $\epsilon_0(\lambda)$ is the ``bare energy'' of the particle, i.e. the energy an isolated particle with momentum $\lambda$ would have in an infinite system. The kernel of the integral equation is given by the derivative of the scattering phase $K(\lambda,\mu)=\partial_\lambda\theta(\lambda,\mu)$. The value of the Fermi level is computed self-consistently from this equation and the requirement that $\epsilon(\pm q)=0$. Once $q$ has been determined, the density $\rho(\lambda)$ is the solution of a similar integral equation \cite{Lieb1963a, *Lieb1963b}
\begin{equation} \label{liebEq2}
\rho(\lambda) -\frac{1}{2\pi} \int_{-q}^{q} K(\lambda,\mu) \rho(\mu) \d\mu = \frac{1}{2\pi}.
\end{equation}
The total density and energy density are given by
\begin{align} \label{density}
D = \int_{-q}^{q} \rho(\lambda)\d\lambda \qquad \text{and} \qquad E = \frac{1}{2\pi} \int_{-q}^q \epsilon(\lambda)\d\lambda .
\end{align}
\par
\begin{figure}
\subfigure[]{\includegraphics[width=0.7\columnwidth]{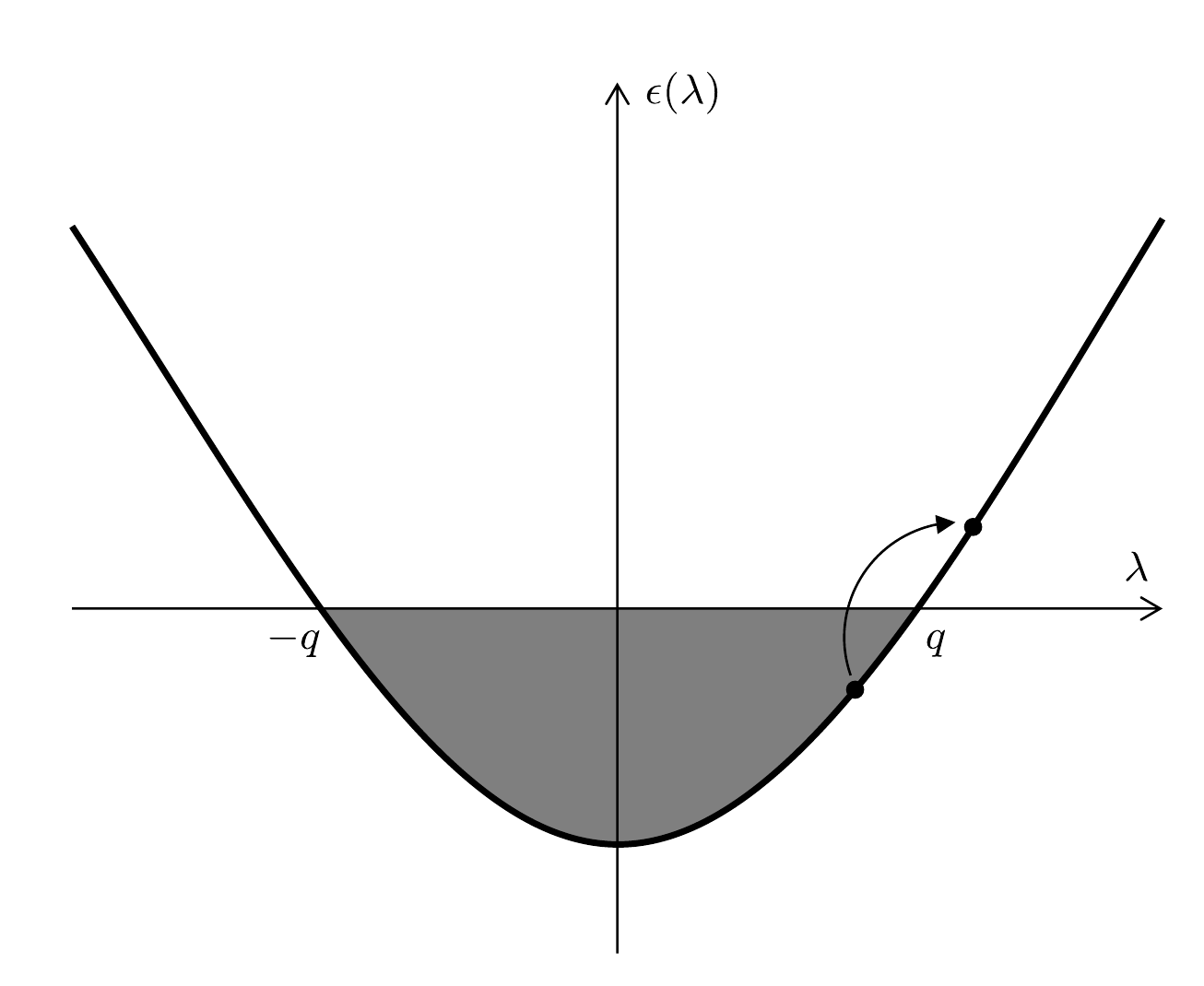}}
\subfigure[]{\includegraphics[width=0.7\columnwidth]{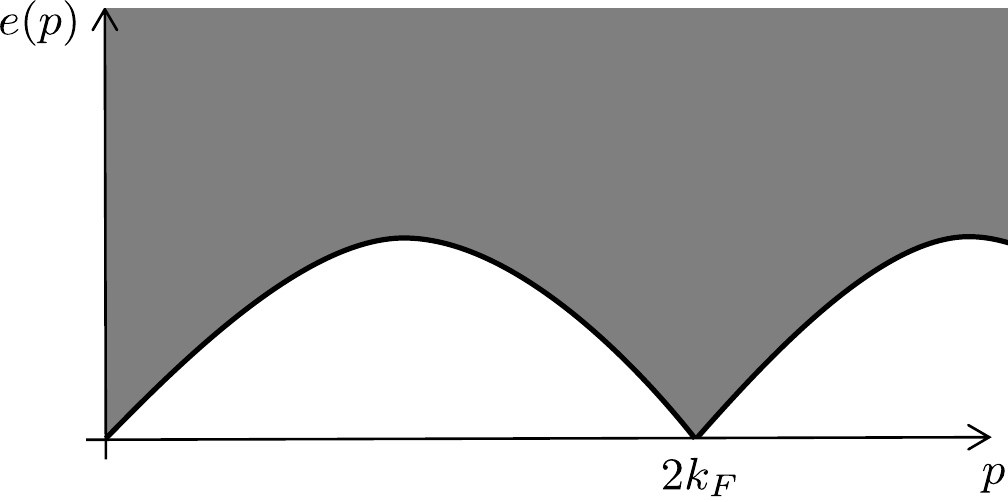}\label{fig:sea}}
\caption{(a) The Fermi sea of pseudo momenta, filled up to the Fermi level $q$. Physical excitations can be pictured as particle-hole excitations close to the Fermi-level. (b) The physical excitation spectrum, the grey area represents a continuum of states. Because of the fact that physical excitations always come in pairs, the spectrum has its minima at momentum 0 and $2k_F$. The slope of the dispersion relation at these momenta is the Fermi velocity $u$.}
\end{figure}

 The excitation spectrum is easily characterized in terms of the pseudo-particles of the Bethe ansatz. We can construct two types of elementary excitations: either we take one particle with momentum $|\lambda|<q$ out of the Fermi sea (hole excitation) or we add one particle with momentum $|\lambda|>q$ (particle excitation). These elementary particle and hole excitations have a topological nature \cite{Korepin1997}, so that the physical excitations -- the ones having a finite overlap with a local operator -- consist of an even number of particles and holes \cite{Lieb1963b}. This gives rise to the physical excitation spectrum as shown in Fig.~\ref{fig:sea}.
\par This critical one-dimensional bose gas can be described as a Luttinger liquid (LL) \cite{*Haldane1981b, Giamarchi2004}. A first important quantity is the Fermi momentum $k_F$, the physical momentum of the gapless particle and hole excitations. It is given by the dressed momentum of the Fermi level and is directly related to the density as (see appendix)
\begin{align} \label{eq:luttingerTheorem}
k_F = \pi D.
\end{align}
Since we have gapless excitations at $0$ and $\pm 2k_F$, correlation functions will have their oscillation periods at these values. The slope of the dispersion relation is the Fermi velocity $u$ and can be calculated from the Bethe ansatz. The third important characteristic quantity is the LL parameter $K$ which determines the power-law decay of correlation functions. In order to calculate it, we define the function $S_R(\lambda)$ as ($h$ is a chemical potential for the particles)
\begin{equation*}
S_R(\lambda) = -  \frac{\partial \epsilon(\lambda)}{\partial h}
\end{equation*}
which (from Eq.~\eqref{liebEq1}) follows the integral equation
\begin{equation}
S_R(\lambda) - \frac{1}{2\pi} \int_{-q}^q K(\lambda,\mu) S_R(\mu) \d\mu = 1.
\end{equation}
In the context of a dilute gas of magnons (see Sec.~\ref{sec:magnProcess}) $S_R(q)$ can be interpreted as the renormalized spin of the magnon close to the Fermi surface. With the low-energy excitations just above the Fermi sea behaving as free fermions \cite{Lesage1997} (i.e. their S matrix is -1), one can show that the LL parameter $K$ is related to $S_R(q)$ as \cite{Konik2002}
\begin{equation} \label{eq:LuttK}
K = S_R(q)^2.
\end{equation}
By thus making the connection between the approximate Bethe ansatz and the LL description, we can infer information on the critical correlations in a system where a finite density of excitations forms on top of a strongly-correlated vacuum state. More specifically, we can infer the long-range behaviour of one-particle and pair correlation functions as \cite{Haldane1981a, Cazalilla2011a}
\begin{equation} \label{eq:correlators} \begin{split} 
& g_1(x) = A_0 \frac{1}{x^{1/2K}} - A_1 \frac{\cos(2\pi Dx)}{x^{2K+1/2K}} + \dots \\
& D_2(x) = D^2 - \frac{K}{2\pi^2x^2} + B_1 \frac{\cos(2\pi Dx)}{x^{2K}} + \dots
\end{split} \end{equation}
where $D$ is the density, $A_0$, $A_1$, and $B_1$ are non-universal constants and the dots denote higher order terms. Depending on whether the operator targets a particle or a pair, the corresponding correlation functions will decay according to one of these two forms.

\subsection{Limiting cases}
\label{sec:limiting}

The Bethe ansatz equations of the previous section can be greatly simplified if we assume that we work at very low densities. Indeed, assuming that only the lowest pseudo-momentum states are occupied, we can approximate the full dispersion relation by its quadratic form $\epsilon_0(\lambda)\approx c\lambda^2-h$, and the full two-particle S matrix by its limiting value of $S(\theta,\mu)\approx-1$. With the kernel of the integral equation zero, we find easily the density and the (physical) Fermi momentum
\begin{align*}
D=\frac{1}{\pi}\sqrt\frac{h}{c}, \qquad k_F = \sqrt\frac{h}{c}
\end{align*}
and the LL parameters
\begin{align*}
u = 2\pi c D, \qquad  K =  1.
\end{align*}
\par Upon increasing the density, the limiting value of the S matrix will no longer apply. From Sec.~\ref{sec:scatBound} we know that the first order correction to the scattering phase is given by the scattering length, so we can insert the form \eqref{linearTheta} into the Bethe equations, while still assuming a quadratic dispersion relation. The first order correction to the Fermi level is linear in the scattering length
\begin{equation}
q = q_\text{FF} + \delta q = q_\text{FF} - \frac{ah}{3\pi c},
\end{equation}
so that the correction to the density is given by
\begin{align} \label{densityScatLength}
D = \frac{1}{\pi} \sqrt{\frac{h}{c}} - \frac{4ha}{3\pi^2c} + \mathcal{O}(a^2).
\end{align}
This result coincides with the one in Ref.~\onlinecite{Lou2000}. The LL parameters in first order in $a$ are given by \cite{Affleck2005, Affleck2004}
\begin{align*}
 u = 2 c \sqrt{\frac{h}{c}} + \frac{4ah}{3\pi} + \mathcal{O}(a^2)
\end{align*}
and 
\begin{align*}
K = 1 - 2 a D + \mathcal{O}(a^2).
\end{align*}

\subsection{Thermodynamic Bethe ansatz}

At zero temperature, the coordinate Bethe ansatz describes an integrable system in its ground state by filling up a Fermi sea of quasi-momentum states; its excitations are holes and particles above this Fermi sea. When a finite temperature $T$ is applied, these particles and holes will have finite distribution densities. By associating an entropy to these distributions and minimizing the free energy, one arrives at the celebrated Yang-Yang equation \cite{Yang1969}
\begin{multline*}
\epsilon(\lambda) = \epsilon_0(\lambda) \\ -\frac{T}{2\pi} \int_{-\infty}^{+\infty} K(\lambda,\mu) \ln\left( 1+\e^{-\epsilon(\mu)/T} \right) \d\mu,
\end{multline*}
a non-linear integral equation for the dressed energy $\epsilon(\lambda)$ of the quasi momentum states; the equation can be solved by iteration \cite{Takahashi2005}. The density of occupied vacancies $\rho(\lambda)$ is given by
\begin{equation*}
\theta(\lambda)=\frac{\rho(\lambda)}{\rho_v(\lambda)} = \frac{1}{1+\e^{\epsilon(\lambda)/T}}
\end{equation*}
with $\rho_v(\lambda)$ the density of all (occupied and empty) vacancies. Through this equation the density of occupied vacancies satisfies the integral equation
\begin{equation}
\rho(\lambda) = \frac{\theta(\lambda)}{2\pi}  \left( 1+ \int_{-\infty}^{+\infty} K(\lambda,\mu) \rho(\mu)\d\mu  \right),
\end{equation}
such that the total density can be calculated as
\begin{equation}
 D = \int_{-\infty}^{+\infty} \rho(\lambda) \d\lambda.
\end{equation}

\subsection{Effective integrable field theories}

Another way of dealing with a finite density of excitations, based on information on the one-particle dispersion and the two-particle S matrix, consists of mapping the system to an effective integrable field theory. The parameters in this effective theory should be tuned to fit the variational information as good as possible. This approach has the advantage that integrability is exact for the effective model, but the mapping is typically only valid in some small region (e.g. low density and/or low temperature).
\par One possible field theory is obtained by making the approximation that the particles interact through a contact potential \cite{Okunishi1999, Okunishi1999a}, so that we end up with a Lieb-Liniger model \cite{Lieb1963a, *Lieb1963b}. The first-quantized Hamiltonian for a collection of $N$ bosons is given by
\begin{equation}
 H = -\frac{1}{2m} \sum_{j=1}^N \frac{\partial^2}{\partial x_j^2} + 2c \sum_{j<k=1}^N \delta(x_j-x_k)
\end{equation}
with the mass $m$ of the bosons and the interaction strength $c$ as the two tunable parameters. The two-boson S matrix is given by $S(\lambda_1,\lambda_2) = -\e^{i\theta(\lambda_1-\lambda_2)}$ with
\begin{equation}
 \theta(\lambda) = 2\arctan\left(\frac{\lambda}{c}\right),
\end{equation}
so that the scattering length for a $\delta$ potential is $a_\delta=-2/c$. The boson dispersion relation is just quadratic, i.e. $\Delta(\lambda) = \lambda^2/(2m)$. By variationally calculating the dispersion relation and the scattering length of the relevant excitations, we can fix the two parameters and map the density of excitations to a Lieb-Liniger model. At low densities, we expect that this mapping is quantitatively correct.
\par Another possibility is the non-linear sigma model, which has proven to capture the qualitative behaviour of Haldane-gapped spin chains such as the spin-1 Heisenberg model \cite{Haldane1983a} or two-leg spin-1/2 ladders. In contrast to the Lieb-Liniger model, however, we can not tune any parameters to fit the exactly known \cite{Zamolodchikov1979} two-particle S matrix. The universal behaviour of e.g. the magnon condensation of a gapped spin chain in a magnetic field \cite{Konik2002}, can nonetheless be captured with this mapping.

\section{Application to spin ladders}
\label{sec:section4}

\begin{figure}
\includegraphics[width=0.7\columnwidth]{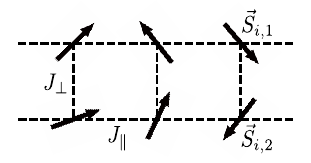}
\caption{The ladder geometry with $J_\parallel$ and $J_\perp$ the couplings along the leg, resp. rung. We will always put $J_\parallel=1$ and define the coupling ratio $\gamma=J_\perp/J_\parallel$.}
\label{fig:ladder}
\end{figure}

We will study the spin-1/2 Heisenberg antiferromagnetic (HAF) two-leg ladder in a magnetic field, defined by the Hamiltonian
\begin{equation}
H = \sum_{i,l} \vec{S}_{i,l} \cdot \vec{S}_{i+1,l} + \gamma \sum_{i} \vec{S}_{i,1} \cdot \vec{S}_{i,2} - h \sum_{i,l}S^z_{i,l}
\end{equation}
where $l=1,2$ denote the two legs of the ladder and $\vec{S}_{i,l}$ denotes the spin operator at site $i$ in the $l$'th leg (see Fig.~\ref{fig:ladder}).
\par The two-leg HAF ladder and its excitation spectrum have been studied intensively for many reasons. First of all, it is the first step to study the transition from one-dimensional systems to higher-dimensional versions.  Secondly, the excitation spectrum has a lot of interesting features, such as the presence of a gap \cite{Haldane1983} and the existence of bound states, and can be studied with a variety of methods depending on the parameter regime. These features can also be observed experimentally \cite{Masuda2006, Notbohm2007, Shapiro2007, Schmidiger2013a}, so that ladders provide an ideal test for these theoretical methods \cite{Bouillot2011, Schmidiger2013}. Finally, the experimental realization of magnetized spin ladders provides an ideal quantitative test of the Luttinger liquid model \cite{Giamarchi2008,Klanjsek2008,Ruegg2008}.
\par In this section we will test our variational method on the two-leg ladder. An MPS approximation for the ground state can be found by first blocking two spins on a rung into one four-level system and applying an MPS optimization algorithm (we have used the TDVP algorithm \cite{Haegeman2011d,Haegeman2014a}). In this representation (to every rung there corresponds one MPS tensor $A$) we find a ground state that is invariant under translations over one site in the leg direction; all momenta in the following subsections are defined with respect to this translation operator. The Hamiltonian and the ground state are invariant under the reflection operator $\mathcal{P}$ which flips the two legs of the ladder. We impose no additional symmetries (e.g. SU(2) invariance) on the MPS, but our variational solution will of course have the right symmetries to high precision.
\par In the first three subsections we will investigate the low-lying spectrum of the ladder without magnetic field. In the following two subsections we will apply the approximate Bethe ansatz to the magnetization process, at zero and finite temperature.

\subsection{One-particle excitations: elementary spectrum and bound states}
\label{sec:ladder1p}

The nature of the elementary excitations in the ladder can be understood starting from different limits. 
\par At zero coupling ($\gamma\rightarrow0$), we have two independent spin-1/2 Heisenberg chains where the elementary excitations are spinons (carrying spin 1/2). These spinons are topologically non-trivial excitations and can only be created in pairs by the action of a local operator. Upon coupling the chains, the spinons are confined into magnons carrying integer spin. This picture has been studied with bosonization techniques \cite{Shelton1996}, showing that the interchain coupling opens up a gap to a triplet of massive magnons (triplons) and a higher-up singlet.
\par At infinite coupling ($\gamma\rightarrow\infty$) we have a collection of independent rungs with antiferromagnetic interaction. In the ground state all rungs are in a singlet state and an elementary excitation is constructed by promoting one rung to a triplet state. When the leg coupling $J_\perp$ is turned on, this triplet obtains a kinetic energy and we get a non-trivial dispersion. This qualitative picture survives for intermediate couplings: through perturbative continuous unitary transformations an effective particle picture can be established and very accurate results on e.g. the elementary dispersion relation and bound states can be obtained \cite{Trebst2000, *Knetter2001, *Schmidt2001, *Knetter2003, *Coester2014, Schmidt2005}.
\par
\begin{figure} 
\centering 
\includegraphics[width=\columnwidth]{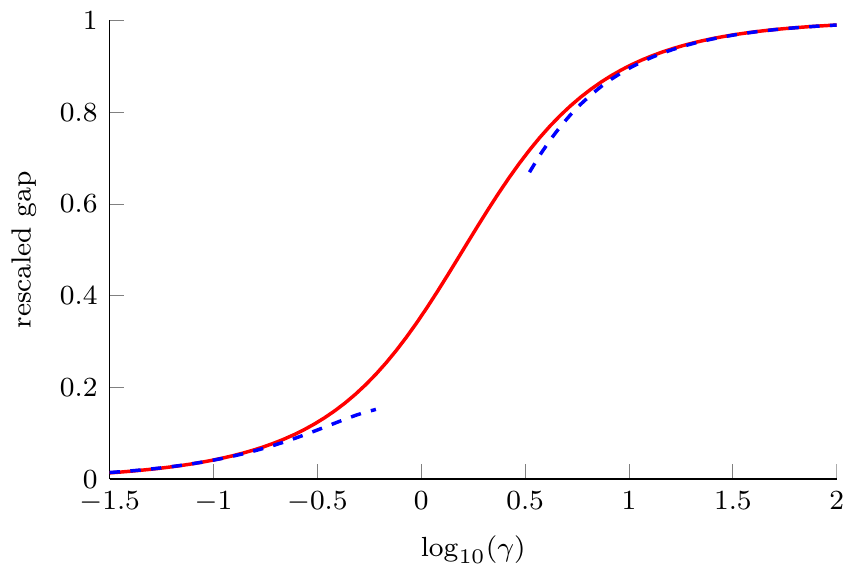}
\caption{The rescaled gap $\frac{\Delta}{\sqrt{1+\gamma^2}}$ in function of the interchain coupling $\gamma$. The blue dashed lines are the first order correction from the strong-coupling limit ($\gamma\rightarrow\infty$) and results from bosonization in the weak-coupling limit ($\gamma\rightarrow0$) \cite{Shelton1996, Larochelle2004}.} 
\label{fig:gap} 
\end{figure}

 In Fig.~\ref{fig:gap} we have plotted the gap in function of the interchain coupling. One can observe that the gap goes to zero in the weak-coupling limit, while it grows to the constant value that one expects from a strong-coupling expansion. Our variational results smoothly interpolates between these two limits.
\par
\begin{figure}
\centering
\includegraphics[width=\columnwidth]{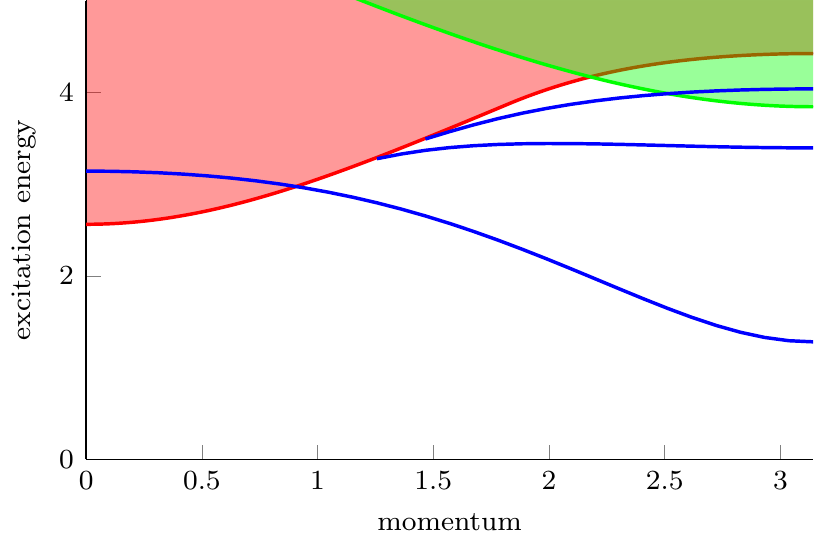}
\caption{The one-particle spectrum consists of a triplet (magnon) which is stable over the whole Brillouin zone (lowest lying blue curve), an singlet (bound state) which is stable for momenta between $\kappa_\text{BS1}\approx0.39\pi$ and $\pi$ (second blue curve), and a triplet (bound state) which is stable for momenta between $\kappa_\text{BS2}\approx0.46\pi$ and $\pi$ (third blue curve). Note that the determination of $\kappa_\text{BS1}$ and $\kappa_\text{BS2}$ is not very precise because the one-particle ansatz is not accurate near the transition. The red region is the two-magnon continuum, the green region is the three-magnon continuum; the other continua (e.g. triplet-singlet continuum) are not shown. }
\label{fig:spectrum}
\end{figure}

\begin{figure}
\includegraphics[width=\columnwidth]{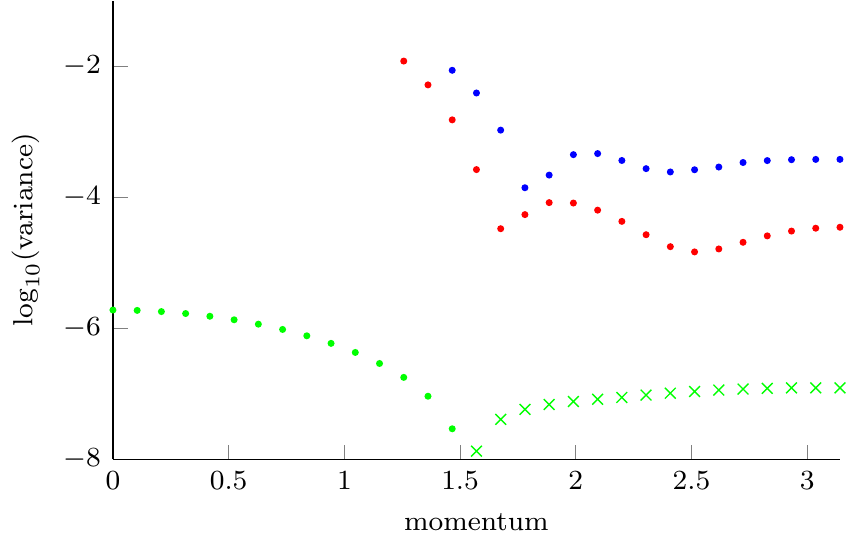}
\caption{The ($\log_{10}$ of the modulus of the) variance of the one-particle excitations; dots, resp. crosses are positive, resp. negative variances (see Appendix \ref{sec:variance} for the meaning of a negative variance). The magnon (green) is clearly a well-defined particle excitation in the whole Brillouin zone. The singlet (red) and triplet (blue) get larger variances as they come closer to the two-particle band, until they actually dive in and are no longer stable. Calculations were done at $\gamma=2$ with bond dimension $D=30$; the ground state variance density is $2.27\times10^{-8}$ at that bond dimension.} 
\label{fig:variance}
\end{figure}

 A typical excitation spectrum in the intermediate region ($\gamma=2$) is shown in Fig.~\ref{fig:spectrum}. The lowest-energy state is an elementary triplet excitation (magnon) with a minimum at momentum $\pi$. The magnon has odd parity under the reflection operator $\mathcal{P}$. The lowest-energy state around momentum zero is a two-magnon scattering state and has even parity. Because the one- and two-magnon state have different parity, the elementary magnon cannot decay and is stable in the whole Brillouin zone. From Fig.~\ref{fig:variance}, where we have plotted the variance of the excitation ansatz, we can indeed see that the magnon is a bona fide particle excitation for all momenta. Note that under a parity-breaking interaction the stability of the magnon inside the continuum breaks down \cite{Fischer2010} and it might prove an interesting question whether we can capture its decay within our framework.
\par The elementary excitation spectrum at $\gamma=2$ has two more elementary particle excitations, a singlet and a triplet, which are stable in a limited region around momentum $\pi$. Both are even under the parity operator $\mathcal{P}$. From the strong-coupling expansion, we can interpret them as two-magnon bound states \cite{Trebst2000}, hence the even parity (without a well-defined particle number, we cannot make this interpretation, so we regard these branches as elementary particles). The variance of the bound states is sufficiently small in the stable region, but it grows larger as the momentum approaches the continuum. From Ref.~\onlinecite{Haegeman2013a} we know that the localized nature of an elementary excitation is related to the gap below and above the excitation branch, so we expect the bound state to become wider as the gap to the continuum closes. This explains the increasing variance of the bound states in Fig.~\ref{fig:variance}. Upon entering the continuum, the bound state has become completely delocalized and no longer exists as a stationary eigenstate of the Hamiltonian.
\par As a last illustration of the one-particle ansatz we have included Table~\ref{table:variance} with excitation energies and variances in the weak-coupling region, showing the elementary triplet and singlet excitations that we expect from a bosonization calculation. We observe that the variances are some orders of magnitude larger in this weak-coupling region. Since the gaps above and below these excitations are a lot smaller at small $\gamma$, this is not unexpected. Note that both the energies and the variances have the right degeneracies, even though we never imposed the corresponding symmetries explicitly.
\begin{table}
\centering
\begin{tabular}{|c|c|c|c|}
\hline
 energy & variance  \\
 \hline
 0.081841224772803 & -0.000178252361115 \\
 \hline
 0.081841224779434 & -0.000178252351941 \\
 \hline
 0.081841224792513 & -0.000178252347304 \\
 \hline
 0.331378942771407 & 0.000337897356458 \\
 \hline
 0.367322866763615 & 0.029803975299627 \\
 \hline
 0.410460620351393 & 0.044970779553592 \\
 \hline 
 \dots & \dots \\
 \hline
 0.513408977989184 & 0.014052233372105 \\
 \hline
 0.513408978649963 & 0.014052233100514 \\
 \hline
 0.513408978939573 & 0.014052232922150 \\
 \hline 
 \dots & \dots \\
 \hline
\end{tabular}
\caption{Excitation energy and variance of the first 6 solutions of the one-particle problem for the HAF ($\gamma=0.2$) at momentum $\pi$ with bond dimension $D=108$. The variance density of the ground state is $9.28.10^{-6}$. The first triplet has negative variance, which shows that this excitation is closer to an exact eigenstate locally than the ground state (see Appendix \ref{sec:variance}). The fourth solution is also a true one-particle (singlet) excitation. All other solutions have a considerably larger variance and correspond to artificial two-particle states. Further up in the continuum, however, we have another triplet with quite small variance, although it is difficult to say whether this corresponds to a true bound state.}
\label{table:variance}
\end{table}

\subsection{Two-particle S matrix}

In this section we will look at the two-magnon S matrix; the scattering of, e.g., an elementary magnon with a bound state will not be considered. The S matrix was defined in Secs.~\ref{sec:norm} and \ref{sec:moller}; in our setting we have three types of particles (the three components of the magnon triplet) and they all have the same dispersion relation. This implies that, for every combination of total momentum $K$ and total energy $\omega$ within the two-magnon continuum, we can build 9 scattering states. The relative coefficients of the asymptotic modes in these scattering states give rise to a ($9\times9$) unitary S matrix (the group velocities will factor out, as all particles have the same dispersion). Furthermore, instead of labeling these scattering states with momentum and energy $(K,\omega)$, we can equally well label them with total and relative momentum $(K,\kappa_1-\kappa_2)$ where $\kappa_1$ and $\kappa_2$ are the two momenta that show up in the asymptotic modes (there is still an ambiguity in the ordering of the momenta, we will always take the convention that $\kappa_1>\kappa_2$, i.e. positive relative momentum).
\par We can simplify the S matrix by making use of SU(2) invariance. Indeed, if we make linear combinations of the asymptotic modes that diagonalize the total spin $S_T^2$ and its projection $S_T^z$, the S matrix should be diagonal. Moreover, since the magnon interactions are SU(2) invariant (both Hamiltonian and ground state are), the S matrix elements should be constant within every sector of total spin. This means that the general expression for the magnon-magnon S matrix in this representation should reduce to
\begin{equation*}
S = \begin{pmatrix} -\e^{i\theta_0} \times \one_{1\times1} & & \\ & -\e^{i\theta_1} \times \one_{3\times3} & \\ & & -\e^{i\theta_2} \times \one_{5\times5} \end{pmatrix},
\end{equation*}
i.e. the S matrix reduces to three phases for every sector of total spin. In our simulations, we always found this reduced form to high precision, so in the following we can restrict to plotting these three phases.
\begin{figure}
\includegraphics[width=\columnwidth]{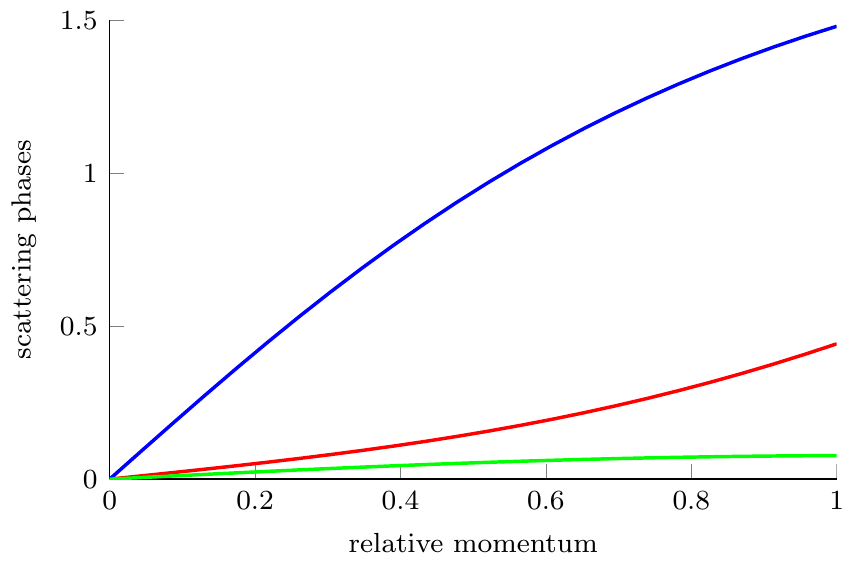}
\caption{The S matrix in function of relative momentum $\kappa_1-\kappa_2$ at total momentum $K=0$. Plotted are the phases of the S matrix in the $S=0$ (red), $S=1$ (blue) and $S=2$ (green) sector. Calculations were done at $\gamma=2$ and with bond dimension $D=32$.}
\label{fig:Smatrix}
\end{figure}
\par
\begin{figure}
\includegraphics[width=\columnwidth]{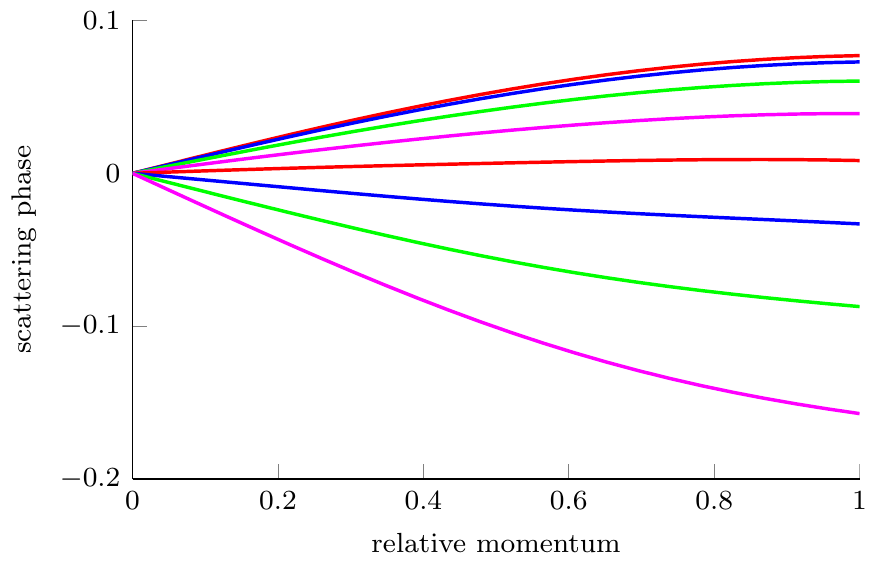}
\caption{The scattering phase in the $S=2$ sector for 8 equally spaced values of the total momentum between $K=0$ (upper line) and $K=\pi/3$ (lower line). Around $K=0$ there is a region where the S matrix is independent of total momentum, which points to some Galilean invariance around the minimum of the dispersion relation. Calculations were done at $\gamma=2$ and with bond dimension $D=32$}
\label{fig:SmatrixDispersion} 
\end{figure}

\begin{figure}
\includegraphics[width=\columnwidth]{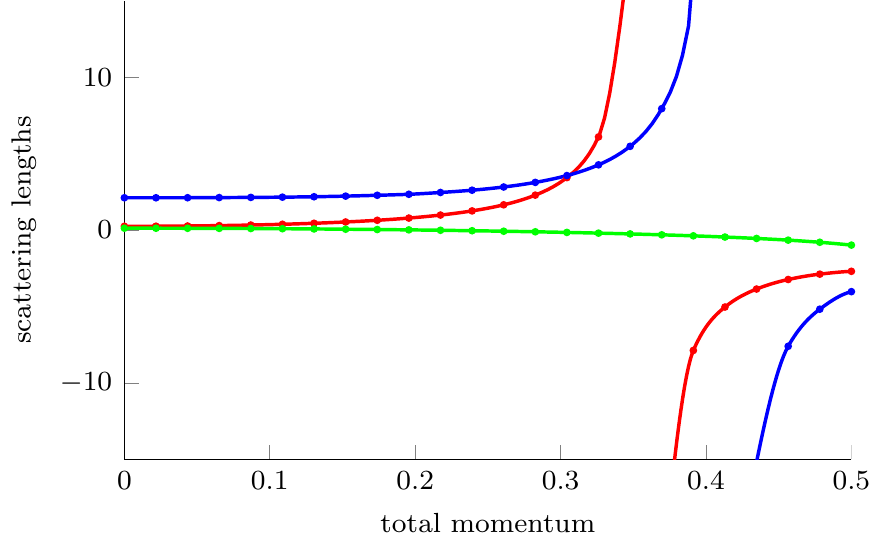}
\caption{The scattering lengths $a_0$ (red), $a_1$ (blue) and $a_2$ (green) in function of the total momentum $K$. In the $S=2$ sector nothing spectacular happens, although it does change sign. In the other sectors we see a divergence at the momentum where a bound state forms. The plotted range does not show all data points around the divergences, the full lines are a guide to the eye and give an indication on where the other points are situated. Calculations were done at $\gamma=2$ and bond dimension $D=32$.}
\label{fig:scatLengthFull}
\end{figure}

 In Fig.~\ref{fig:Smatrix} we have plotted the S matrix in function of the relative momentum $\kappa_1-\kappa_2$ for total momentum $K=0$. One can observe (i) the limit $S=-\one$ for the relative momentum going to zero, and (ii) the linear region around this limit (the slope is the scattering length). The sign of the phase is positive for all three sectors (although this does not have to be the case, see Figs.~\ref{fig:SmatrixDispersion} and \ref{fig:scatLengthFull}).
\par In Fig.~\ref{fig:SmatrixDispersion} we have plotted the S matrix in the $S=2$ sector for different values of the total momentum. We observe that the S matrix depends strongly on $K$ in a non-trivial way, but there seems to be a small region around $K=0$ where it is quasi-constant. This points to the presence of a region around the minimum of the dispersion relation where the interaction is Galilean invariant (note that the dispersion should be quadratic in this region). At larger momenta, this Galilean invariance is broken, as one expects in a lattice system.
\par Even more spectacular things can happen when we vary the total momentum, such as the formation of a bound state. In Fig.~\ref{fig:scatLengthFull} we have plotted the scattering lengths in all three sectors in function of the total momentum. We can see that the scattering lengths in the $S=0$ and $S=1$ sectors diverge, signalling the formation of the singlet and triplet bound states (in agreement with the discussion in Sec.~\ref{sec:scatBound}).

\subsection{Spectral function}
\label{sec:spectral}

Since we have a two-leg ladder system, we can look at spectral functions with transversal momentum $q$ equal to $0$ or $\pi$. We define the two rung operators (defined on rung $i$)
\begin{align}
& (S^z_0)_i = S^z_{i,1} + S^z_{i,2} \label{evenP} \\
& (S^z_\pi)_i = S^z_{i,1} - S^z_{i,2} . \label{oddP}
\end{align}
These operators have even, resp. odd parity under the action of the reflection operator $\mathcal{P}$. We will look at spectral functions $S_{0/\pi}(\kappa,\omega)$ with respect to these two operators,
\begin{multline}
 S_{0/\pi}(\kappa,\omega) = \sum_{n} \int\d t \; \e^{i(\omega t-\kappa n)} \\ \times  \bra{\Psi_0} \e^{-iHt} (S^z_{0/\pi})_n\dag \e^{iHt} (S^z_{0/\pi})_0 \ket{\Psi_0}
\end{multline}
where $\sum_n$ represents a sum over rungs.
\par
\begin{figure}
\includegraphics[width=\columnwidth]{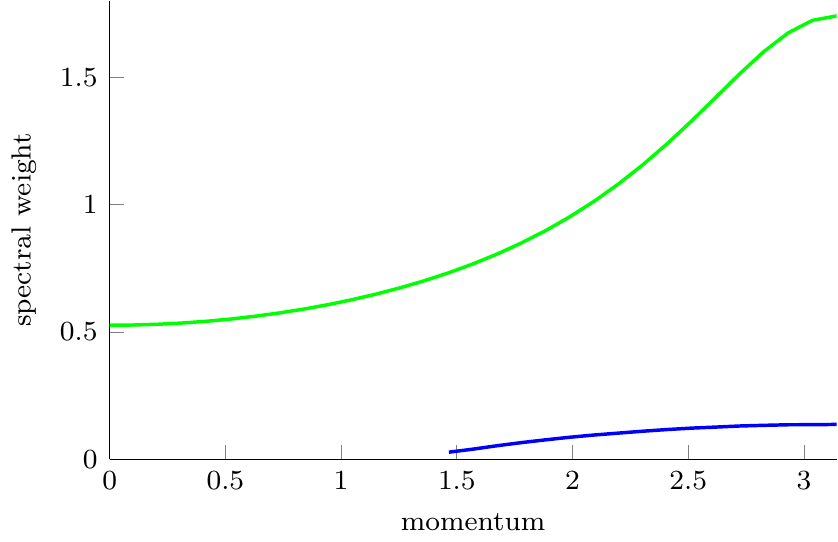}
\caption{The one-particle spectral weights; these appear in the spectral functions $S_{0/\pi}(\kappa,\omega)$ as the prefactor of the $2\pi\delta(\omega-\Delta(\kappa))$ function (where $\Delta(\kappa)$ is the dispersion relation of the particle). We have plotted the magnon weights w.r.t. to the odd operator (green) and the weight of the triplet bound state w.r.t. to the even operator (blue). All the other one-particle spectral weights are identically zero. These results are in accordance with Ref. \onlinecite{Schmidt2005}. Note that the one-particle description of the bound state gets worse when coming closer to the continuum, so that the calculation of its spectral weight loses accuracy in this region. It is nevertheless clear that the spectral weight goes to zero as the bound state loses stability.}
\label{fig:spectralOne} 
\end{figure}

 Let us first look at the one-particle contributions. Since the elementary magnon is odd under $\mathcal{P}$, it can only carry spectral weight with respect to the odd operator. From SU(2) symmetry we know that the singlet bound state does not carry any spectral weight with respect to both operators (they are both spin-1 operators). Lastly, the triplet bound state is even under $\mathcal{P}$, so it only contributes to the even operator spectral function $S_0(\kappa,\omega)$. These considerations lead to the picture in Fig.~\ref{fig:spectralOne}. One can see that the spectral weight of the bound state goes to zero as it approaches the continuum.
\par
\begin{figure}
\includegraphics[width=\columnwidth]{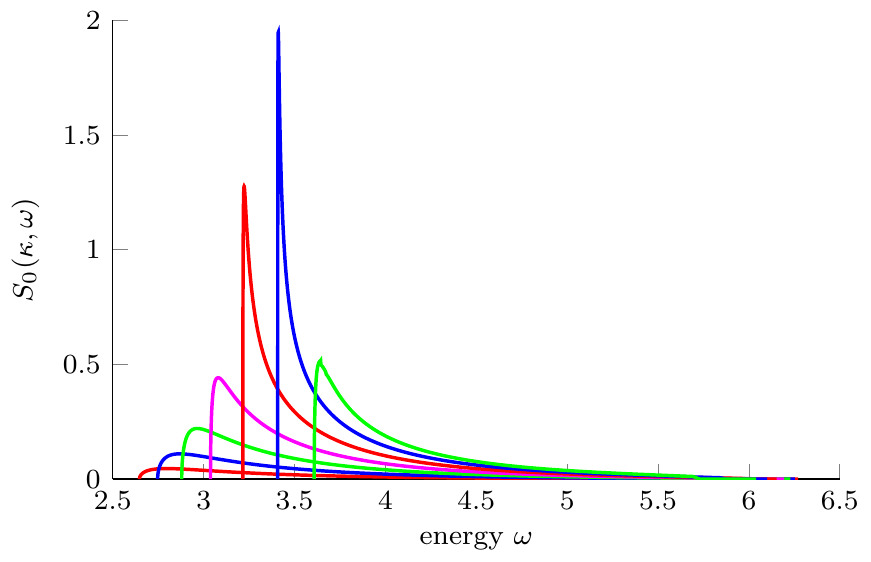}
\caption{The two-particle contribution to the spectral function $S_0(\kappa,\omega)$ for equally spaced values of the momentum between $\kappa=0$ and $\kappa=\pi/2$. The $\kappa=0$ curve is not shown as it is equal to zero everywhere. Calculations were done at $\gamma=2$ with bond dimension $D=32$.}
\label{fig:spectralTwo}
\end{figure}

 Next we look at the two-magnon contribution, which has only overlap with the even parity operator. In Fig.~\ref{fig:spectralTwo} we have plotted different momentum slices of the spectral function. At momentum zero, the spectral function is identically zero (the ground state is a singlet) and grows for small momenta as $\propto\kappa^2$ (cfr. Ref.~\onlinecite{Affleck1992}). For larger momenta, we see that the spectral function gets strongly peaked at some value for $\kappa$, after which the peak again disappears. The origin of this resonance is of course the formation of the bound state: before it is stable, the bound state is already visible in the spectral function as a resonance. 
\par
\begin{figure}
\includegraphics[width=\columnwidth]{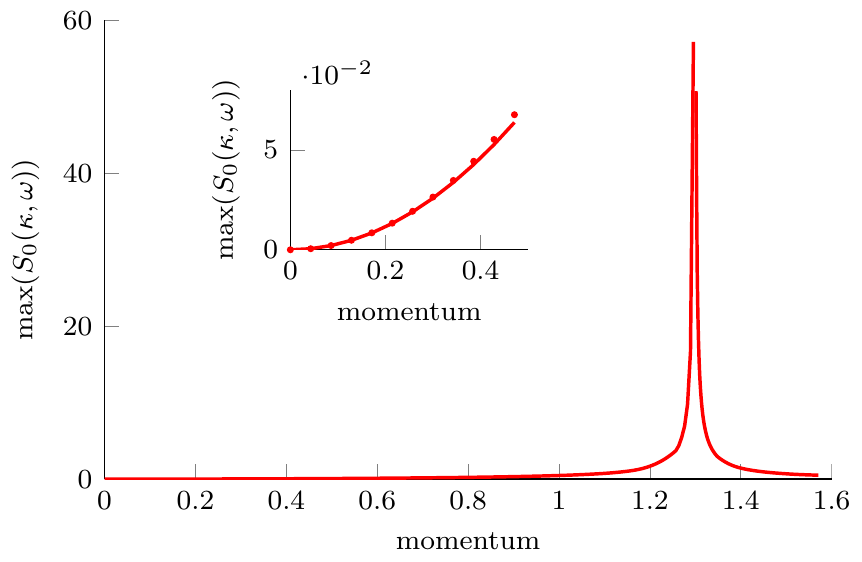}
\caption{The maximum of the two-particle contribution to the spectral function $S_{0}(\kappa,\omega)$ for different momentum slices. The full line is a guide to the eye. In the inset we show a close-up of the small momentum region, the full line is quadratic fit. Calculations were done at $\gamma=2$ with bond dimension $D=32$.}
\label{fig:specMax}
\end{figure}

 To further confirm this picture, we have plotted the maximum of the peak in function of the momentum in Fig.~\ref{fig:specMax}. One can see the resonance clearly diverging at the point where the bound state reaches stability: from that point on the stable bound state contributes a delta peak to the spectral function.
\par
\begin{figure}
\includegraphics[width=\columnwidth]{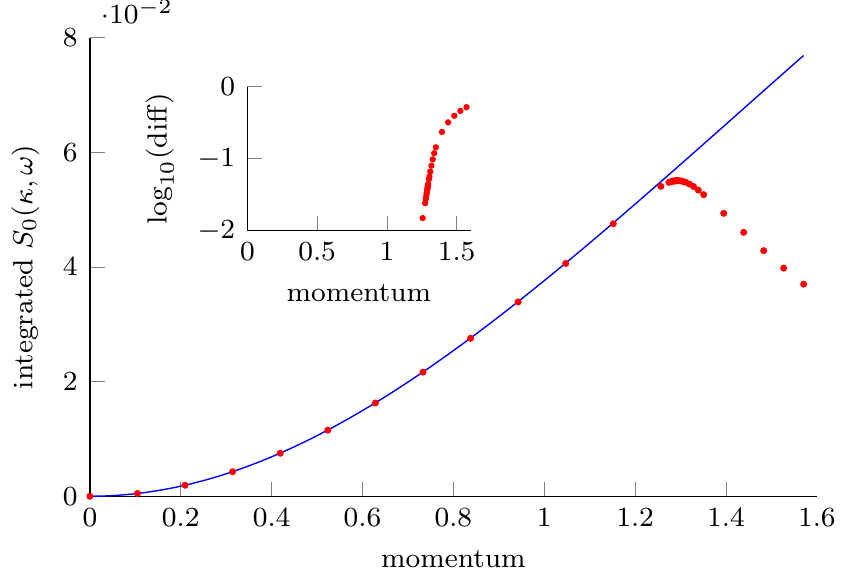}
\caption{The integrated spectral function $\int\d\omega/2\pi S(\kappa,\omega)$ in function of the momentum $\kappa$ (red dots) compared with the momentum space correlation function $s_0(\kappa)$ (blue line). In the inset we plot the (log10 of the) difference between the two; values below $10^{-2}$ are not shown. Calculations were done at $\gamma=2$ with bond dimension $D=32$.}
\label{fig:integrated}
\end{figure}

 We have also plotted the integrated spectral function in Fig.~\ref{fig:integrated}. Before the formation of the bound state, we see that the sum rules are completely satisfied (up to numerical errors), which shows that the one- and two-particle sectors indeed capture the full spectral function, at least in this momentum range (see also Ref. \onlinecite{Schmidt2005}). Again, we clearly see the $\propto\kappa^2$ dependence at small momenta. After the bound state has formed, however, the two-magnon part loses increasing spectral weight to the bound state.

\subsection{Magnetization process}
\label{sec:magnProcess}

Let us now turn on the magnetic field. For SU(2) invariant systems, this perturbation does not affect the singlet ground state and induces a Zeeman splitting of the elementary magnon excitation. When the magnetic field reaches the value of the gap, one of the components of the triplet forms a pseudo-condensate (no real condensate can form in one dimension); the system undergoes a continuous phase transition from a commensurate phase with zero magnetization to an incommensurate phase with non-zero magnetization \cite{Schulz1980}.
\par The physical picture of this condensation can be understood from the approximate Bethe ansatz that was developed in Sec.~\ref{sec:aba}. Indeed, once it crosses the gap, the magnetic field serves as a chemical potential for the +1 component of the magnon triplet (the other components remain gapped, so we will not consider them in our calculations). The information on the magnon dispersion relation and the magnon-magnon S matrix we have gathered in the previous sections will allow us to compute both thermodynamic properties and correlation functions for the magnetized chain.
\par We start very close to the phase transition, where only the momenta around the minimum will be occupied, so that we can approximate them as free fermions. If we introduce a characteristic velocity $v$ for the magnon dispersion around its minimum as
\begin{equation*}
 \Delta(\kappa) = \Delta + \frac{v^2}{2\Delta} (\kappa-\kappa_\text{min})^2,
\end{equation*}
the magnetization (i.e. the density of condensed magnons) will be given by \cite{Tsvelik1990, Affleck1991, Sorensen1993}
\begin{equation} \label{freeFermion}
 m(h) = \frac{\sqrt{2\Delta}}{\pi v} \sqrt{(h-h_c)}.
\end{equation}
When more pseudo-momentum levels are filled up, the two-particle S matrix will deviate from its limiting value of $-1$ and the free-fermion approximation will no longer hold. As a first order correction, we can assume a linear scattering phase with the scattering length $a$ as the slope (and still a quadratic dispersion). From Eq.~\eqref{densityScatLength} it follows that the correction to the magnetization curve is given by
\begin{equation} \label{mScatLength}
m(h) = \frac{\sqrt{2\Delta}}{\pi v} \sqrt{(h-h_c)} - \frac{8\Delta a}{3\pi^2v^2} (h-h_c),
\end{equation}
a result which was obtained in Ref.~\onlinecite{Lou2000} by a similar reasoning.
\par
\begin{figure}
\includegraphics[width=\columnwidth]{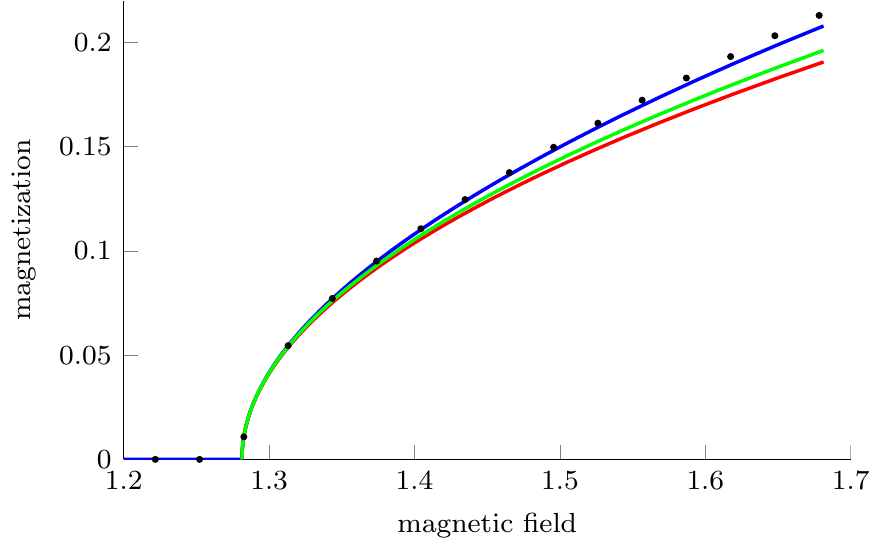}
\caption{The magnetization of the ladder ($\gamma=2$) in function of the applied magnetic field $h$. The dots are calculated with a direct MPS optimization (using an adapted version of Ref.~\onlinecite{Haegeman2011d}), the red line is the free-fermion result [Eq.~\eqref{freeFermion}], the green one is with the scattering length correction [Eq.~\eqref{mScatLength}], and the blue line is a full approximate Bethe ansatz calculation. }
\label{fig:magnetization}
\end{figure}

 When even higher momenta are occupied these approximations (quadratic dispersion relation, linear scattering phase and Galilean invariance) will get worse and only a full Bethe ansatz calculation will give the correct magnetization curve. In Fig.~\ref{fig:magnetization} we have plotted this.
\par
\begin{figure}
\includegraphics[width=\columnwidth]{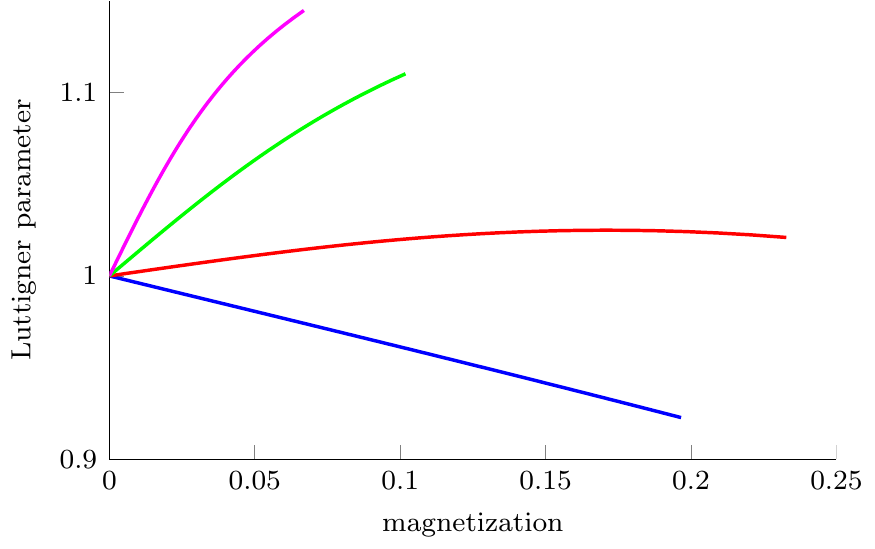}
\caption{The LL parameter in function of the magnetization for $\gamma=5$ (blue), $\gamma=2$ (red), $\gamma=1$ (green) and $\gamma=1/2$ (magenta).}
\label{fig:luttinger}
\end{figure}

 Next we look at correlation functions of the magnetized ladder. With our methods, we have no direct access to these correlation functions, but we can infer their form by combining the Luttinger liquid formalism with the thermodynamic properties computed from the approximate Bethe ansatz. Indeed, since we have seen in Sec.~\ref{sec:spectral} that the $S^x_\pi$ operator essentially creates a magnon out of the vacuum at momentum $\pi$ and the $S^z_0$ operator creates a two-magnon state at momentum $0$, we can translate the expressions for the Bose gas correlators [Eq.~\eqref{eq:correlators}] to the magnetized ladder as
\begin{align} 
& \braket{(S_\pi^x)_{i'}(S_\pi^x)_{i}} = A_x \frac{(-1)^{i-i'}}{|i-i'|^{1/2K}}  \nonumber \\ & \hspace{2.5cm} - B_x (-1)^{i-i'} \frac{\cos(2\pi m (i-i'))}{|i-i'|^{2K+1/2K}} \label{xx} \\
& \braket{(S_0^z)_{i'}(S_0^z)_{i}} = m^2 - \frac{K}{2\pi^2|i-i'|^2} \nonumber \\ & \hspace{2.5cm} + A_z  \frac{\cos(2\pi m (i-i'))}{|i-i'|^{2K}}, \label{zz}
\end{align}
in accordance with Ref.~\onlinecite{Hikihara2001}. The power-law decay of these correlation functions is controlled by the LL parameter $K$. In Fig.~\ref{fig:luttinger} we have plotted $K$ in function of the magnetization $m$ for the ladder at different values of $\gamma$. At very low magnetization $m\rightarrow0$ the LL parameter reaches the universal value of 1, but it appears that, beyond this limiting value, $K(m)$ changes qualitatively as we vary $\gamma$. The same behaviour was observed in Ref.~\onlinecite{Hikihara2001} by fitting the analytic form of the correlation functions \eqref{xx} and \eqref{zz} with numerical calculations.
\par
\begin{figure}
\includegraphics[width=\columnwidth]{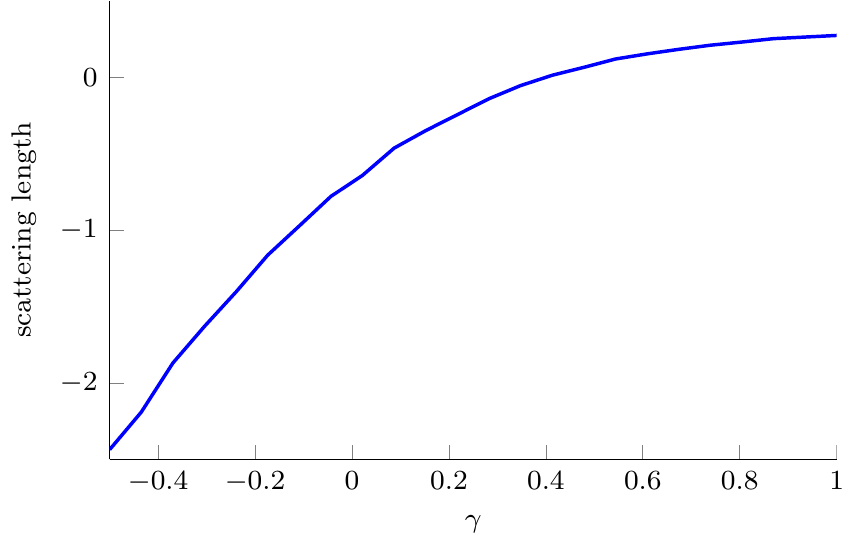}
\caption{The scattering length for different values of the interchain coupling $\gamma$.}
\label{fig:scatLengthRange}
\end{figure}

 This behaviour can again be explained by starting with the free-fermion limit at very low densities. In Sec.~\ref{sec:limiting} we have shown that the LL parameter equals $K=1$ in this case. The first order correction on this value is determined by the magnon-magnon scattering length; in first order in $m$ the LL parameters is given by \cite{Affleck2005}
\begin{equation} \label{LLa}
K(m) = 1 - 2am. 
\end{equation}
In Fig.~\ref{fig:scatLengthRange} we have plotted the scattering length in function of the interchain coupling $\gamma$. Based on Eq.~\eqref{LLa}, the change of the sign of $a$ confirms the varying qualitative behaviour of $K(m)$ as observed in Fig.~\ref{fig:luttinger} and in Ref.~\onlinecite{Hikihara2001}.
\par
\begin{figure}
\includegraphics[width=\columnwidth]{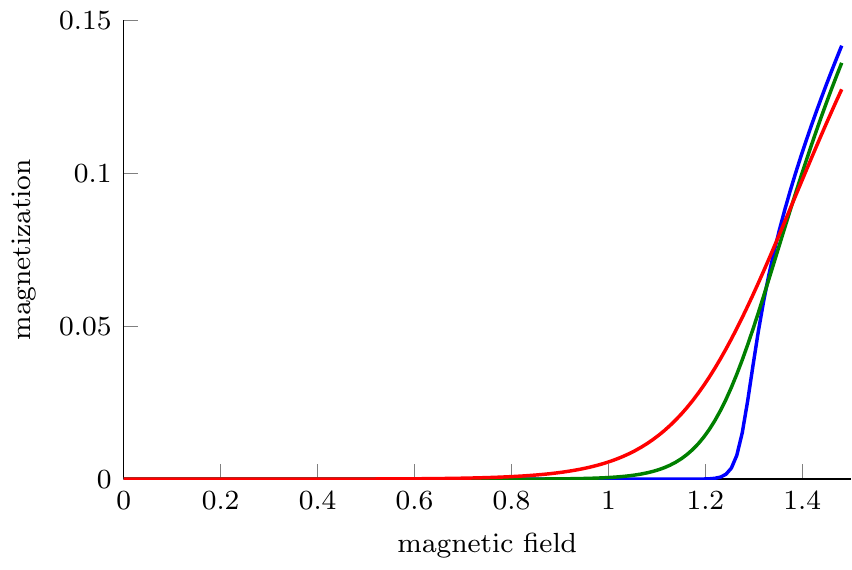}
\caption{The magnetization in function of the magnetic field $h$ for three values of the temperature: $T=.01\Delta$ (blue), $T=.045\Delta$ (green) and $T=.08\Delta$ (red).}
\label{fig:tba}
\end{figure}

 Finally, we can study the magnetization process at finite temperatures using the thermodynamic Bethe ansatz.  In Fig.~\ref{fig:tba} we have plotted the magnetization curve for different temperatures, showing that the zero-temperature square-root dependence around the phase transition is smoothed out at finite temperature. Note that we have included the other components of the magnon triplet -- they are thermally excited as well -- in a decoupled fashion. In a more correct analysis we would have to solve the fully coupled Bethe equations for the three components, but this falls outside the scope of this paper.

\section{Future directions}
\label{sec:section5}

In the previous sections we have shown how to variationally determine all properties of one- and two-particle excitations of generic quantum spin chains. In this last section we show how our framework can be extended to study domain wall excitations and bound states and how to compute spectral functions at finite temperature. Since we believe that our work provides a crucial step towards the construction of an effective Fock space of interacting, particle-like excitations, we provide some further steps in this direction. Lastly, we reflect shortly on the application of our methods to two dimensional systems.

\subsection{Topological excitations and bound states}

In the previous sections we have restricted our framework to the case where we have a unique ground state. We can easily extend the framework, however, to situations where we have symmetry breaking and the elementary excitations are domain walls rather than localized particles.
\par Suppose we have a doubly degenerate ground state, approximated by two MPS $\ket{\Psi[A_1]}$ and $\ket{\Psi[A_2]}$. The obvious ansatz for a domain wall excitation is
\begin{multline} \label{domainWall}
    \ket{\Phi_\kappa[B]} = \sum_n \e^{i\kappa n} \sum_{\{s\}} \lv \left[ \prod_{m<n} A_1^{s_m} \right] \\
    \times B^{s_n} \left[ \prod_{m>n} A_2^{s_m} \right] \rv \spst,
\end{multline}
i.e. the domain wall interpolates between the two ground states. The ansatz has been successfully applied to the gapped XXZ model in Ref.~\onlinecite{Haegeman2012a}, where the elementary excitations are spinons, and to the Lieb-Liniger model in Ref.~\onlinecite{Draxler2013}, where topological excitations are elementary. 
\par Strictly speaking, however, the momentum of the ansatz [Eq.~\eqref{domainWall}] is not well defined: multiplying the tensor $A_2$ with an arbitrary phase factor $A_2\leftarrow A_2\e^{i\phi}$ shifts the momentum with $\kappa\leftarrow\kappa+\phi$. The origin of this ambiguity is the fact that one domain wall cannot be properly defined when using periodic boundary conditions.
\par Physically, however, domain walls should come in pairs. The procedure for constructing a scattering state of two domain walls is completely analogous as in Sec.~\ref{sec:section2}. For these states the total momentum is well-defined, although the individual momenta can be arbitrarily transferred between the two domain walls. Scattering states of two domain walls are especially relevant as they are the first excitations that carry any spectral weight. Consequently, a first non-trivial contribution to dynamical correlation functions asks for a solution of the scattering problem.
\par A second extension of the scattering formalism is towards the study of bound states. As we explained above, a bound state should be interpreted as a one-particle excitation and described by a one-particle ansatz. Yet, in the case where the bound state becomes very wide -- e.g. when it is close to a two-particle continuum -- the one-particle ansatz is not able to capture its delocalized nature. One possible extension consists of working on multiple MPS tensors at once, leading to the ansatz \cite{Haegeman2013a,Haegeman2013b}
\begin{multline}
\ket{\Phi_\kappa[B]} = \sum_n \e^{i\kappa n} \sum_{\{s\}} \lv \left[ \prod_{m<n} A^{s_m} \right] \\
    \times B^{s_n,s_{n+1},\dots , s_{n+N}} \left[ \prod_{m>N+n} A^{s_m} \right] \rv \spst.
\end{multline}
The number of the variational parameters in the big $B$ tensor grows exponentially in the number of sites, so that we cannot systematically grow the block as the bound state gets wider. 
\begin{figure*}
\includegraphics[width=1.3\columnwidth]{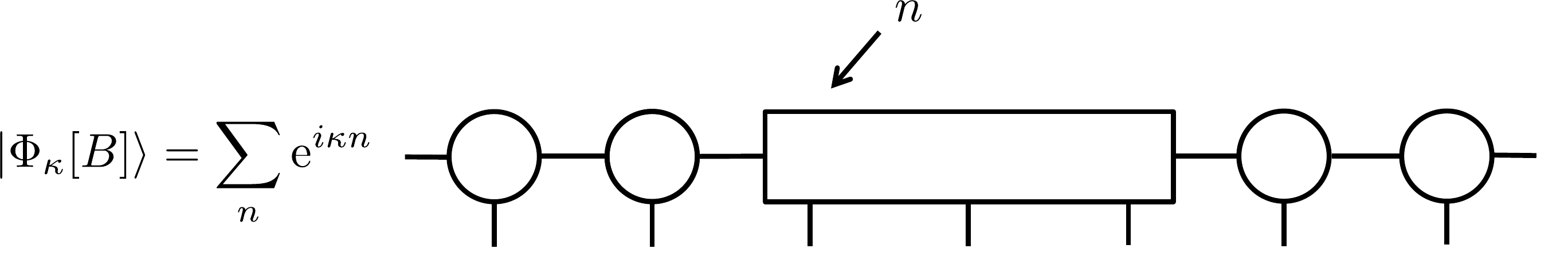}
\caption{Graphical representation of the bound state ansatz. The $B$ tensor of the one-particle ansatz in Fig.~\ref{fig:ansatz} is spread over more than one site.}
\label{fig:block}
\end{figure*}
\par As a more systematic way to study wide bound states, we should use the two-particle ansatz \eqref{eq:ansatz} to describe them. In contrast to a scattering state the energy of a bound state is not known from the one-particle dispersions, so that we will have to scan a certain energy range in search of bound state solutions -- of course, with the one-particle ansatz we can get a pretty good idea where to look. A bound state corresponds to solutions for the eigenvalue equation \eqref{eig} with only decaying modes in the asymptotic regime. In principle we should even be able to find bound state solutions within a continuum of scattering states (i.e. a stationary bound-state, not a resonance within the continuum) by the presence of additional localized solutions for the scattering problem.

\subsection{Spectral functions at finite temperature}

At finite temperatures, the thermally excited density of excitations already present in the thermal state destroys the perfect coherence of one-particle contributions to spectral functions: the delta peaks at zero temperature will get smeared out in finite temperature spectral functions. It appears that this thermal broadening depends heavily on the interactions between the particles \cite{Essler2008,Tennant2012}, so that a full quantum mechanical treatment is needed to accurately resolve it.
\par At zero temperature the spectral function $S(\kappa,\omega)$ can be expressed in terms of the spectral weights of the low-energy excitations of the system. At finite temperatures, this is no longer true as we generally need form factors corresponding to states with arbitrarily high energies. In gapped integrable systems -- where the higher energy states can be labelled with a particle number $n$ and have an energy of the order $n\Delta$ -- the higher-energy form factors are suppressed with a Boltzmann factor $\mathcal{O}\left(\e^{-n\Delta/T}\right)$, so one can restrict to low-particle form factors at low enough temperatures (compared to the gap) \cite{Konik2003, Essler2008, Tennant2012}.
\par In this paper we have shown that, even in non-integrable systems, the particle picture remains valid at low densities (low temperatures), which makes the low-temperature expansion in $\mathcal{O}\left(\e^{-\Delta/T}\right)$ possible for the non-integrable case as well (see also Ref.~\onlinecite{Fauseweh2014} for a similar expansion for non-integrable systems). So we can associate a particle number to higher excitations and we can write down the finite temperature expression for the spectral function in the Lehmann representation as
\begin{multline} \label{LehmannT}
S(\kappa,\omega) = \frac{1}{Z} \sum_{mn} \sum_{\left\{\alpha\right\}\left\{\beta\right\}} 2\pi\delta\Big(E(\{\alpha\}) - E(\{\beta\}) - \omega\Big) \\ 2\pi\delta\Big(K(\{\alpha\}) - K (\{\beta\}) -\kappa\Big) \\ \e^{-\beta E(\{\alpha\})}  \left|\bra{m,\{\alpha_m\}} O \ket{n,\{\beta_n\}} \right|^2
\end{multline}
where $\sum_{mn}$ is a double sum over particle numbers ranging to $\infty$ and $\{\alpha_m\}$ is a set of $m$ particle types: the states $\ket{m,\{\alpha_m\}}$ can then be identified with the multi-particle states in the approximate Bethe ansatz picture of Sec.~\ref{sec:aba}. We can see that, for gapped systems, the Boltzmann factor provides a small parameter, so that excitations with many particles only play a limited role at low temperatures. In the thermodynamic limit, two difficulties remain: (i) when coming close to the one-particle dispersion curve (where the zero-temperature spectral function has its $\delta$ peak divergence) we have to perform a resummation in order to take into account an infinite number of terms, and (ii) the form factors appearing in Eq.~\eqref{LehmannT} can be divergent in the thermodynamic limit. A careful analysis shows that both difficulties can be overcome in the case of integrable (free and interacting) massive field theories \cite{Essler2009}. Within our framework, it should prove possible to calculate finite-temperature spectral functions for generic spin chains (non-integrable) and go beyond the perturbative approaches of Refs.~\onlinecite{James2008} and \onlinecite{Goetze2010a}.

\subsection{Effective field theory}

Whereas the approximate Bethe ansatz provides a way to construct an effective first-quantized wave function for a finite density of excitations, a systematic construction of an interacting many-particle model should be formulated in second quantization \cite{Haegeman2013b, Vanderstraeten2014, Keim2015}. We introduce momentum space creation and annihilation operators that act on the ground state as
\begin{equation*} \begin{split}
 & c_\alpha\dag(\kappa) \ket{\Psi[A]} = \ket{\Phi_\alpha(\kappa)} \\
 & c_\alpha(\kappa) \ket{\Psi[A]} = 0
\end{split} \end{equation*}
and write down an effective interacting theory
\begin{multline} \label{eq:effective}
H = \sum_\alpha\int\frac{\d\kappa}{2\pi} \Delta_\alpha(\kappa) c_\alpha\dag(\kappa) c_\alpha(\kappa) \\ + \sum_{\alpha'\beta'\alpha\beta} \int \frac{\d\kappa}{2\pi}\frac{\d\kappa_1}{2\pi}\frac{\d\kappa_2}{2\pi} V_{\alpha'\beta',\alpha\beta}(\kappa,\kappa_1,\kappa_2) \\ \times c_{\alpha'}\dag(\kappa_1+\kappa_2-\kappa) c_{\beta'}\dag(\kappa) c_\beta(\kappa_2) c_\alpha(\kappa_1).
\end{multline}
Since we only have explicit access to the operator acting on the ground state and not the operator itself, it is a priori not clear how to determine the $c_\alpha\dag(\kappa)$ and $c_\alpha(\kappa)$ in a unique way. Moreover, there seems to be no trivial way for imposing the correct commutation relations. Thirdly, because these operators will be momentum-dependent, the transition to a local, real-space representation of the Fock operators might not be well-defined. The construction of Wannier states out of the momentum eigenstates might provide a good starting point \cite{Keim2015}, although it is still not clear how to find the unique real-space operators that are essential for computing the interaction term in Eq.~\eqref{eq:effective}.
\par A different approach can be taken by starting from a free theory of particles with generalized statistics that match the two-particle S matrix. The following effective Hamiltonian
\begin{equation*}
 H_0 = \sum_\alpha\int\frac{\d\kappa}{2\pi} \Delta_\alpha(\kappa) Z_\alpha\dag(\kappa) Z_\alpha(\kappa)
\end{equation*}
indeed captures the low-lying spectrum of the original Hamiltonian if the $Z_\alpha$ and $Z_\alpha\dag$ are the so-called Faddeev-Zamolodchikov (FZ) operators obeying the following commutation relations
\begin{align*}
& Z_\alpha(\kappa_1)Z_\beta(\kappa_2) = S_{\alpha\beta}^{\gamma\delta}(\kappa_1,\kappa_2) Z_\delta(\kappa_2)Z_\gamma(\kappa_1) \\
& Z_\alpha(\kappa_1)\dag Z_\beta(\kappa_2)\dag = S_{\alpha\beta}^{\gamma\delta}(\kappa_1,\kappa_2) Z_\delta(\kappa_2)\dag Z_\gamma(\kappa_1)\dag \\
& Z_\alpha(\kappa_1) Z_\beta(\kappa_2)\dag = 2\pi\delta(\kappa_1-\kappa_2)\delta_{\alpha\beta} \\ & \hspace{3cm} +  S_{\beta\gamma}^{\delta\alpha}(\kappa_1,\kappa_2) Z_\delta(\kappa_2)\dag Z_\gamma(\kappa_1).
\end{align*}
The idea is to look at perturbations of $H_0$ and express them in terms of these FZ operators. Indeed, when applying a non-commuting perturbation, we could have a new Hamiltonian of the form
\begin{multline} \label{perturbationFZ}
H = H_0+\sum_{\alpha\beta} \int\frac{\d\kappa}{2\pi} \left( M^{\alpha\beta}_p Z\dag_\alpha(\kappa)Z_\beta(\kappa) \right. \\ + \left. M_n^{\alpha\beta} Z_\alpha(-\kappa)Z_\beta(\kappa) + h.c. \right)
\end{multline}
where
\begin{align*}
& M_p^{\alpha\beta}(\kappa) = \bra{\Phi_\alpha(\kappa)} \hat{M} \ket{\Phi_\beta(\kappa)} \\
& M_n^{\alpha\beta}(\kappa) = \bra{\Psi[A]} \hat{M} \ket{\Upsilon_{\beta\alpha}(\kappa,-\kappa)} 
\end{align*}
are the particle preserving, resp. particle non-preserving parts of the perturbation. For small perturbations, we can assume that only small momentum states will be occupied and that the S matrix is approximately $-\one$. In that case, the FZ operators reduce to fermion creation and annihilation operators and we can diagonalize the Hamiltonian [Eq.~\eqref{perturbationFZ}] with a Bogoliubov rotation. In general, this proves not to be possible \cite{Sotiriadis2012} and a more sophisticated strategy will have to be developed.
\par When studying the time evolution of integrable systems, the occupation numbers $n_\alpha(\kappa) = Z_\alpha\dag(\kappa)Z_\alpha(\kappa)$ corresponding to the FZ operators are integrals of motion \cite{Essler2014}. For non-integrable systems this is no longer the case, although the observation of so-called prethermalization plateaus might point to the fact that they are almost preserved. Indeed, the mode occupation numbers $n_\alpha(\kappa)$ provide a way to distinguish a thermal Gibbs ensemble from a generalized Gibbs ensemble \cite{Essler2014a,*Bertini2015a}. Consequently, by finding an explicit (real-space) representation of the FZ operators we could follow the occupation numbers $n_\alpha(\kappa)$ through time, also when starting from an interacting theory.

\subsection{Breaking of integrability and Yang-Baxter equation}

Integrable systems possess a number of interrelated properties -- diffractionless scattering, local conservation laws, etc. -- that makes them amenable to a number of analytical techniques. Once the integrability is broken, these techniques are no longer applicable. An important question is to what extent the different manifestations of integrability survive in an approximate way close to an integrable point.
\par One simple consistency condition for integrability is the Yang-Baxter equation \cite{Yang1967,*Yang1968,Baxter1982}, expressing that three-particle scattering should be indepedent of the order in which it is decomposed into consecutive two-particle processes. As such it is a condition on the two-particle S matrix. Our methods provide a way to test this condition for non-integrable systems, and, more specifically, to study the breaking of the Yang-Baxter equation for systems close to integrable points \cite{preparation}.

\subsection{Higher dimensions}

Matrix product states have a higher-dimensional generalization called projected entangled-pair states (PEPS) \cite{Verstraete2004b}. Just as in one dimension, it has been shown that PEPS are able to capture the ground state properties of generic two-dimensional quantum spin systems \cite{Jordan2008}, so it should be able to straightforwardly generalize the one-particle ansatz of Eq.~\eqref{oneparticle} to the PEPS formalism. Compared to the MPS setting, however, the computation of the effective one-particle Hamiltonian is a lot more involved, because of the fact that the environment in a PEPS contraction is a one-dimensional object itself (compared to the zero-dimensional environment in MPS).
\par In Refs.~\onlinecite{Zauner2015} and \onlinecite{Haegeman2014b} elementary particle excitations in two dimensions were studied by looking at the spectrum of the transfer matrix. The next step, i.e. a full variational calculation of the effective Hamiltonian matrix, should lead to quantitative estimates of the gap and full dispersion relations of generic two-dimensional spin systems \cite{Vanderstraeten2015b}.

\begin{acknowledgments}
The authors would like to thank Sven Bachmann, Fabian Essler, Paul Fendley, Tobias Osborne, and Didier Poilblanc for inspiring discussions. Research supported by the Research Foundation Flanders (LV,JH), the Austrian FWF SFB grants FoQuS and ViCoM, and the European grants SIQS and QUTE (FV).
\end{acknowledgments}

\bibliography{bibliography}

\appendix
\onecolumngrid

\section{Ground state and one-particle excitations: technical details}

\subsection{Uniform matrix product states}

Consider a one-dimensional quantum spin system with $N$ sites and physical dimension $d$. In the thermodynamic limit ($N\rightarrow\infty$) we can define a uniform matrix product state (uMPS) as
\begin{equation} \label{uMPSA}
\ket{\Psi[A]} = \sum_{\{s\}=1}^d \lv \left[ \prod_{m=-\infty}^{+\infty} A^{s_m} \right] \rv \ket{ \{s\} },
\end{equation}
which is parametrized by the set of $D\times D$ complex matrices $A^s$ for $s=1,\dots,d$ or, equivalently, the tensor $A \in \CC^{D\times d \times D}$. It can be shown that all local expectation values are independent of the $D$-dimensional boundary vectors $\lv$ and $\rv$ if the MPS is injective or pure \cite{Fannes1992,Haegeman2014}. This is the case if the transfer matrix, which is defined as $E = \sum_s A^s \otimes \ol{A}^s$ and acts as an operator in a $D\times D$ dimensional vector space, has a non-degenerate largest eigenvalue $\omega$ and the corresponding left and right eigenvectors $(l|$ and $|r)$ are full rank when written as semi-positive definite Hermitian ($D \times D$) matrices $l$ and $r$. A proper normalization of the uMPS amounts to rescaling the $A$ tensor as $A^s\rightarrow A^s/\sqrt{\omega}$, so the spectral radius of the transfer matrix rescales to unity. Indeed, the norm of the uMPS can be formally computed as
\begin{equation*}
\braket{\Psi[A]|\Psi[A]} = \Big( \lv \otimes \overline{\lv} \Big) E^\infty  \Big( \rv \otimes \overline{\rv} \Big) \propto (l|r) = \tr(lr)
\end{equation*}
so that a proper rescaling of $l$ and $r$ suffices to fix the norm to unity (the proportionality factor is unimportant as all expectation values will contain this same factor). The parametrization of \eqref{uMPSA} has a redundancy: the state $\ket{\Psi[A]}$ is invariant under a \textit{gauge transformation} $A^s\rightarrow G^{-1} A^s G$ with $G$ an invertible matrix. There are different ways for fixing this gauge freedom and we will not specify which one to choose.
\par The other eigenvalues of the transfer matrix $E$ have significance as well; the second eigenvalue $\omega^{(2)}$ determines the correlation length $\xi$ of the uMPS as
\begin{equation} \label{corrLength}
\xi = - \frac{1}{\log(\omega^{(2)})}.
\end{equation}
Uniform matrix product states prove to offer a very accurate description of ground states of gapped, translation invariant Hamiltonians in the thermodynamic limit. For simplicity's sake, we will restrict to nearest neighbour interactions, so that $\hat{H}=\sum_n \hat{h}_{n,n+1}$. Having found a variationally optimal tensor $A$ for this Hamiltonian (with variational energy density $e_0$), we can calculate its variance with respect to the state $\ket{\Psi[A]}$ to get an idea of how well it approximates the true ground state. This variance scales with the system size, however, so we should define a local state error as
\begin{align*}
 \epsilon_{\text{GS}}  = \frac{1}{|\ZZ|} \Delta H_\text{GS} = \frac{1}{|\ZZ|} \bra{\Psi[A]} \hat{H}^2 \ket{\Psi[A]} ,
\end{align*}
where $|\ZZ|$ represents the diverging system size and we have redefined the Hamiltonian as $h_{n,n+1} \rightarrow h_{n,n+1}-e_0$. A simple calculation shows that the local state error is equal to
\begin{align*}
\epsilon_{\text{GS}} &= \frac{1}{|\ZZ|} \sum_n\sum_{n'} \bra{\Psi[A]} \hat{h}_{n,n+1} \hat{h}_{n',n'+1} \ket{\Psi[A]} \\
&= \sum_{n'} \bra{\Psi[A]} \hat{h}_{0,1} \hat{h}_{n',n'+1} \ket{\Psi[A]} \\
&= 2\times (l|\H{A}{A}{A}{A} \sum_{n=0}^{+\infty} E^n \H{A}{A}{A}{A} |r) + (l|\HH{A}{A}{(A)}{(A)}{A}{A}|r) + (l|\HH{(A)}{A}{A}{A}{A}{(A)} |r) + (l|\J{A}{A}{A}{A}|r)
\end{align*}
where we have used the following notations
\begin{align*}
&\H{A}{B}{C}{D} = \sum_{ss'tt'} A^s B^t \otimes \ol{C}^{s'} \ol{D}^{t'} \bra{s't'}\hat{h}\ket{st} \\
&\HH{(A)}{B}{C}{D}{E}{(F)} = \sum_{ss'tt'uu'} A^s B^t C^u \otimes \ol{D}^{s'} \ol{E}^{t'} \ol{F}^{u'} \times \sum_{v} \bra{vu'}\hat{h}\ket{tu} \bra{s't'}\hat{h}\ket{sv} \\
&\J{A}{B}{C}{D} = \sum_{ss'tt'} A^s B^t \otimes \ol{C}^{s'} \ol{D}^{t'} \times\sum_{vw} \bra{vw}\hat{h}\ket{st} \bra{s't'}\hat{h}\ket{vw} = \sum_{ss'tt'} A^s B^t \otimes \ol{C}^{s'} \ol{D}^{t'} \times\bra{s't'}\hat{h}^2\ket{st}.
\end{align*}
\par As the transfer matrix has spectral radius 1, the infinite sum does not converge. On every encounter of a geometric sum over $E$, we will seperate it into a \textit{disconnected} part corresponding to the rank 1 projector $Q=|r)(l|$ onto its eigenspace with eigenvalue 1, and a \emph{connected} part corresponding to $\tilde{E}=E-Q = P E = E P = P E P$ with $P=1-Q$ the complementary projector. Since $\tilde{E}$ has a spectral radius smaller than 1, the geometric series over the latter can be safely calculated and we obtain
\begin{equation*}
\sum_{n=0}^{+\infty} E^n = \sum_{n=0}^{+\infty} |r)(l| + P \sum_{n=0}^{+\infty} \tilde{E}^n P = \sum_{n=0}^{+\infty} |r)(l| + (1 - E )^P
\end{equation*}
where the extra projector $P$ in the second term is only necessary to ensure the correct results for the $n=0$ term, and we have introduced the notation
\begin{equation}
(1-E)^P = P (1-\tilde{E})^{-1} P = (1-\tilde{E})^{-1}P.
\end{equation}
$(1-E)^P$ is zero in the eigenspace of $(1-E)$ with eigenvalue zero, and acts as the inverse of $(1-E)$ in the complementary space. It thus acts as a kind of pseudo-inverse, although we will also use the $(\dots)^P$ notation more generally below as $(1-e^{i\kappa} E)^P = P (1-e^{i\kappa}\tilde{E})^{-1} P$. Now using that $(l|\H{A}{A}{A}{A}|r)=0$ through the redefinition of the Hamiltonian, we can conclude that any geometric series of $E$ which has $(l|\H{A}{A}{A}{A}$ to its left or $\H{A}{A}{A}{A}|r)$ to its right will have no contribution from the disconnected part, and yield a convergent (finite) result. In particular, the result for the ``state error density'' is
\begin{equation*}
\epsilon_{\text{GS}} = 2\times (l|\H{A}{A}{A}{A} (1-E)^P \H{A}{A}{A}{A} |r) + (l|\HH{A}{A}{(A)}{(A)}{A}{A}|r) + (l|\HH{(A)}{A}{A}{A}{A}{(A)} |r) + (l|\J{A}{A}{A}{A}|r).
\end{equation*}

\subsection{The particle ansatz} \label{sec:1pA}

The ansatz for an elementary excitation on top of the uMPS ground state, parametrized by the tensor $A$, is given by
\begin{equation} \label{oneparticleA}
    \ket{\Phi_\kappa[B]} = \sum_{n=-\infty}^{+\infty} \e^{i\kappa n} \sum_{\{s\}} \lv \left[ \prod_{m<n} A^{s_m} \right] B^{s_n} \left[ \prod_{m>n} A^{s_m} \right] \rv \spst.
\end{equation}
It is the momentum superposition of a localized disturbance, parametrized by the tensor $B$ (same dimensions as $A$). At zero momentum, this excitation lives in the tangent space of the uMPS manifold with fixed bond dimension, at the point $\ket{\Psi[A]}$ (see Ref.~\onlinecite{Haegeman2013b} for more details). The gauge freedom within this manifold has its reflection in the tangent plane: the state $\ket{\Phi_\kappa[B]}$ is invariant under the transformation
\begin{equation*}
  B^s \rightarrow B^s + X A^s - \e^{i\kappa} A^s X
\end{equation*}
with $X$ a general ($D\times D$) matrix. The tensors $\tilde{B}^s = X A^s - \e^{i\kappa} A^s X$ give rise to so-called null modes. Getting rid of them is possible by imposing a \textit{gauge fixing condition} on the tensors $B$ and introducing a corresponding restricted parametrization. Two choices are especially convenient:
\begin{enumerate}
\item \textit{Left gauge}. We construct the $(qD\times D)$-matrix $L_{a,(b,s)}=((A\dag)^s l^{1/2})_{a,b}$ and find the right null space $V_L$ of $L$, so that $LV_L=0$. This matrix $V_L$ has dimensions $qD\times (q-1)D$ and is orthonormalized: $V_L\dag V_L=\one$. The left gauge fixing condition and its reduced parametrization in terms of the ($D(d-1)\times D$) matrix $X$ are then given by
\begin{equation*}
(l| \E{B}{A} = 0 \qquad \rightarrow \quad B_L[X] = l^{-1/2} V_L^s X r^{-1/2}.
\end{equation*}
\item \textit{Right gauge}. We construct the $(qD\times D)$-matrix $R_{(a,s),b}=(r^{1/2} (A\dag)^s)_{a,b}$ and find the left null space $V_R$ of $R$, so that $V_RR=0$. This matrix $V_R$ has dimensions $(q-1)D\times qD$ and is orthonormalized $V_RV_r\dag = \one$. The right gauge fixing condition and its reduced parametrization in terms of the ($D\times D(d-1)$) matrix $X$ are then given by
\begin{equation*}
 \E{B}{A} |r) = 0 \qquad \rightarrow \quad B_R[X] = l^{-1/2} X V_R^s r^{-1/2}.
\end{equation*}
\end{enumerate}
The expression for the norm of the state $\ket{\Phi_\kappa[B]}$ is simplified with one of these choices to be just the Euclidian norm in terms of the parameters $X$ (up to momentum $\delta$ factor)
\begin{equation} \label{euclideanA}
\braket{\Phi_{\kappa'}[B_{L/R}(X')]|\Phi_{\kappa}[B_{L/R}(X)]} = 2\pi\delta(\kappa'-\kappa) \times \tr(X'X)
\end{equation}
Moreover, with either of these gauge conditions the excitation is orthogonal to the ground state, so that $\braket{\Phi_\kappa[B]|\Psi[A]}=2\pi\delta(\kappa)(l|\E{A}{B}|r) = 0 $. The overlap of the Hamiltonian between two excited states can be calculated to be (see Ref.~\onlinecite{Haegeman2013b})
\begin{equation} \label{H1lA} \begin{split}
  \bra{\Phi_{\kappa'}[B']}\hat{H}\ket{\Phi_{\kappa}[B]} &= 2\pi\delta(\kappa-\kappa') \Big[ (l|\H{B}{A}{B'}{A}|r) + (l|\H{A}{B}{A}{B'}|r) + \e^{-i\kappa}(l|\H{B}{A}{A}{B'}|r) + \e^{i\kappa}(l|\H{A}{B}{B'}{A}|r) \\
  & \quad + (l|\E{B}{B'}(1-E)^P\H{A}{A}{A}{A}|r) + (l|\H{A}{A}{A}{A}(1-E)^P\E{B}{B'}|r) \\
  & \quad + \e^{-i\kappa} (l|\H{A}{B}{A}{A} (1-\e^{-i\kappa}E)^P \E{A}{B'} |r) + \e^{-2i\kappa} (l|\H{B}{A}{A}{A} (1-\e^{-i\kappa}E)^P \E{A}{B'} |r) \\
  & \quad + \e^{i\kappa} (l|\H{A}{A}{A}{B'} (1-\e^{i\kappa}E)^P \E{B}{A} |r) + \e^{2i\kappa} (l|\H{A}{A}{B'}{A} (1-\e^{i\kappa}E)^P \E{B}{A} |r) \\
  & \quad + \e^{-i\kappa} (l|\H{A}{A}{A}{A} (1-E)^P \E{B}{A} (1-\e^{-i\kappa}E)^P \E{A}{B'} |r) \\
  & \quad + \e^{i\kappa} (l|\H{A}{A}{A}{A} (1-E)^P \E{A}{B'} (1-\e^{i\kappa}E)^P \E{B}{A} |r) \Big]
\end{split} \end{equation}
for $B$ and $B'$ in the left gauge and
\begin{equation} \label{H1rA} \begin{split}
  \bra{\Phi_{\kappa'}[B']}\hat{H}\ket{\Phi_{\kappa}[B]} &= 2\pi\delta(\kappa-\kappa') \Big[ (l|\H{B}{A}{B'}{A}|r) + (l|\H{A}{B}{A}{B'}|r) + \e^{-i\kappa}(l|\H{B}{A}{A}{B'}|r) + \e^{i\kappa}(l|\H{A}{B}{B'}{A}|r) \\
  & \quad + (l|\E{B}{B'}(1-E)^P\H{A}{A}{A}{A}|r) + (l|\H{A}{A}{A}{A}(1-E)^P\E{B}{B'}|r) \\
  & \quad + \e^{-i\kappa} (l|\E{B}{A} (1-\e^{-i\kappa}E)^P \H{A}{A}{B'}{A} |r) + \e^{-2i\kappa} (l|\E{B}{A} (1-\e^{-i\kappa}E)^P \H{A}{A}{A}{B'} |r) \\
  & \quad + \e^{i\kappa} (l|\E{A}{B'} (1-\e^{i\kappa}E)^P \H{B}{A}{A}{A} |r) + \e^{2i\kappa} (l|\E{A}{B'} (1-\e^{i\kappa}E)^P \H{A}{B}{A}{A} |r) \\
  & \quad + \e^{-i\kappa} (l|\E{B}{A} (1-\e^{-i\kappa}E)^P \E{A}{B'} (1-E)^P \H{A}{A}{A}{A} |r) \\
  & \quad + \e^{i\kappa} (l|\E{A}{B'} (1-\e^{i\kappa}E)^P \E{B}{A} (1-E)^P \H{A}{A}{A}{A} |r) \Big].
\end{split} \end{equation}
for a right-gauge fixed $B$ and $B'$. We have introduced the notation for a ``generalized'' transfer matrix $\E{A}{B} = \sum_s A^s \otimes \ol{B}^s $.
\par Because of the linear parametrization of \eqref{oneparticleA} in terms of $B$, variationally optimizing this ansatz can be reformulated as an eigenvalue problem
\begin{equation*}
\min_{X} \frac{\bra{\Phi_\kappa[B_{L/R}(X)]}\hat{H}\ket{\Phi_\kappa[B_{L/R}(X)]}}{\braket{\Phi_\kappa[B_{L/R}(X)]|\Phi_\kappa[B_{L/R}(X)]}} \qquad \rightarrow \quad  \mathsf{H}_\text{1p,eff} X = \lambda \mathsf{N}_\text{eff,1p} X
\end{equation*}
with $\mathsf{H}_\text{1p,eff}$ the Hamiltonian overlap matrix between two excited states (Eqs.~\eqref{H1lA} and \eqref{H1rA}) and the effective norm matrix $\mathsf{N}_\text{1p,eff}$ equal to the identity matrix because of Eq.~\eqref{euclideanA}. The eigenvalue $\lambda$ is the excitation energy. By repeating this procedure for different momenta, we can trace out the excitation spectrum. Note that the interpretation of the solutions in terms of one- and multi-particle excitations can be made on the basis of the computation of the variance, as explained in Sec. \ref{sec:variance}. Indeed, it might very well be that the lowest eigenvalue at a certain momentum corresponds to a two-particle scattering state.

\subsection{One-particle form factors}
\label{sec:form1p}

The states \eqref{oneparticleA} provide a variational approximation for the true low-lying excitations of the full Hamiltonian. Their overlaps with a local operator acting on the ground state (their spectral weights) provide an important contribution to the spectral function
\begin{equation*}
S(\kappa,\omega) = \sum_{n=-\infty}^{+\infty} \int_{-\infty}^{+\infty} \d t \, \e^{i(\omega t - \kappa n)} \bra{\Psi_0} O_n\dag(t) O_0(0) \ket{\Psi_0}.
\end{equation*}
By inserting a projector on the one-particle subspace, the one-particle contribution can be written as ($\Gamma_1(\kappa)$ denotes the set of one-particle states at momentum $\kappa$)
\begin{equation*}
S(\kappa,\omega)_\text{1p} = \sum_{\alpha\in\Gamma_1(\kappa)} 2\pi \delta(\Delta_\alpha(\kappa)-\omega) \left| \bra{\Phi_\kappa[B_\alpha]} \hat{O}_0 \ket{\Psi[A]} \right|^2.
\end{equation*}
The overlap is given by (with $B_\alpha$ in the left gauge)
\begin{equation*}
\bra{\Phi_\kappa[B_\alpha]} \hat{O}_0 \ket{\Psi[A]} = (l| \O{A}{B_\alpha} |r) + (l| \O{A}{A} (1-E)^P \E{A}{B_\alpha} |r)
\end{equation*}
where we have again generalized our notation to an ``operator transfer matrix'' $\O{A}{B} = \sum_{s,t} A^s \otimes \ol{B}^t \bra{t}\hat{O}\ket{s}$.

\subsection{Variance of the one-particle ansatz}
\label{sec:variance}

If we write the one-particle ansatz as
\begin{equation*}
\ket{\Phi_\kappa[B]} = \sum_n \e^{i\kappa n} \ket{\chi(n)} \qquad \text{with} \quad \ket{\chi(n)} = \sum_{\{s\}} \lv \left[ \prod_{m<n} A^{s_m} \right] B^{s_n} \left[ \prod_{m>n} A^{s_m} \right] \rv \spst,
\end{equation*}
where $B$ is in the left gauge, such that the site dependent states are orthonormalized as $\braket{\chi(n')|\chi(n)}=\delta_{nn'}$. The variance of the Hamiltonian with respect to this state can be calculated as (we denote $\Delta(\kappa)=\bra{\Phi_\kappa[B]} H \ket{\Phi_\kappa[B]}$)
\begin{align*}
\epsilon_\text{EX} &= \bra{\Phi_{\kappa'}[B]} (\hat{H}-\Delta(\kappa))^2 \ket{\Phi_\kappa[B]} \\
&= \sum_{n} \e^{i\kappa n} \sum_{n'} \e^{-i\kappa'n'}\bra{\chi(n')} \hat{H}^2 \ket{\chi(n)} - \Delta(\kappa)^2 \braket{\Phi_{\kappa'}[B]|\Phi_\kappa[B]} \\
& = 2\pi\delta(\kappa-\kappa') \left( \sum_{n'=-\infty}^{+\infty} \e^ {-i\kappa n' } \bra{\chi(n')} \hat{H}^2 \ket{\chi(0)} - \Delta(\kappa)^2 \right) .
\end{align*}
Does this expression make sense? First of all, the sum breaks off for high enough $n'$, i.e. $\bra{\chi(n')} \hat{H}^2 \ket{\chi(0)}\rightarrow0$ for $n'$ large enough, as we will see later on. The infinite $\delta$-prefactor is also no problem as this is just the norm of the momentum superposition state. The contribution $\bra{\chi(0)} \hat{H}^2 \ket{\chi(0)}$ is somewhat more problematic however, as it contains an infinite contribution from the ground state error. Therefore, we subtract the (infinite) ground state variance from this expression. We get the following
\begin{align*}
\epsilon_\text{EX} = 2\pi\delta(\kappa-\kappa') \left[ \bra{\chi(0)} \hat{H}^2 - \Delta H_\text{GS} \ket{\chi(0)} + \sum_{n'=1}^{+\infty} \left( \e^ {-i\kappa n' } \bra{\chi(n')} \hat{H}^2 \ket{\chi(0)} + c.c. \right) - \Delta(\kappa)^2 \right].
\end{align*}
In the calculations it will become clear that this is indeed a finite expression.
\par \textit{The first contribution.} We will first calculate the contribution $\bra{\chi(0)} \hat{H}^2 - \Delta H_\text{GS} \ket{\chi(0)}$. We have two infinite sums and one infinite quantity in this contribution, so we have to be precise in our summations. We have
\begin{align*}
\bra{\chi(0)} \hat{H}^2 - \Delta H_\text{GS} \ket{\chi(0)}  = \sum_{n,n'} \left(  \bra{\chi(0)} \hat{h}_{n,n+1} \hat{h}_{n',n'+1} \ket{\chi(0)} - \bra{\Psi[A]} \hat{h}_{n,n+1} \hat{h}_{n',n'+1} \ket{\Psi[A]} \braket{\chi(0)|\chi(0)}\right).
\end{align*}
Every term for $n$ can be calculated individually, making sure that the right number of ground state errors $\epsilon_\text{GS}$ is subtracted
\begin{align*}
\bra{\chi(0)} & \hat{H}^2 - \Delta H_\text{GS} \ket{\chi(0)}  \\
&= \sum_{n=-\infty}^{-3} \Big[ (l|\left(\H{A}{A}{A}{A}(1-E)^P \H{A}{A}{A}{A} \E{A}{A} + \HH{A}{A}{(A)}{(A)}{A}{A} \E{A}{A} + \J{A}{A}{A}{A} \E{A}{A} + \HH{(A)}{A}{A}{A}{A}{(A)} \right) (\E{A}{A})^{|n|-3} \E{B}{B}|r) \\
& \qquad + (l| \H{A}{A}{A}{A} \sum_{n'=n+2}^{-2} (\E{A}{A})^{|n|-|n'|-2} \H{A}{A}{A}{A} (\E{A}{A})^{|n'|-2}
\E{B}{B}|r) \\
& \qquad  + (l| \H{A}{A}{A}{A} (\E{A}{A})^{|n|-3} \left( \H{A}{B}{A}{B} + \E{A}{A} \H{B}{A}{B}{A} + \E{A}{A} \E{B}{B} (1-E)^P \H{A}{A}{A}{A} \right) |r) - \epsilon_\text{GS}(l|\E{B}{B}|r) \Big] \\
& + (l| \left(\H{A}{A}{A}{A}(1-E)^P \H{A}{A}{A}{A} \E{B}{B} + \HH{(A)}{A}{A}{A}{A}{(A)} \E{B}{B} + \J{A}{A}{A}{A} \E{B}{B} + \HH{A}{A}{(B)}{(A)}{A}{B} \right. \\
& \hspace{5cm} + \left. \H{A}{A}{A}{A} \H{B}{A}{B}{A} + \H{A}{A}{A}{A} \E{B}{B} (1-E)^P \H{A}{A}{A}{A} \right) |r) - \epsilon_\text{GS}(l|\E{B}{B}|r) \\
& + (l| \left(\H{A}{A}{A}{A}(1-E)^P \H{A}{B}{A}{B} + \HH{(A)}{A}{B}{A}{A}{(B)} + \J{A}{B}{A}{B} + \HH{A}{B}{(A)}{(A)}{B}{A} + \H{A}{B}{A}{B}(1-E)^P \H{A}{A}{A}{A} \right) |r) - \epsilon_\text{GS}(l|\E{B}{B}|r) \\
& + (l| \left( \H{A}{A}{A}{A}(1-E)^P \H{B}{A}{B}{A} + \HH{(A)}{B}{A}{A}{B}{(A)} + \J{B}{A}{B}{A} + \HH{B}{A}{(A)}{(B)}{A}{A} + \H{B}{A}{B}{A} (1-E)^P\H{A}{A}{A}{A} \right) |r) - \epsilon_\text{GS}(l|\E{B}{B}|r) \\
& + (l| \left( \H{A}{A}{A}{A}(1-E)^P \E{B}{B} \H{A}{A}{A}{A} + \H{A}{B}{A}{B} \H{A}{A}{A}{A} + \HH{(B)}{A}{A}{B}{A}{(A)} + \E{B}{B} \J{A}{A}{A}{A} \right. \\
& \hspace{5cm} +  \left. \E{B}{B}\HH{A}{A}{(A)}{(A)}{A}{A} + \E{B}{B} \H{A}{A}{A}{A} (1-E)^P\H{A}{A}{A}{A} \right) |r) - \epsilon_\text{GS}(l|\E{B}{B}|r)  \\
& + \sum_{n=2}^{+\infty} \Big[ (l| \left( \H{A}{A}{A}{A}(1-E)^P \E{B}{B} \E{A}{A} + \H{A}{B}{A}{B} \E{A}{A} + \H{B}{A}{B}{A} \right) (\E{A}{A})^{n-2} \H{A}{A}{A}{A} |r) \\
& \qquad + (l| \E{B}{B} \sum_{n'=1}^{n-2} (\E{A}{A})^{n'-1} \H{A}{A}{A}{A} (\E{A}{A})^{n-n'-2} \H{A}{A}{A}{A} |r) \\
& \qquad + (l| \E{B}{B} (\E{A}{A})^{n-2} \left( \HH{(A)}{A}{A}{A}{A}{(A)} + \E{A}{A} \J{A}{A}{A}{A} + \E{A}{A}\HH{A}{A}{(A)}{(A)}{A}{A} + \E{A}{A} \H{A}{A}{A}{A}(1-E)^P\H{A}{A}{A}{A} \right) |r) - \epsilon_\text{GS}(l|\E{B}{B}|r) \Big].
\end{align*}
The infinite sums on the first and last three lines need to be investigated further. Separating all powers of $\E{A}{A}$ into connected and disconnected parts, the connected parts will yield finite results. This also enables to interchange the sums (with appropriate redefinition of the summation boundaries) in the double sum on the second and second to last line, so as to obtain for e.g. the latter
\begin{align*}
(l| \E{B}{B} (1-E)^P \H{A}{A}{A}{A} (1-E)^P \H{A}{A}{A}{A} |r).
\end{align*}
The disconnected and potentially diverging contributions that survive in e.g. the last three lines are given by
\begin{align*}
&(l| \E{B}{B}|r) \sum_{n=2}^{+\infty} \Big[  \sum_{n'=1}^{n-2} (l| \H{A}{A}{A}{A} (\E{A}{A})^{n-n'-2} \H{A}{A}{A}{A} |r) + (l| \left( \HH{(A)}{A}{A}{A}{A}{(A)} + \J{A}{A}{A}{A} + \HH{A}{A}{(A)}{(A)}{A}{A} +  \H{A}{A}{A}{A}(1-E)^P\H{A}{A}{A}{A} \right) |r) - \epsilon_\text{GS} \Big].
\end{align*}
By writing the $\sum_{n'=1}^{n-2} = \sum_{n'=-\infty}^{n-2} - \sum_{n'=-\infty}^{0}$ and substituting $n'\rightarrow-n'$ in the last sum, we obtain
\begin{align*}
&(l| \E{B}{B}|r) \sum_{n=2}^{+\infty} \Big[  (l| \H{A}{A}{A}{A} (1-E)^P \H{A}{A}{A}{A} |r) + (l| \left( \HH{(A)}{A}{A}{A}{A}{(A)} + \J{A}{A}{A}{A} + \HH{A}{A}{(A)}{(A)}{A}{A} +  \H{A}{A}{A}{A}(1-E)^P\H{A}{A}{A}{A} \right) |r) - \epsilon_\text{GS} \Big]\\
& - (l| \E{B}{B}|r)(l| \H{A}{A}{A}{A} (1-E)^P(1-E)^P \H{A}{A}{A}{A} |r).
\end{align*}
The terms in the remaining infinite sum exactly cancel thanks to presence of $\epsilon_{\text{GS}}$ and the finite result of the second line is obtained. A similar result is obtained from the disconnected part of the first three lines. Inserting this in the complete expression yields
\begin{align*}
 \bra{\chi(0)} \hat{H}^2 - \Delta H_\text{GS} \ket{\chi(0)}  &= (l| \left(\H{A}{A}{A}{A}(1-E)^P \H{A}{A}{A}{A} \E{B}{B} + \HH{(A)}{A}{A}{A}{A}{(A)} \E{B}{B} + \J{A}{A}{A}{A} \E{B}{B} + \HH{A}{A}{(B)}{(A)}{A}{B} \right. \\
& \hspace{5cm} + \left. \H{A}{A}{A}{A} \H{B}{A}{B}{A} + \H{A}{A}{A}{A} \E{B}{B} (1-E)^P \H{A}{A}{A}{A} \right) |r)  \\
& \qquad + (l| \left(\H{A}{A}{A}{A}(1-E)^P \H{A}{B}{A}{B} + \HH{(A)}{A}{B}{A}{A}{(B)} + \J{A}{B}{A}{B} + \HH{A}{B}{(A)}{(A)}{B}{A} + \H{A}{B}{A}{B}(1-E)^P \H{A}{A}{A}{A} \right) |r) \\
& \qquad + (l| \left( \H{A}{A}{A}{A}(1-E)^P \H{B}{A}{B}{A} + \HH{(A)}{B}{A}{A}{B}{(A)} + \J{B}{A}{B}{A} + \HH{B}{A}{(A)}{(B)}{A}{A} + \H{B}{A}{B}{A} (1-E)^P\H{A}{A}{A}{A} \right) |r) \\
& \qquad + (l| \left( \H{A}{A}{A}{A}(1-E)^P \E{B}{B} \H{A}{A}{A}{A} + \H{A}{B}{A}{B} \H{A}{A}{A}{A} + \HH{(B)}{A}{A}{B}{A}{(A)} + \E{B}{B} \J{A}{A}{A}{A} \right. \\
& \hspace{5cm} +  \left. \E{B}{B}\HH{A}{A}{(A)}{(A)}{A}{A} + \E{B}{B} \H{A}{A}{A}{A} (1-E)^P\H{A}{A}{A}{A} \right) |r) \\
& \qquad - 4\times\epsilon_\text{GS} \\
& \qquad + (l| \left( \H{A}{A}{A}{A}(1-E)^P \H{A}{A}{A}{A} \E{A}{A} + \H{A}{A}{A}{A}(1-E)^P \H{A}{A}{A}{A} + \HH{(A)}{A}{A}{A}{A}{(A)} \E{A}{A} \right. \\
& \hspace{5cm} \left. + \J{A}{A}{A}{A} \E{A}{A} + \HH{A}{A}{(A)}{(A)}{A}{A} \right) (1-E)^P \E{B}{B}|r) \\
& \qquad  + (l| \H{A}{A}{A}{A} (1-E)^P \left( \H{A}{B}{A}{B} + \E{A}{A} \H{B}{A}{B}{A} + \E{A}{A} \E{B}{B} (1-E)^P \H{A}{A}{A}{A} \right) |r) \\
& \qquad + (l| \E{B}{B} (1-E)^P \left( \HH{(A)}{A}{A}{A}{A}{(A)} + \E{A}{A} \J{A}{A}{A}{A} + \E{A}{A} \HH{A}{A}{(A)}{(A)}{A}{A} \right. \\
& \hspace{5cm} \left. + \H{A}{A}{A}{A}(1-E)^P\H{A}{A}{A}{A} + \E{A}{A} \H{A}{A}{A}{A}(1-E)^P\H{A}{A}{A}{A} \right) |r) \\
& \qquad + (l| \left( \H{A}{A}{A}{A}(1-E)^P \E{B}{B} \E{A}{A} + \H{A}{B}{A}{B} \E{A}{A} + \H{B}{A}{B}{A} \right) (1-E)^P \H{A}{A}{A}{A} |r) \\
& \qquad - 2\times (l| \H{A}{A}{A}{A} (1-E)^P (1-E)^P \H{A}{A}{A}{A} |r).
\end{align*}
\par \textit{All other contributions.} Next we calculate $\bra{\chi(0)} \hat{H}^2 \ket{\chi(1)}$. No problems with subtracting an infinite amount of ground state errors is present here, so we have
\begin{align*}
\bra{\chi(1)} \hat{H}^2 \ket{\chi(0)}  &= (l|\left(2\times\H{A}{A}{A}{A}(1-E)^P \H{A}{A}{A}{A} (1-E)^P \E{B}{A} + \HH{(A)}{A}{A}{A}{A}{(A)} (1-E)^P \E{B}{A} + \J{A}{A}{A}{A} (1-E)^P \E{B}{A} \right. \\
& \qquad \left. \HH{A}{A}{(A)}{(A)}{A}{A} (1-E)^P \E{B}{A} + 2\times \H{A}{A}{A}{A} (1-E)^P \H{A}{B}{A}{A} + \HH{A}{A}{(B)}{(A)}{A}{A} + \HH{(A)}{A}{B}{A}{A}{(A)} + \J{A}{B}{A}{A} \right) \E{A}{B} |r) \\
& + (l|\left( 2\times \H{A}{A}{A}{A} (1-E)^P \H{B}{A}{A}{B} + \HH{A}{B}{(A)}{(A)}{A}{B} + \J{B}{A}{A}{B} + \HH{(A)}{B}{A}{A}{A}{(B)} \right) |r) \\
& + (l| \left( 2\times \H{A}{A}{A}{A}(1-E)^P \E{B}{A} \H{A}{A}{B}{A} + 2\times \H{A}{B}{A}{A} \H{A}{A}{B}{A} + \HH{B}{A}{(A)}{(A)}{B}{A} + \HH{(B)}{A}{A}{A}{B}{(A)} \right) |r) \\
& + 2\times (l| \left( \H{A}{A}{A}{A}(1-E)^P \E{BA}{AB} + \H{A}{B}{A}{A} \E{A}{B} + \H{B}{A}{A}{B} \right)(1-E)^P \H{A}{A}{A}{A} |r).
\end{align*}
Analogously,
\begin{align*}
\bra{\chi(2)} \hat{H}^2 \ket{\chi(0)}  &= (l|\left(2\times\H{A}{A}{A}{A}(1-E)^P \H{A}{A}{A}{A} (1-E)^P \E{B}{A} + \HH{(A)}{A}{A}{A}{A}{(A)} (1-E)^P \E{B}{A} \right.\\
& \qquad \qquad + \J{A}{A}{A}{A} (1-E)^P \E{B}{A} + \HH{A}{A}{(A)}{(A)}{A}{A} (1-E)^P \E{B}{A} \\
& \qquad \qquad  + \left.2\times \H{A}{A}{A}{A} (1-E)^P \H{A}{B}{A}{A} + \HH{A}{A}{(B)}{(A)}{A}{A} + \HH{(A)}{A}{B}{A}{A}{(A)} + \J{A}{B}{A}{A} \right) (\E{A}{A}) \E{A}{B} |r) \\
& \qquad + (l|\left( 2\times \H{A}{A}{A}{A} (1-E)^P \H{B}{A}{A}{A} + \HH{A}{B}{(A)}{(A)}{A}{A} + \J{B}{A}{A}{A} + \HH{(A)}{B}{A}{A}{A}{(A)} \right) \E{A}{B} |r) \\
& \qquad + (l| \left(2\times\H{A}{A}{A}{A}(1-E)^P\E{B}{A}\H{A}{A}{A}{B}+2\times\H{A}{B}{A}{A}\H{A}{A}{A}{B} + \HH{B}{A}{(A)}{(A)}{A}{B} + \HH{(B)}{A}{A}{A}{A}{(B)} \right) |r) \\
& \qquad + 2\times (l| \left( \H{A}{A}{A}{A}(1-E)^P \E{BA}{AA} + \H{A}{B}{A}{A} \E{A}{A} + \H{B}{A}{A}{A} \right) \left( \H{A}{A}{B}{A} + \E{A}{B} (1-E)^P \H{A}{A}{A}{A} \right) |r)
\end{align*}
and for $n>2$
\begin{align*}
\bra{\chi(n)} \hat{H}^2 \ket{\chi(0)}  &= (l|\left(2\times\H{A}{A}{A}{A}(1-E)^P \H{A}{A}{A}{A} (1-E)^P \E{B}{A} + \HH{(A)}{A}{A}{A}{A}{(A)} (1-E)^P \E{B}{A} \right. \\
& \qquad \qquad + \left. \J{A}{A}{A}{A} (1-E)^P \E{B}{A} + \HH{A}{A}{(A)}{(A)}{A}{A} (1-E)^P \E{B}{A} \right. \\
& \qquad \qquad +  \left. 2\times \H{A}{A}{A}{A} (1-E)^P \H{A}{B}{A}{A} + \HH{A}{A}{(B)}{(A)}{A}{A} + \HH{(A)}{A}{B}{A}{A}{(A)} + \J{A}{B}{A}{A} \right) (\E{A}{A})^{n-1} \E{A}{B} |r) \\
& \qquad +(l|\left( 2\times \H{A}{A}{A}{A} (1-E)^P \H{B}{A}{A}{A} + \HH{A}{B}{(A)}{(A)}{A}{A} + \J{A}{B}{A}{A} + \HH{(A)}{B}{A}{A}{A}{(A)} \right) (\E{A}{A})^{n-2} \E{A}{B} |r) \\
& \qquad + (l| \left( 2\times\H{A}{A}{A}{A}(1-E)^P\E{B}{A}\H{A}{A}{A}{A}+2\times\H{A}{B}{A}{A}\H{A}{A}{A}{A}+\HH{B}{A}{(A)}{(A)}{A}{A} + \HH{(B)}{A}{A}{A}{A}{(A)} \right) (\E{A}{A})^{n-3} \E{A}{B} |r) \\
& \qquad + 2\times (l| \left( \H{A}{A}{A}{A}(1-E)^P \E{BA}{AA} + \H{A}{B}{A}{A} \E{A}{A} + \H{B}{A}{A}{A} \right) \\
& \qquad \qquad \qquad \times  \left( \sum_{i=0}^{n-4} (\E{A}{A})^i \H{A}{A}{A}{A} (\E{A}{A})^{n-4-i} \E{A}{B} + (\E{A}{A})^{n-3} \H{A}{A}{A}{B} \right. \\
& \hspace{5cm}  \left. \phantom{\sum_{i=0}^{n-4}} +   (\E{A}{A})^{n-2} \H{A}{A}{B}{A} + (\E{A}{A})^{n-2} \E{A}{B} (1-E)^P \H{A}{A}{A}{A} \right) |r).
\end{align*}
We can throw everything together in order to obtain
\begin{align*}
\sum_{n=1}^\infty & e^{-i\kappa n}\bra{\chi(n)}\hat{H}^2\ket{\chi(0)} = \\
& \e^{-i\kappa} (l|\left(2\times\H{A}{A}{A}{A}(1-E)^P \H{A}{A}{A}{A} (1-E)^P + \HH{(A)}{A}{A}{A}{A}{(A)} (1-E)^P \right. \\
& \qquad \left. + \J{A}{A}{A}{A} (1-E)^P + \HH{A}{A}{(A)}{(A)}{A}{A} (1-E)^P \right) \E{B}{A} (1-\e^{-i\kappa}E)^{-1} \E{A}{B} |r) \\
& + \e^{-i\kappa} (l| \left( 2\times \H{A}{A}{A}{A} (1-E)^P \H{A}{B}{A}{A} + \HH{A}{A}{(B)}{(A)}{A}{A} + \HH{(A)}{A}{B}{A}{A}{(A)} + \J{A}{B}{A}{A} \right) (1-\e^{-i\kappa}E)^{-1} \E{A}{B} |r) \\
& + \e^{-i\kappa}(l|\left( 2\times \H{A}{A}{A}{A} (1-E)^P \H{B}{A}{A}{B} + \HH{A}{B}{(A)}{(A)}{A}{B} + \J{B}{A}{A}{B} + \HH{(A)}{B}{A}{A}{A}{(B)} \right) |r) \\
& + \e^{-2i\kappa} (l|\left( 2\times \H{A}{A}{A}{A} (1-E)^P \H{B}{A}{A}{A} + \HH{A}{B}{(A)}{(A)}{A}{A} + \J{A}{B}{A}{A} + \HH{(A)}{B}{A}{A}{A}{(A)} \right) (1-\e^{-i\kappa}E)^{-1} \E{A}{B} |r) \\
& + \e^{-i\kappa} (l| \left( 2\times \H{A}{A}{A}{A}(1-E)^P \E{B}{A} \H{A}{A}{B}{A} + 2\times \H{A}{B}{A}{A} \H{A}{A}{B}{A} + \HH{B}{A}{(A)}{(A)}{B}{A} + \HH{(B)}{A}{A}{A}{B}{(A)} \right) |r) \\
& + \e^{-2i\kappa} (l| \left( 2\times\H{A}{A}{A}{A} (1-E)^P \H{A}{A}{A}{B} \E{B}{A} + 2\times \H{A}{B}{A}{A} \H{A}{A}{A}{B} + \HH{B}{A}{(A)}{(A)}{A}{B} + \HH{(B)}{A}{A}{A}{A}{(B)} \right) |r) \\
& + \e^{-3i\kappa} (l| \left( 2\times \H{A}{A}{A}{A} (1-E)^P \E{B}{A} \H{A}{A}{A}{A} + 2\times \H{A}{B}{A}{A} \H{A}{A}{A}{A} + \HH{B}{A}{(A)}{(A)}{A}{A} + \HH{(B)}{A}{A}{A}{A}{(A)} \right) (1-\e^{-i\kappa}E)^{-1} \E{A}{B} |r) \\
& + 2\times \e^{-i\kappa}  (l| \left( \H{A}{A}{A}{A}(1-E)^P \E{BA}{AB} + \H{A}{B}{A}{A} \E{A}{B} + \H{B}{A}{A}{B} \right)(1-E)^P \H{A}{A}{A}{A} |r) \\
& + 2\times \e^{-2i\kappa} (l| \left( \H{A}{A}{A}{A}(1-E)^P \E{BA}{AA} + \H{A}{B}{A}{A} \E{A}{A} + \H{B}{A}{A}{A} \right) (1-\e^{-i\kappa}E)^{-1} \left( \H{A}{A}{B}{A} + \E{A}{B} (1-E)^P \H{A}{A}{A}{A} \right) |r) \\
& + 2\times \e^{-3i\kappa} (l| \left( \H{A}{A}{A}{A}(1-E)^P \E{BA}{AA} + \H{A}{B}{A}{A} \E{A}{A} + \H{B}{A}{A}{A} \right) (1-\e^{-i\kappa}E)^{-1} \H{A}{A}{A}{B} |r) \\
& + 2\times \e^{-4i\kappa} (l| \left( \H{A}{A}{A}{A}(1-E)^P \E{BA}{AA} + \H{A}{B}{A}{A} \E{A}{A} + \H{B}{A}{A}{A} \right) (1-\e^{-i\kappa}E)^{-1} \H{A}{A}{A}{A} (1-\e^{-i\kappa}E)^{-1} \E{A}{B}|r).
\end{align*}
Note that the infinite sum could give rise to one potential divergence coming from the disconnected contribution of the last line of $\braket{\chi(n)|{\ham}^2 |\chi(0)}$ corresponding to
\begin{align*}
 \sum_{n=3}^{\infty} \sum_{i=0}^{n-4} (l| \left( \H{A}{A}{A}{A}(1-E)^P \E{B}{A} + \H{A}{B}{A}{A} + \H{B}{A}{A}{A} \right) |r)(l| \left(\H{A}{A}{A}{A} (\E{A}{A})^{n-4-i} \E{A}{B} + \H{A}{A}{A}{B} + \H{A}{A}{B}{A}\right)|r).
 \end{align*}
However, the first factor is automatically zero if $\ket{\Psi[A]}$ is a variational minimum within the MPS manifold, as it corresponds exactly to the directional derivative of the energy expectation value in the direction of $B$.

\section{Two-particle excitations: technical details}

In this appendix we give all technical details concerning the two-particle ansatz that was defined as
\begin{align*}
 &  \ket{\Upsilon(K,\omega)} = \sum_{n=0}^{+\infty} \sum_{j=1}^{M_n} c_{K,\omega}^j(n) \ket{\chi_{j,K}(n)}
 \end{align*}
 with
 \begin{align*}
   & \qquad \ket{\chi_{j,K}(0)} = \sum_{n=-\infty}^{+\infty} \e^{i K n} \sum_{\{s\}=1}^d \lv \left[ \prod_{m<n} A^{s_m} \right] B_j^{s_n} \left[ \prod_{m>n} A^{s_m} \right] \rv \spst \\
  & \qquad \ket{\chi_{(j_1,j_2),K}(n)} = \sum_{n_1=-\infty}^{+\infty} \e^{i K n_1} \sum_{\{s\}=1}^d \lv \left[ \prod_{m<n_1} A^{s_m} \right] B_{j_1}^{s_{n_1}} \left[ \prod_{n_1<m<n_1+n} A^{s_m} \right]  B_{j_2}^{s_{n_1+n}} \left[ \prod_{m>n_1+n} A^{s_m} \right] \rv \spst.
\end{align*}
Just as in the case of a one-particle excitation, there is a gauge freedom in this ansatz. We can again choose a left or right gauge fixing condition on the tensors $B_j$, depending on the situation. We will choose to put all $B$ tensors in the left gauge fixing condition, which has the consequence that the states $\ket{\chi_j(n)}$ are not orthogonal for different $n$ (see further). As was argued in the main body, this choice allows for the strictly local term, for which we keep all variational parameters, to correct for the inability of the other terms to describe the deformation of the particles as they approach. Alternatively, one could choose the left tensor $B_{j_1}$ to be in the left gauge and the right tensor $B_{j_2}$ in the right gauge; this would make the states $\ket{\chi_j(n)}$ orthogonal for different $n$. When studying bound states with the two-particle ansatz, this might prove to be a better choice.
\par Similar to the one-particle case, we can enforce the gauge fixing conditions by implementing an effective parametrization in terms of a matrix $X$ with $D^2(d-1)$ elements. As we keep all variational freedom in the strictly local term $\ket{\chi_{j,K}(0)}$, this will correspond to $D^2(d-1)$ variational parameters. In the non-local terms $\ket{\chi_{(j_1,j_2),K}(n)}$ we insert a basis of left-gauged tensors $B_{j_1}$ and $B_{j_2}$ which both describe the (relevant part of the) one-particle spectrum. If we have $L$ particles in the system and we need $M$ basis vectors to describe the dispersion of each, we will have $(L\times M)\times(L\times M)$ basis states $\ket{\chi_{(j_1,j_2),K}(n)}$. The gauge fixing and normalization conditions on all the $B$ tensors can be summarized as
\begin{align*}
(l| \E{B_j}{A} = (l| \E{B_{j_1}}{A} = (l|\E{B_{j_2}}{A} = 0 \qquad \text{and} \qquad (l| \E{B_{j_1}}{B_{j_2}} |r) = \delta_{j_1,j_2}.
\end{align*}

\subsection{Effective norm matrix}

The effective norm matrix $(\mathsf{N}_\text{eff})_{n'j',nj} = \braket{\chi_{j',K}(n') | \chi_{j,K}(n)}$ has matrix elements
\begin{align*}
  & \braket{\chi_{j',K}(0) | \chi_{j,K}(0)} = 2\pi\delta(K-K') (l|\E{B_j}{B_{j'}}|r) = 2\pi\delta(K-K') \delta_{j,j'} \\
  & \braket{\chi_{j',K}(n') | \chi_{j,K}(0)} = 2\pi\delta(K-K') (l|\E{B_j}{B_{j_1'}} (\E{A}{\tA})^{n'-1} \E{A}{B_{j_2'}} |r)  \\
  & \braket{\chi_{j',K}(n) | \chi_{j,K}(n)} = 2\pi\delta(K-K') (l|\E{B_{j_1}}{B_{j_1'}} (\E{A}{\tA})^{n-1} \E{B_{j_2}}{B_{j_2'}} |r)  \\
  & \braket{\chi_{j',K}(n') | \chi_{j,K}(n)} = 2\pi\delta(K-K') (l|\E{B_{j_1}}{B_{j_1'}} (\E{A}{\tA})^{n-1} \E{B_{j_2}}{A} (\E{A}{\tA})^{n'-n-1} \E{A}{B_{j_2'}} |r)  \qquad (n'>n).
\end{align*}

\subsection{Effective Hamiltonian matrix}

The effective Hamiltonian matrix $(\mathsf{H}_\text{eff})_{n'j',nj} = \bra{\chi_{j',K}(n')}\hat{H} \ket{ \chi_{j,K}(n)}$ has matrix elements
\begin{align*}
  & \hspace{-1cm} \bra{\chi_{j',K'}(0)}\hat{H} \ket{ \chi_{j,K}(0)} = 2\pi\delta(K-K') \Big[ \\
  & (l|\H{B_j}{A}{B_{j'}}{A}|r) + (l|\H{A}{B_j}{A}{B_{j'}}|r) + \e^{-iK}(l|\H{B_j}{A}{A}{B_{j'}}|r) + \e^{iK}(l|\H{A}{B_j}{B_{j'}}{A}|r) \\
  & \quad + (l|\E{B_j}{B_{j'}}(1-E)^P\H{A}{A}{A}{A}|r) + (l|\H{A}{A}{A}{A}(1-E)^P\E{B_j}{B_{j'}}|r) \\
  & \quad + \e^{-iK} (l|\H{A}{B_j}{A}{A} (1-\e^{-iK}E)^P \E{A}{B_{j'}} |r) + \e^{-2iK} (l|\H{B_j}{A}{A}{A} (1-\e^{-iK}E)^P \E{A}{B_{j'}} |r) \\
  & \quad + \e^{iK} (l|\H{A}{A}{A}{B_{j'}} (1-\e^{iK}E)^P \E{B_j}{A} |r) + \e^{2iK} (l|\H{A}{A}{B_{j'}}{A} (1-\e^{iK}E)^P \E{B_j}{A} |r) \\
  & \quad + \e^{-iK} (l|\H{A}{A}{A}{A} (1-E)^P \E{B_j}{A} (1-\e^{-iK}E)^P \E{A}{B_{j'}} |r) \\
  & \quad + \e^{iK} (l|\H{A}{A}{A}{A} (1-E)^P \E{A}{B_{j'}} (1-\e^{iK}E)^P \E{B_j}{A} |r) \Big] \\
  & \hspace{-1cm} \bra{\chi_{(j_1',j_2'),K'}(1)}\hat{H} \ket{ \chi_{j,K}(0)} = 2\pi\delta(K-K') \Big[ \\
  & (l| \H{A}{A}{A}{A}(1-E)^P \E{B_jA}{B_{j_1'} B_{j_2'}} |r) + (l| \H{A}{B_j}{A}{B_{j_1'}} \E{A}{B_{j_2'}} |r) + (l|\H{B_j}{A}{B_{j_1'}}{B_{j_2'}} |r) \\
  & \qquad + (l| \E{B_j}{B_{j_1}} \H{A}{A}{B_{j_2'}}{A} |r) + (l| \E{B_jA}{B_{j_1'} B_{j_2'}} (1-E)^P \H{A}{A}{A}{A} |r) \\
  & + \e^{-iK} \Big( (l| \H{A}{A}{A}{A} (1-E)^P \E{B_j}{A} \E{AA}{B_{j_1'} B_{j_2'}} |r) + (l| \H{A}{B_j}{A}{A} \E{AA}{B_{j_1'} B_{j_2'}} |r) + (l| \H{B_j}{A}{A}{B_{j_1'}} \E{A}{B_{j_2'}} |r) \Big) \\
  & + \e^{-2iK} (l| \Big( \H{A}{A}{A}{A} (1-E)^P \E{B_j}{A} \E{A}{A} + \H{A}{B_j}{A}{A} \E{A}{A} + \H{B_j}{A}{A}{A} \Big) (1-\e^{-iK}E)^P \E{AA}{B_{j_1'} B_{j_2'}} |r) \\
  & + \e^{iK} \Big( (l| \H{A}{A}{A}{A}(1-E)^P \E{AB_j}{B_{j_1'} B_{j_2'}} |r) + (l|\H{A}{A}{A}{B_{j_1'}}\E{B_j}{B_{j_2'}} |r) + (l| \H{A}{B_j}{B_{j_1'}}{B_{j_2'}} |r) \Big) \\
  & + \e^{2iK} (l| \Big( \H{A}{A}{A}{A}(1-E)^P \E{AA}{B_{j_1'} B_{j_2'}} + \H{A}{A}{A}{B_{j_1'}} \E{A}{B_{j_2'}} + \H{A}{A}{B_{j_1'}}{B_{j_2'}} \Big) (1-\e^{iK}E)^P \E{B_j}{A} |r) \Big] \\
  & \hspace{-1cm} \bra{\chi_{(j_1',j_2'),K'}(n')}\hat{H} \ket{ \chi_{j,K}(0)} = 2\pi\delta(K-K') \Big[ \\
  & (l| \Big( \H{A}{A}{A}{A}(1-E)^P \E{B_j}{B_{j_1'}} \E{A}{\tA} + \H{A}{B_j}{A}{B_{j_1'}} \E{A}{\tA} + \H{B_j}{A}{B_{j_1'}}{\tA}  \Big) (\E{A}{\tA})^{n'-2} \E{A}{B_{j_2'}} |r)  \\
   & \qquad + (l| \E{B_j}{B_{j_1'}} \Big( \sum_{i=0}^{n'-3} (\E{A}{\tA})^i \H{A}{A}{\tA}{\tA} (\E{A}{\tA})^{n'-3-i} \Big) \E{A}{B_{j_2'}}  |r) \\
  & \qquad + (l| \E{B_j}{B_{j_1'}} (\E{A}{\tA})^{n'-2} \Big( \H{A}{A}{\tA}{B_{j_2'}} + \E{A}{\tA} \H{A}{A}{B_{j_2'}}{A} + \E{A}{\tA} \E{A}{B_{j_2'}} (1-E)^P \H{A}{A}{A}{A} \Big) |r) \\
  & + \e^{-iK} (l| \Big( \H{A}{A}{A}{A} (1-E)^P \E{B_jA}{AB_{j_1'}} + \H{A}{B_j}{A}{A} \E{A}{B_{j_1'}} + \H{B_j}{A}{A}{B_{j_1'}} \Big) (\E{A}{\tA})^{n'-1} \E{A}{B_{j_2'}} |r) \\
  & + \e^{-2iK} (l| \Big( \H{A}{A}{A}{A} (1-E)^P \E{B_j}{A}\E{A}{A} + \H{A}{B_j}{A}{A} \E{A}{A} + \H{B_j}{A}{A}{A} \Big) (1-\e^{-iK}E)^P \E{A}{B_{j_1'}} (\E{A}{\tA})^{n'-1} \E{A}{B_{j_2}p} |r) \\
  & + \e^{iK} (l| \Big( \H{A}{A}{A}{A}(1-E)^P \E{A}{B_{j_1'}}\E{B_j}{\tA} + \H{A}{A}{A}{B_{j_1'}} \E{B_j}{\tA} + \H{A}{B_j}{B_{j_1'}}{\tA} \Big) (\E{A}{\tA})^{n'-2} \E{A}{B_{j_2'}} |r) \\
  & + (l| \Big( \H{A}{A}{A}{A} (1-E)^P \E{A}{B_{j_1'}} \E{A}{\tA} + \H{A}{A}{A}{B_{j_1'}} \E{A}{\tA} + \H{A}{A}{B_{j_1'}}{\tA} \Big) \Big( \\
  & \qquad \sum_{j=2}^{n'-1} \e^{ijK} (\E{A}{\tA})^{j-2} \E{B_j}{\tA} (\E{A}{\tA})^{n'-j-1} \E{A}{B_{j_2'}} \\
  & \qquad + \e^{in'K} (\E{A}{\tA})^{n'-2} \E{B_j}{B_{j_2'}} + \e^{i(n'+1)K} (\E{A}{\tA})^{n'-2} \E{A}{B_{j_2'}} (1-\e^{iK}E)^P \E{B_j}{A} \Big)|r) \Big]  \\
   & \hspace{-1cm} \bra{\chi_{(j_1',j_2'),K'}(1)}\hat{H} \ket{ \chi_{(j_1,j_2),K}(1)} = 2\pi\delta(K-K') \Big[ \\
  & (l| \H{A}{A}{A}{A}(1-E)^P \E{B_{j_1}}{B_{j_1'}} \E{B_{j_2}}{B_{j_2'}}|r) + (l| \H{A}{B_{j_1}}{A}{B_{j_1'}} \E{B_{j_2}}{B_{j_2'}} |r) + (l| \H{B_{j_1}}{B_{j_2}}{B_{j_1'}}{B_{j_2'}} |r) \\
  & \qquad + (l| \E{B_{j_1}}{B_{j_1'}} \H{B_{j_2}}{A}{B_{j_2'}}{A} |r) + (l| \E{B_{j_1}}{B_{j_1'}} \E{B_{j_2}}{B_{j_2'}} (1-E)^P \H{A}{A}{A}{A} |r) \\
  & + \e^{-iK} (l| \Big( \H{A}{A}{A}{A} (1-E)^P \E{B_{j_1}}{A} \E{B_{j_2}}{B_{j_1'}} + \H{A}{B_{j_1}}{A}{A} \E{B_{j_2}}{B_{j_1'}} + \H{B_{j_1}}{B_{j_2}}{A}{B_{j_1'}} \Big) \E{A}{B_{j_2'}} |r) \\
  & + \e^{-i2K} (l| \Big( \H{A}{A}{A}{A} (1-E)^P \E{B_{j_1}}{A} \E{B_{j_2}}{A} + \H{A}{B_{j_1}}{A}{A} \E{B_{j_2}}{A} + \H{B_{j_1}}{B_{j_2}}{A}{A} \Big) (1-\e^{-iK}E)^P \E{A}{B_{j_1'}} \E{A}{B_{j_2'}} |r) \\
  & + \e^{iK} (l| \Big( \H{A}{A}{A}{A}(1-E)^P \E{A}{B_{j_1'}}\E{B_{j_1}}{B_{j_2'}} + \H{A}{A}{A}{B_{j_1'}} \E{B_{j_1}}{B_{j_2'}} + \H{A}{B_{j_1}}{B_{j_1'}}{B_{j_2'}} \Big) \E{B_{j_2}}{A} |r) \\
  & + \e^{i2K} (l| \Big( \H{A}{A}{A}{A} (1-E)^P \E{A}{B_{j_1'}} \E{A}{B_{j_2'}} + \H{A}{A}{A}{B_{j_1'}} \E{A}{B_{j_2'}} + \H{A}{A}{B_{j_1'}}{B_{j_2'}} \Big) (1-\e^{iK}E)^P \E{B_{j_1}}{A} \E{B_{j_2}}{A}\Big)|r) \Big] \\
  &  \hspace{-1cm} \bra{\chi_{(j_1',j_2'),K'}(n)}\hat{H} \ket{ \chi_{(j_1,j_2),K}(n)} = 2\pi\delta(K-K') \Big[ \\
  & (l| \Big( \H{A}{A}{A}{A}(1-E)^P \E{B_{j_1}}{B_{j_1'}} (\E{\tA}{\tA})^{n-1} + \H{A}{B_{j_1}}{A}{B_{j_1'}} (\E{\tA}{\tA})^{n-1} + \H{B_{j_1}}{\tA}{B_{j_1'}}{\tA} (\E{\tA}{\tA})^{n-2} \\
  & \qquad + \E{B_{j_1}}{B_{j_1'}} \sum_{i=0}^{n-3} (\E{\tA}{\tA})^{i} \H{\tA}{\tA}{\tA}{\tA} (\E{\tA}{\tA})^{n-i-3} \Big) \E{B_{j_2}}{B_{j_2'}} |r) \\
  & + (l| \E{B_{j_1}}{B_{j_1'}} (\E{\tA}{\tA})^{n-2} \Big( \H{\tA}{B_{j_2}}{\tA}{B_{j_2'}} + \E{\tA}{\tA} \H{B_{j_2}}{A}{B_{j_2'}}{A} + \E{\tA}{\tA} \E{B_{j_2}}{B_{j_2'}} (1-E)^P \H{A}{A}{A}{A} \Big) |r) \\
  & + \e^{-iK} (l| \Big( \H{A}{A}{A}{A} (1-E)^P \E{B_{j_1}}{A} \E{\tA}{B_{j_1'}} + \H{A}{B_{j_2}}{A}{A} \E{\tA}{B_{j_1'}} + \H{B_{j_1}}{\tA}{A}{B_{j_1'}} \Big) (\E{\tA}{\tA})^{n-2} \E{B_{j_2}}{\tA} \E{A}{B_{j_2'}} |r) \\
  & + (l| \Big( \H{A}{A}{A}{A} (1-E)^P \E{B_{j_1}}{A} \E{\tA}{A} + \H{A}{B_{j_1}}{A}{A} \E{\tA}{A} + \H{B_{j_1}}{\tA}{A}{A} \Big) \Big( \\
  & \qquad \sum_{j=2}^{n-1} \e^{-ijK} (\E{\tA}{A})^{j-2} \E{\tA}{B_{j_1'}} (\E{\tA}{\tA})^{n-j-1} \E{B_{j_2}}{\tA} (\E{A}{\tA})^{j-1} \\
  & \qquad + \e^{-inK} (\E{\tA}{A})^{n-2} \E{B_{j_2}}{B_{j_1'}} (\E{A}{\tA})^{n-1} \\
  & \qquad + \e^{-i(n+1)K} (\E{\tA}{A})^{n-2} \E{B_{j_2}}{A} (1-\e^{-iK}E)^P \E{A}{B_{j_1'}} (\E{A}{\tA})^{n-1} \Big) \E{A}{B_{j_2'}} |r) \\
  & + \e^{iK} (l| \Big( \H{A}{A}{A}{A}(1-E)^P \E{A}{B_{j_1'}}\E{B_{j_1}}{\tA} + \H{A}{A}{A}{B_{j_1'}} \E{B_{j_1}}{\tA} + \H{A}{B_{j_1}}{B_{j_1'}}{\tA} \Big) (\E{\tA}{\tA})^{n-2} \E{\tA}{B_{j_2'}} \E{B_{j_2}}{A} |r) \\
  & + (l| \Big( \H{A}{A}{A}{A} (1-E)^P \E{A}{B_{j_1'}} \E{A}{\tA} + \H{A}{A}{A}{B_{j_1'}} \E{A}{\tA} + \H{A}{A}{B_{j_1'}}{\tA} \Big) \Big( \\
  & \qquad \sum_{j=2}^{n-1} \e^{ijK} (\E{A}{\tA})^{j-2} \E{B_{j_1}}{\tA} (\E{\tA}{\tA})^{n-j-1} \E{\tA}{B_{j_2'}}  (\E{\tA}{A})^{j-1} \\
  & \qquad + \e^{inK} (\E{A}{\tA})^{n-2} \E{B_{j_1}}{B_{j_2'}} (\E{\tA}{A})^{n-1} \\
  & \qquad + \e^{i(n+1)K} (\E{A}{\tA})^{n-2} \E{A}{B_{j_2'}} (1-\e^{iK}E)^P \E{B_{j_1}}{A} (\E{\tA}{A})^{n-1} \Big) \E{B_{j_2}}{A} |r) \Big] \\
   & \hspace{-1cm} \bra{\chi_{(j_1',j_2'),K'}(2)}\hat{H} \ket{ \chi_{(j_1,j_2),K}(1)} = 2\pi\delta(K-K') \Big[ \\
  & (l| \Big( \H{A}{A}{A}{A}(1-E)^P \E{B_{j_1}}{B_{j_1'}} \E{B_{j_2}}{\tA} + \H{A}{B_{j_1}}{A}{B_{j_1'}} \E{B_{j_2}}{\tA} + \H{B_{j_1}}{B_{j_2}}{B_{j_1'}}{\tA} \Big) \E{A}{B_{j_2'}} |r) \\
  & \qquad + (l| \E{B_{j_1}}{B_{j_1'}} \Big( \H{B_{j_2}}{A}{\tA}{B_{j_2'}} + \E{B_{j_2}}{\tA} \H{A}{A}{B_{j_2'}}{A} + \E{B_{j_2}}{\tA} \E{A}{B_{j_2'}} (1-E)^P \H{A}{A}{A}{A} \Big) |r) \\
  & + \e^{-iK} (l| \Big( \H{A}{A}{A}{A} (1-E)^P \E{B_{j_1}}{A} \E{B_{j_2}}{B_{j_1'}} + \H{A}{B_{j_1}}{A}{A} \E{B_{j_2}}{B_{j_1'}} + \H{B_{j_1}}{B_{j_2}}{A}{B_{j_1'}} \Big) \E{A}{\tA} \E{A}{B_{j_2'}} |r) \\
  & + \e^{-i2K} (l| \Big( \H{A}{A}{A}{A} (1-E)^P \E{B_{j_1}}{A} \E{B_{j_2}}{A} + \H{A}{B_{j_1}}{A}{A} \E{B_{j_2}}{A} + \H{B_{j_1}}{B_{j_2}}{A}{A} \Big)  (1-\e^{-iK}E)^P \E{A}{B_{j_1'}} \E{A}{\tA} \E{A}{B_{j_2'}} |r) \\
  & + \e^{iK} (l| \Big( \H{A}{A}{A}{A}(1-E)^P \E{A}{B_{j_1'}}\E{B_{j_1}}{\tA} + \H{A}{A}{A}{B_{j_1'}} \E{B_{j_1}}{\tA} + \H{A}{B_{j_1}}{B_{j_1'}}{\tA} \Big) \E{B_{j_2}}{B_{j_2'}} |r) \\
  & + (l| \Big( \H{A}{A}{A}{A} (1-E)^P \E{A}{B_{j_1'}} \E{A}{\tA} + \H{A}{A}{A}{B_{j_1'}} \E{A}{\tA} + \H{A}{A}{B_{j_1'}}{\tA} \Big) \Big( \\
  & \qquad \e^{i2K} \E{B_{j_1}}{B_{j_2}p} \E{B_{j_2}}{A} + \e^{i3K} \E{A}{B_{j_2}p} (1-\e^{iK}E)^P \E{B_{j_1}}{A} \E{B_{j_2}}{A}\Big)|r) \Big] \\
  & \hspace{-1cm} \bra{\chi_{(j_1',j_2'),K'}(n+1)}\hat{H} \ket{ \chi_{(j_1,j_2),K}(1)} = 2\pi\delta(K-K') \Big[ \\
  & (l| \Big( \H{A}{A}{A}{A}(1-E)^P \E{B_{j_1}}{B_{j_1'}} (\E{\tA}{\tA})^{n-1} + \H{A}{B_{j_1}}{A}{B_{j_1'}} (\E{\tA}{\tA})^{n-1} + \H{B_{j_1}}{\tA}{B_{j_1'}}{\tA} (\E{\tA}{\tA})^{n-2}  \Big) \E{B_{j_2}}{\tA} \E{A}{B_{j_2'}} |r) \\
  & \qquad + (l| \E{B_{j_1}}{B_{j_1'}} \Big( \sum_{i=0}^{n-3} (\E{\tA}{\tA})^{i} \H{\tA}{\tA}{\tA}{\tA} (\E{\tA}{\tA})^{n-i-3} \Big) \E{B_{j_2}}{\tA} \E{A}{B_{j_2'}} |r) \\
  & \qquad + (l| \E{B_{j_1}}{B_{j_1'}} (\E{\tA}{\tA})^{n-2} \Big( \H{\tA}{B_{j_2}}{\tA}{\tA} \E{A}{B_{j_2'}} + \E{\tA}{\tA} \H{B_{j_2}}{A}{\tA}{B_{j_2'}} \\
  & \qquad\hspace{4cm} + \E{\tA}{\tA} \E{B_{j_2}}{\tA} \H{A}{A}{B_{j_2'}}{A} + \E{\tA}{\tA} \E{B_{j_2}}{\tA} \E{A}{B_{j_2'}} (1-E)^P \H{A}{A}{A}{A} \Big) |r) \\
  & + \e^{-iK} (l| \Big( \H{A}{A}{A}{A} (1-E)^P \E{B_{j_1}}{A} \E{\tA}{B_{j_1'}} + \H{A}{B_{j_1}}{A}{A} \E{\tA}{B_{j_1'}} + \H{B_{j_1}}{\tA}{A}{B_{j_1'}} \Big) (\E{\tA}{\tA})^{n-2} \E{B_{j_2}}{\tA} \E{A}{\tA} \E{A}{B_{j_2}p} |r) \\
  & + (l| \Big( \H{A}{A}{A}{A} (1-E)^P \E{B_{j_1}}{A} \E{\tA}{A} + \H{A}{B_{j_1}}{A}{A} \E{\tA}{A} + \H{B_{j_1}}{\tA}{A}{A} \Big) \Big( \\
  & \qquad  \sum_{j=2}^{n-1} \e^{-ijK} (\E{\tA}{A})^{j-2} \E{\tA}{B_{j_1'}} (\E{\tA}{\tA})^{n-j-1} \E{B_{j_2}}{\tA} (\E{A}{\tA})^{n'-n+j-1} \E{A}{B_{j_2'}} \\
  & \qquad + \e^{-inK} (\E{\tA}{A})^{n-2} \E{B_{j_2}}{B_{j_1'}} (\E{A}{\tA})^{n'-1} \E{A}{B_{j_2'}} \\
  & \qquad + \e^{-i(n+1)K} (\E{\tA}{A})^{n-2} \E{B_{j_2}}{A} (1-\e^{-iK}E)^P \E{A}{B_{j_1'}} (\E{A}{\tA})^{n'-1} \E{A}{B_{j_2'}} \Big) |r) \\
  & + \e^{iK} (l| \Big( \H{A}{A}{A}{A}(1-E)^P \E{A}{B_{j_1'}}\E{B_{j_1}}{\tA} + \H{A}{A}{A}{B_{j_1'}} \E{B_{j_1}}{\tA} + \H{A}{B_{j_1}}{B_{j_1'}}{\tA} \Big) (\E{\tA}{\tA})^{n-1} \E{B_{j_2}}{B_{j_2'}} |r) \\
  & + (l| \Big( \H{A}{A}{A}{A} (1-E)^P \E{A}{B_{j_1'}} \E{A}{\tA} + \H{A}{A}{A}{B_{j_1'}} \E{A}{\tA} + \H{A}{A}{B_{j_1'}}{\tA} \Big) \Big( \\
  & \qquad \sum_{j=2}^{n'-1} \e^{ijK} (\E{A}{\tA})^{j-2} \E{B_{j_1}}{\tA} (\E{\tA}{\tA})^{n'-j-1} \E{\tA}{B_{j_2'}}  (\E{\tA}{A})^{n-n'+j-1} \E{B_{j_2}}{A}  \\
  & \qquad + \e^{in'K} (\E{A}{\tA})^{n'-2} \E{B_{j_1}}{B_{j_2'}} (\E{\tA}{A})^{n-1} \E{B_{j_2}}{A} \\
  & \qquad + \e^{i(n'+1)K} (\E{A}{\tA})^{n'-2} \E{A}{B_{j_2'}} (1-\e^{iK}E)^P \E{B_{j_1}}{A} (\E{\tA}{A})^{n-1} \E{B_{j_2}}{A}\Big)|r) \Big] \\
   & \hspace{-1cm} \bra{\chi_{(j_1',j_2'),K'}(n')}\hat{H} \ket{ \chi_{(j_1,j_2),K}(1)} = 2\pi\delta(K-K') \Big[ \\
  & (l| \Big( \H{A}{A}{A}{A}(1-E)^P \E{B_{j_1}}{B_{j_1'}} \E{B_{j_2}}{\tA} + \H{A}{B_{j_1}}{A}{B_{j_1'}} \E{B_{j_2}}{\tA} + \H{B_{j_1}}{B_{j_2}}{B_{j_1'}}{\tA} \Big) (\E{A}{\tA})^{n'-2} \E{A}{B_{j_2'}} |r) \\
  & \qquad + (l| \E{B_{j_1}}{B_{j_1'}} \H{B_{j_2}}{A}{\tA}{\tA} (\E{A}{\tA})^{n'-3} \E{A}{B_{j_2'}} |r) \\
  & \qquad + (l| \E{B_{j_1}}{B_{j_1'}} \E{B_{j_2}}{\tA} \sum_{i=0}^{n'-4} (\E{A}{\tA})^i \H{A}{A}{\tA}{\tA} (\E{A}{\tA})^{n'-4-i} \E{A}{B_{j_2}'} |r) \\
  & \qquad + (l| \E{B_{j_1}}{B_{j_1'}} \E{B_{j_2}}{\tA} (\E{A}{\tA})^{n'-3} \Big( \H{A}{A}{\tA}{B_{j_2'}} + \E{A}{\tA} \H{A}{A}{B_{j_2'}}{A} + \E{A}{\tA} \E{A}{B_{j_2'}} (1-E)^P \H{A}{A}{A}{A} \Big) |r) \\
  & + \e^{-iK} (l| \Big( \H{A}{A}{A}{A} (1-E)^P \E{B_{j_1}}{A} \E{B_{j_2}}{B_{j_1'}} + \H{A}{B_{j_1}}{A}{A} \E{B_{j_2}}{B_{j_1'}} + \H{B_{j_1}}{B_{j_2}}{A}{B_{j_1'}} \Big) (\E{A}{\tA})^{n'-1} \E{A}{B_{j_2'}} |r) \\
  & + \e^{-i2K} (l| \Big( \H{A}{A}{A}{A} (1-E)^P \E{B_{j_1}}{A} \E{B_{j_2}}{A} + \H{A}{B_{j_1}}{A}{A} \E{B_{j_2}}{A} + \H{B_{j_1}}{B_{j_2}}{A}{A} \Big) \\
  & \hspace{6cm} \times (1-\e^{-iK}E)^P \E{A}{B_{j_1'}}(\E{A}{\tA})^{n'-1} \E{A}{B_{j_2'}} |r) \\
  & + \e^{iK} (l| \Big( \H{A}{A}{A}{A}(1-E)^P \E{A}{B_{j_1'}}\E{B_{j_1}}{\tA} + \H{A}{A}{A}{B_{j_1'}} \E{B_{j_1}}{\tA} + \H{A}{B_{j_1}}{B_{j_1'}}{\tA} \Big) \E{B_{j_2}}{\tA} (\E{A}{\tA})^{n'-3} \E{A}{B_{j_2'}} |r)\\
  & + (l| \Big( \H{A}{A}{A}{A} (1-E)^P \E{A}{B_{j_1'}} \E{A}{\tA} + \H{A}{A}{A}{B_{j_1'}} \E{A}{\tA} + \H{A}{A}{B_{j_1'}}{\tA} \Big) \Big( \\
  & \qquad \sum_{j=2}^{n'-2} \e^{ijK} (\E{A}{\tA})^{j-2} \E{B_{j_1}}{\tA} \E{B_{j_2}}{\tA} (\E{A}{\tA})^{n'-j-2} \E{A}{B_{j_2}'} \\
  & \qquad + \e^{i(n'-1)K} (\E{A}{\tA})^{n'-3} \E{B_{j_1}}{\tA} \E{B_{j_2}}{B_{j_2'}} \\
  & \qquad + \e^{in'K} (\E{A}{\tA})^{n'-2} \E{B_{j_1}}{B_{j_2'}} \E{B_{j_2}}{A} \\
  & \qquad + \e^{i(n'+1)K} (\E{A}{\tA})^{n'-2} \E{A}{B_{j_2'}} (1-\e^{iK}E)^P \E{B_{j_1}}{A} \E{B_{j_2}}{A}\Big)|r) \Big] \\
   & \hspace{-1cm} \bra{\chi_{(j_1',j_2'),K'}(n')}\hat{H} \ket{ \chi_{(j_1,j_2),K}(n)} = 2\pi\delta(K-K') \Big[ \\
  & (l| \Big( \H{A}{A}{A}{A}(1-E)^P \E{B_{j_1}}{B_{j_1'}} \E{\tA}{\tA} + \H{A}{B_{j_1}}{A}{B_{j_1'}} \E{\tA}{\tA} + \H{B_{j_1}}{\tA}{B_{j_1'}}{\tA} \Big) (\E{\tA}{\tA})^{n-2} \E{B_{j_2}}{\tA} \E{A}{\tA} (\E{A}{\tA})^{n'-n-2} \E{A}{B_{j_2'}} |r) \\
  & + (l| \E{B_{j_1}}{B_{j_1'}} \Big( \sum_{i=0}^{n-3} (\E{\tA}{\tA})^{i} \H{\tA}{\tA}{\tA}{\tA} (\E{\tA}{\tA})^{n-i-3} \Big) \E{B_{j_2}}{\tA} \E{A}{\tA} (\E{A}{\tA})^{n'-n-2} \E{A}{B_{j_2'}} |r)\\
  & + (l| \E{B_{j_1}}{B_{j_1'}} (\E{\tA}{\tA})^{n-2} \Big( \H{\tA}{B_{j_2}}{\tA}{\tA} \E{A}{\tA} + \E{\tA}{\tA} \H{B_{j_2}}{A}{\tA}{\tA} \Big) (\E{A}{\tA})^{n'-n-2} \E{A}{B_{j_2'}} |r) \\
  & + (l| \E{B_{j_1}}{B_{j_1'}} (\E{\tA}{\tA})^{n-1} \E{B_{j_2}}{\tA} \Big( \sum_{i=0}^{n'-n-3} (\E{A}{\tA})^i \H{A}{A}{\tA}{\tA} (\E{A}{\tA})^{n'-n-3-i} \E{A}{B_{j_2'}} + (\E{A}{\tA})^{n'-n-2} \H{A}{A}{\tA}{B_{j_2'}} \\
  & \hspace{4cm} + (\E{A}{\tA})^{n'-n-1} \H{A}{A}{B_{j_2'}}{A} + (\E{A}{\tA})^{n'-n-1} \E{A}{B_{j_2'}} (1-E)^P \H{A}{A}{A}{A} \Big) |r) \\
  & + \e^{-iK} (l| \Big( \H{A}{A}{A}{A} (1-E)^P \E{B_{j_1}}{A} \E{\tA}{B_{j_1'}} + \H{A}{B_{j_1}}{A}{A} \E{\tA}{B_{j_1'}} + \H{B_{j_1}}{\tA}{A}{B_{j_1'}} \Big) \\
  & \hspace{6cm} \times (\E{\tA}{\tA})^{n-2} \E{B_{j_2}}{\tA} (\E{A}{\tA})^{n'-n} \E{A}{B_{j_2'}} |r) \\
  & + (l| \Big( \H{A}{A}{A}{A} (1-E)^P \E{B_{j_1}}{A} \E{\tA}{A} + \H{A}{B_{j_1}}{A}{A} \E{\tA}{A} + \H{B_{j_1}}{\tA}{A}{A} \Big) \Big( \\
  & \qquad \Big( \sum_{j=2}^{n-1} \e^{-ijK} (\E{\tA}{A})^{j-2} \E{\tA}{B_{j_1'}} (\E{\tA}{\tA})^{n-j-1} \E{B_{j_2}}{\tA} (\E{A}{\tA})^j \Big) (\E{A}{\tA})^{n'-n-1} \E{A}{B_{j_2'}} \\
  & \qquad + \e^{-inK} (\E{\tA}{A})^{n-2} \E{B_{j_2}}{B_{j_1'}} (\E{A}{\tA})^{n'-1} \E{A}{B_{j_2'}} \\
  & \qquad + \e^{-i(n+1)K} (\E{\tA}{A})^{n-2} \E{B_{j_2}}{A} (1-\e^{-iK}E)^P \E{A}{B_{j_1'}} (\E{A}{\tA})^{n'-1} \E{A}{B_{j_2'}} \Big) |r) \\
  & + \e^{iK} (l| \Big( \H{A}{A}{A}{A}(1-E)^P \E{A}{B_{j_1'}}\E{B_{j_1}}{\tA} + \H{A}{A}{A}{B_{j_1'}} \E{B_{j_1}}{\tA} + \H{A}{B_{j_1}}{B_{j_1'}}{\tA} \Big) \\
  & \hspace{6cm} \times (\E{\tA}{\tA})^{n-1} \E{B_{j_2}}{\tA} (\E{A}{\tA})^{n'-n-2} \E{A}{B_{j_2'}} |r) \\
  & + (l| \Big( \H{A}{A}{A}{A} (1-E)^P \E{A}{B_{j_1'}} \E{A}{\tA} + \H{A}{A}{A}{B_{j_1'}} \E{A}{\tA} + \H{A}{A}{B_{j_1'}}{\tA} \Big) \Big( \\
  & \qquad \sum_{j=2}^{n'-n-1} \e^{ijK} (\E{A}{\tA})^{j-2} \E{B_{j_1}}{\tA} (\E{\tA}{\tA})^{n-1} \E{B_{j_2}}{\tA} (\E{A}{\tA})^{n'-n-j-1} \E{A}{B_{j_2'}} \\
  & \qquad + \e^{i(n'-n)K} (\E{A}{\tA})^{n'-n-2} \E{B_{j_1}}{\tA} (\E{\tA}{\tA})^{n-1} \E{B_{j_2}}{B_{j_2'}} \\
  & \qquad + \sum_{j=n'-n+1}^{n'-1} \e^{ijK} (\E{A}{\tA})^{j-2} \E{B_{j_1}}{\tA} (\E{\tA}{\tA})^{n'-j-1} \E{\tA}{B_{j_2'}}  (\E{\tA}{A})^{n-n'+j-1} \E{B_{j_2}}{A}  \\
  & \qquad + \e^{in'K} (\E{A}{\tA})^{n'-2} \E{B_{j_1}}{B_{j_2'}} (\E{\tA}{A})^{n-1} \E{B_{j_2}}{A} \\
  & \qquad + \e^{i(n'+1)K} (\E{A}{\tA})^{n'-2} \E{A}{B_{j_2'}} (1-\e^{iK}E)^P \E{B_{j_1}}{A} (\E{\tA}{A})^{n-1} \E{B_{j_2}}{A} \Big) |r) \Big].
\end{align*}

\subsection{Asymptotic regime}
\label{sec:asymptotic}

The expressions for the effective norm and Hamiltonian matrices above are largely determined by powers of the transfer matrices. The power of the transfer matrices behaves as
\begin{align*}
   (\E{A}{A})^n  = |r)(l| + \mathcal{O}\left(\e^{-n/\xi}\right) \qquad \text{as} \quad n\rightarrow\infty,
\end{align*}
where the correlation length $\xi$ of the MPS was defined in Eq.~\eqref{corrLength}. The asymptotic regime in $\mathsf{N}_\text{eff}$ and $\mathsf{H}_\text{eff}$ is reached when the corrections can be safely neglected, i.e. $n>\xi\times\log(1/\epsilon)$ where $\epsilon$ is the allowed error.
\par The effective norm matrix reduces to the unit matrix in this regime
\begin{align*}
\braket{\chi_{(j_1',j_2'),K'}(n') | \chi_{(j_1,j_2),K}(n)} = 2\pi\delta(K-K') \delta_{n',n} \delta_{j_1',j_1} \delta_{j_2',j_2} ,
\end{align*}
and the effective Hamiltonian matrix is greatly simplified:
\begin{align*}
  & \bra{\chi_{(j_1',j_2'),K'}(n)} \hat{H} \ket{\chi_{(j_1,j_2),K}(n)} = 2\pi\delta(K-K') \Big[ \\
  & \qquad \delta_{j_1,j_1'}(l| \Big( \H{\tA}{B_{j_2}}{\tA}{B_{j_2'}} + \H{B_{j_2}}{A}{B_{j_2'}}{A} + \E{B_{j_2}}{B_{j_2'}} (1-E)^P \H{A}{A}{A}{A} + \H{\tA}{\tA}{\tA}{\tA} (1-E)^P \E{B_{j_2}}{B_{j_2'}} \Big) |r) \\
  & \qquad + \delta_{j_2,j_2'} (l| \Big( \H{A}{A}{A}{A} (1-E)^P \E{B_{j_1}}{B_{j_1'}} + \H{A}{B_{j_1}}{A}{B_{j_1'}} + \H{B_{j_1}}{\tA}{B_{j_1'}}{\tA} + \E{B_{j_1}}{B_{j_1'}} (1-E)^P \H{\tA}{\tA}{\tA}{\tA} \Big) |r) \Big] \\
  &\bra{\chi_{(j_1',j_2'),K'}(n+1)} \hat{H} \ket{\chi_{(j_1,j_2),K}(n)} = 2\pi\delta(K-K') \Big[ \\
  & \qquad \delta_{j_1,j_1'} (l| \Big(  \H{\tA}{\tA}{\tA}{\tA} (1-E)^P \E{B_{j_2}}{\tA} \E{A}{B_{j_2'}} + \H{\tA}{B_{j_2}}{\tA}{\tA} \E{A}{B_{j_2'}} + \H{B_{j_2}}{A}{\tA}{B_{j_2'}} \Big) |r) \\
  & \qquad + \delta_{j_2,j_2'} \e^{iK} (l| \Big( \H{A}{A}{A}{A} (1-E)^P \E{A}{B_{j_1'}} \E{B_{j_1}}{\tA} + \H{A}{A}{A}{B_{j_1'}} \E{B_{j_1}}{\tA} + \H{A}{B_{j_1}}{B_{j_1'}}{\tA} \Big) |r)  \Big] \\
  & \bra{\chi_{(j_1',j_2'),K'}(n')} \hat{H} \ket{\chi_{(j_1,j_2),K}(n)} = 2\pi\delta(K-K') \Big[ \\
  & \qquad \delta_{j_1,j_1'} (l| \Big( \H{\tA}{\tA}{\tA}{\tA} (1-E)^P \E{B_{j_2}}{\tA} \E{A}{\tA} + \H{\tA}{B_{j_2}}{\tA}{\tA} \E{A}{\tA} + \H{B_{j_2}}{A}{\tA}{\tA} \Big) (\E{A}{\tA})^{(n'-n-2)} \E{A}{B_{j_2'}} |r) \\
  & \qquad + \e^{iK(n'-n)} \delta_{j_2,j_2'} (l| \Big(\H{A}{A}{A}{A} (1-E)^P \E{A}{B_{j_1'}} \E{A}{\tA} + \H{A}{A}{A}{B_{j_1'}} \E{A}{\tA} + \H{A}{A}{B_{j_1'}}{\tA} \Big) (\E{A}{\tA})^{(n'-n-2)} \E{B_{j_1}}{\tA} |r) \Big] .
\end{align*}
One can observe that the matrix elements indeed form a repeating row of block matrices, centered around the diagonal and exponentially decaying
\begin{align*}
  (\mathsf{H}_\text{eff})_{n'j_1'j_2',nj_1j_2}
  &=  \bra{\chi_{(j_1',j_2'),K'}(n')} \hat{H} \ket{\chi_{(j_1,j_2),K}(n)} \\
  &= (A_{n'-n})_{j_1'j_2',j_1j_2} \\
  &= \mathcal{O}\left(\e^{-|n'-n|/\xi}\right) \qquad \text{as} \quad |n-n'|\rightarrow\infty.
\end{align*}

\subsection{Two-particle form factors}

Again we start from the spectral function
\begin{equation*}
S(\kappa,\omega) = \sum_{n=-\infty}^{+\infty} \int_{-\infty}^{+\infty} \d t \, \e^{i(\omega t - \kappa n)} \bra{\Psi_0} O_n\dag(t) O_0(0) \ket{\Psi_0}.
\end{equation*}
Inserting a projector on the two-particle subspace, the two-particle contribution to this function can be written as ($\Gamma_2(\kappa,\omega)$ is the set of all two-particle states at that momentum-energy combination)
\begin{equation*}
S(\kappa,\omega)_\text{2p} = \sum_{i\in\Gamma_2(K,\omega)} \left| \bra{\Upsilon_\gamma(\kappa,\omega)}\hat{O}_0 \ket{\Psi_0} \right|^2.
\end{equation*}
If we denote the coefficients $c^j(n)$ of the two-particle states as
\begin{equation*}
c^j(n) = c^j_\text{local}(n) + \sum_{\gamma=1}^{2\Gamma} q^\gamma \e^{i\kappa_\gamma n} v^j_\gamma
\end{equation*}
such that $c^j_\text{local}(n)\approx0$ if $n>R$ for some value of $R$. The overlap appearing in the spectral functions can be calculated as
\begin{equation*}
 \bra{\Psi[A]} \hat{O}_0 \ket{\Upsilon(K,\omega)} = \sum_{n=0}^{\infty} \sum_j c^j(n) \bra{\Psi[A]} \hat{O}_0 \ket{\chi_{j,K}(n)}
\end{equation*}
where
\begin{align*}
\bra{\Psi[A]} \hat{O}_0 \ket{\chi_{j,K}(0)} &= (l| \O{B_j}{A} |r) + e^{iK}(l| \O{A}{A} (1-\e^{iK}E)^P \E{B_j}{A} |r) \\
 \bra{\Psi[A]} \hat{O}_0 \ket{\chi_{(j_1,j_2),K}}   & = (l| \O{B_{j_1}}{A} (\E{A}{A})^{n-1} \E{B_{j_2}}{A} |r) + e^{iK}(l| \O{A}{A} (1-\e^{iK}E)^P \E{B_{j_1}}{A} (\E{A}{A})^{n-1} \E{B_{j_2}}{A} |r) \\
& = (l| \left( \O{B_{j_1}}{A} + e^{iK}\O{A}{A} (1-\e^{iK}E)^P \E{B_{j_1}}{A} \right) (\E{A}{A})^{n-1} \E{B_{j_2}}{A} |r).
\end{align*}
We have
\begin{align*}
 \bra{\Psi[A]} \hat{O}_0 \ket{\Upsilon(K,\omega)}  &= \sum_j c^j(0) \left( (l| \O{B_j}{A} |r) + \e^{iK}(l| \O{A}{A} (1-\e^{iK}E)^P \E{B_j}{A} |r) \right) \\
& \qquad + \sum_{n=1}^R \sum_{j_1,j_2} c^{(j_1,j_2)}_\text{local} (n) (l| \left( \O{B_{j_1}}{A} + \e^{iK}\O{A}{A} (1-\e^{iK}E)^P \E{B_{j_1}}{A} \right) (\E{A}{A})^{n-1} \E{B_{j_2}}{A} |r) \\
& \qquad + \sum_{\gamma=1}^{2\Gamma} q^\gamma \sum_{j_1,j_2} v_\gamma^{(j_1,j_2)}(l| \left( \E{B_{j_1}}{A} + \e^{iK}\O{A}{A} (1-\e^{iK}E)^P \E{B_{j_1}}{A} \right) (1-\e^{i\kappa_\gamma n} E)^P \E{B_{j_2}}{A} |r).
\end{align*}

\section{Proof of equation (\ref{conservation})}
\label{proof}

Let us start with the polynomial eigenvalue equation for the asymptotic solutions of the scattering problem, Eq.~\eqref{polynomial}
\begin{equation*}
\sum_{m=-M}^{+M} \mu^{m} \mathsf{A}_{m} \bm{v} = \omega \bm{v}.
\end{equation*}
For a given $\mu\in\mathbb{C}$, this equation is an ordinary eigenvalue problem with eigenvalue $\omega$ and eigenvector $\vect{v}$. Given the property $\mathsf{A}_{m}^{\dagger}=\mathsf{A}_{-m}$, we are only assured of a Hermitian eigenvalue problem (and thus of real eigenvalues $\omega$) if $\mu^{\ast}=\mu^{-1}$, i.e. if $\mu$ is on the unit circle. So for any $\mu_\gamma=\exp(i\kappa)$, let there be eigenvalues $\omega_\gamma(\kappa)$ and corresponding normalized eigenvectors $\vect{v}_\gamma(\kappa)$. The functions $\omega_\gamma(\kappa)$ and $\vect{v}_\gamma(\kappa)$ are assumed to be smooth such that at least the first derivatives are well defined. By taking the derivative of the eigenvalue equation with respect to $\kappa$ we obtain
\begin{equation*}
\sum_{m=-M}^{+M} i m \mathsf{A}_{m} \e^{i\kappa m} \vect{v}_\gamma(\kappa)+ \sum_{m=-M}^{+M} \mathsf{A}_{m} \e^{i\kappa m} \frac{\d \vect{v}_\gamma}{\d\kappa}(\kappa) = \frac{\d \omega_\gamma}{\d\kappa}(\kappa) \vect{v}_\gamma(\kappa) +\omega_\gamma(\kappa) \frac{\d \vect{v}_\gamma}{\d\kappa}(\kappa).
\end{equation*}
By multiplying this equation with $\vect{v}_{\gamma'}(\kappa)\dag$ and using the normalization $\vect{v}_{\gamma'}(\kappa)\dag\vect{v}_{\gamma}(\kappa)=\delta_{\gamma'\gamma}$, we tobtain the following relation for later use
\begin{equation} \label{lateruse}
\sum_{m=-M}^{+M} i m \vect{v}_{\gamma'}(\kappa)\dag \mathsf{A}_{m} \e^{i\kappa m} \vect{v}_\gamma(\kappa)= \delta_{\gamma'\gamma} \frac{\d \omega_\gamma}{\d\kappa}(\kappa).
\end{equation}
Now consider a two particle eigenstate $\ket{\Upsilon(K,\omega)}$, which has the asymptotic form
\begin{equation*}
 \vect{c}(K,\omega) = \sum_{\gamma=1}^{2\Gamma} q_\gamma \e^{i\kappa_\gamma n} \vect{v}_\gamma.
\end{equation*}
We can introduce the projectors (we will omit all dependencies on the total momentum $K$)
\begin{equation*}
P_R = \sum_{n=0}^R \sum_{j=1}^{L_n} \ket{\chi_j(n)}\bra{\chi_j(n)} \qquad \text{and} \qquad P_R^\perp = \sum_{n>R} \sum_{j=1}^{L_n} \ket{\chi_j(n)}\bra{\chi_j(n)}
\end{equation*}
so that we have
\begin{equation*}
 \bra{\Upsilon(\omega)} P_R^\perp H P_R \ket{\Upsilon(\omega)} = \bra{\Upsilon(\omega)} P_R H P_R^\perp \ket{\Upsilon(\omega)}
\end{equation*}
upon the condition that $\ket{\Upsilon(\omega)}$ is an eigenstate. If we choose $R>M$, we can insert the asymptotic form for the effective Hamiltonian
\begin{equation*}
\sum_{n=0}^{R}\sum_{n'>R} \vect{c}(n')\dag \mathsf{A}_{n-n'}\vect{c}(n) - \vect{c}(n)\dag \mathsf{A}_{n'-n} \vect{c}(n')=0.
\end{equation*}
Since $\mathsf{A}_{m}=0$ for $\lvert m\rvert > M$, this allows to restrict the summations and rewrite this equality as
\begin{equation*}
\Im \left[\sum_{m=1}^{M}\sum_{n=R+1-m}^{R} \vect{c}(n)\dag \mathsf{A}_{m}\vect{c}(n+m) \right] = 0.
\end{equation*}
We can insert the asymptotic form for $\vect{c}(n)$ to obtain for the ``diagonal terms'' where $\gamma$ has the same value for both sums
\begin{align*}
\Im \left[\sum_{\gamma=1}^{2\Gamma} \left| q_\gamma \right|^2 \sum_{m=1}^{M}\sum_{n=R+1-m}^{R}  \vect{v}_\gamma\dag \mathsf{A_m} \vect{v}_\gamma \e^{i\kappa_\gamma m} \right]
&= \Im\left[\sum_{\gamma=1}^{2\Gamma} \left| q_\gamma \right|^2 \sum_{m=1}^{M} m  \vect{v}_\gamma\dag \mathsf{A_m} \vect{v}_\gamma \e^{i\kappa_\gamma m} \right]  \\
&=- \sum_{\gamma=1}^{2\Gamma} \left| q_\gamma \right|^2 \sum_{m=1}^{M}  \left( im \vect{v}_\gamma\dag \mathsf{A_m} \vect{v}_\gamma \e^{i\kappa_\gamma m} - im \vect{v}_\gamma\dag \mathsf{A_m}\dag \vect{v}_\gamma \e^{-i\kappa_\gamma m} \right) \\
&=-\sum_{\gamma=1}^{2\Gamma} \left| q_\gamma \right|^2 \sum_{m=-M}^{M}  im \vect{v}_\gamma\dag \mathsf{A_m} \vect{v}_\gamma \e^{i\kappa_\gamma m} \\
&= -\sum_{\gamma=1}^{2\Gamma} \left| q_\gamma \right|^2 \frac{\d\omega_\gamma}{\d\kappa}(\kappa_\gamma)
\end{align*}
and this expression has to equal zero if we can show that the contribution of all ``non-diagonal terms'' ($\gamma\neq \gamma'$ in the two sums) vanish. We look at a single contribution with $\gamma\neq \gamma'$ and the corresponding term with $\gamma$ and $\gamma'$ interchanged, and first assume $\kappa_\gamma\neq \kappa_{\gamma'}$.
\begin{align*}
& \Im \left[ q_\gammap^* q_\gamma  \sum_{m=1}^{M}\sum_{n=R+1-m}^{R}  \vv_\gammap\dag \mathsf{A_m} \vv_\gamma \e^{i\kappa_\gamma m}\e^{i (\kappa_\gamma-\kappa_\gammap) n} + q_\gamma^* q_\gammap  \sum_{m=1}^{M}\sum_{n=R+1-m}^{R}  \vv_\gamma\dag \mathsf{A_m} \vv_\gammap \e^{i\kappa_\gammap m}\e^{i (\kappa_\gammap-\kappa_\gamma) n} \right]\\
& \qquad = \Im \left[ \frac{\e^{ i (\kappa_\gamma-\kappa_\gammap)(R+1)}}{\e^{i (\kappa_\gamma-\kappa_\gammap)}-1} q_\gammap^* q_\gamma  \sum_{m=1}^{M} \vv_\gammap\dag \mathsf{A_m} \vv_\gamma \left( \e^{i\kappa_\gamma m} - \e^{i\kappa_\gammap m} \right) \right. \\
& \qquad \hspace{5cm} + \left. \frac{\e^{ i (\kappa_\gammap-\kappa_\gamma)(R+1)}}{\e^{i (\kappa_\gammap-\kappa_\gamma)}-1} q_\gamma^* q_\gammap  \sum_{m=1}^{M} \vv_\gamma\dag \mathsf{A_m} \vv_\gammap \left( \e^{i\kappa_\gammap m} - \e^{i\kappa_\gamma m} \right) \right] \\
& \qquad =\frac{1}{2\sin((\kappa_\gamma-\kappa_\gammap)/2)} \Re \left[ \e^{ i (\kappa_\gamma-\kappa_\gammap)(R+1/2)} q_\gammap^* q_\gamma  \sum_{m=1}^{M} \vv_\gammap\dag \mathsf{A_m} \vv_\gamma \left( \e^{i\kappa_\gamma m} - \e^{i\kappa_\gammap m} \right) \right.\\
& \qquad \hspace{5cm} - \left. \e^{ i (\kappa_\gammap-\kappa_\gamma)(R+1/2)} q_\gamma^* q_\gammap  \sum_{m=1}^{M} \vv_\gamma\dag \mathsf{A_m} \vv_\gammap \left( \e^{i\kappa_\gammap m} - \e^{i\kappa_\gamma m} \right) \right] \\
& \qquad =\frac{1}{2\sin((\kappa_\gamma-\kappa_\gammap)/2)} \Re \left[ \e^{ i (\kappa_\gamma-\kappa_\gammap)(R+1/2)} q_\gammap^* q_\gamma  \sum_{m=1}^{M} \vv_\gammap\dag \mathsf{A_m} \vv_\gamma \left( \e^{i\kappa_\gamma m} - \e^{i\kappa_\gammap m} \right) \right.\\
& \qquad \hspace{5cm} - \left. \e^{ i (\kappa_\gamma-\kappa_\gammap)(R+1/2)} q_\gammap^* q_\gamma  \sum_{m=1}^{M} \vv_\gammap\dag \mathsf{A_m}\dag \vv_\gamma \left( \e^{-i\kappa_\gammap m} - \e^{-i\kappa_\gamma m} \right) \right] \\
& \qquad = \frac{1}{2\sin((\kappa_\gamma-\kappa_\gammap)/2)} \Re \left[  \e^{ i (\kappa_\gamma-\kappa_\gammap)(R+1/2)} q_\gammap^* q_\gamma  \sum_{m=-M}^{M} \vv_\gammap\dag \mathsf{A_m} \vv_\gamma \left( \e^{i\kappa_\gamma m} - \e^{i\kappa_\gammap m} \right) \right].
\end{align*}
Note that we are missing the term for $m=0$, but that this term is zero anyway because of the factor $(\e^{i \kappa_\gamma m}-\e^{i \kappa_\gammap m})$. Finally noting that $\sum_{m=-M}^{M}   \mathsf{A_m}  \vv_\gamma \e^{i\kappa_\gamma m}=\omega \vv_\gamma$ and  $\sum_{m=-M}^{M} \vv_\gammap\dag \mathsf{A_m}  \e^{i\kappa_\gammap m}=\omega \vv_\gammap\dag$, it is clear that both contributions cancel and the total expression evaluates to zero.
Finally, we consider the case that $\kappa_\gamma=\kappa_\gammap=\kappa$. We obtain
\begin{align*}
& \Im \left[ q_\gammap^*q_\gamma \sum_{m=1}^{M}\sum_{n=R+1-m}^{R} \vv_\gammap\dag \mathsf{A_m} \vv_\gamma\e^{i\kappa m} + q_\gamma^*q_\gammap  \sum_{m=1}^{M}\sum_{n=R+1-m}^{R} \vv_\gamma\dag \mathsf{A_m} \vv_\gammap\e^{i\kappa m} \right]\\
& \hspace{5cm} = \Im \left[ q_\gammap^*q_\gamma  \sum_{m=1}^{M} m \vv_\gammap\dag \mathsf{A_m} \vv_\gamma\e^{i\kappa m} + q_\gamma^*q_\gammap  \sum_{m=1}^{M} m \vv_\gamma\dag \mathsf{A_m} \vv_\gammap\e^{i\kappa m} \right]\\
& \hspace{5cm} =\Im \left[ q_\gammap^*q_\gamma  \sum_{m=-M}^{M} m \vv_\gammap\dag \mathsf{A_m} \vv_\gamma\e^{i\kappa m} \right].
\end{align*}
In the last line, we replaced the second term of the line before by the negative of its complex conjugate, since we are taking the imaginary part of the whole expression anyway. Using that $\vv_\gamma$ and $\vv_\gammap$ correspond to some $\vv_\gamma(\kappa)$ and $\vv_\gammap(\kappa)$ with different $\gamma\neq\gamma'$ but equal $\omega_\gamma(\kappa)=\omega_\gammap(\kappa)$, we can employ Eq.~\eqref{lateruse} to conclude that this term is zero.

\section{M{\o}ller operators, the S matrix and scattering states in one dimension}
\label{sec:mollerA}

In this appendix we will translate some basic notions of single particle scattering theory from an external potential \cite{Taylor1972} to the one-dimensional case where we have different types of particles with general dispersion relations. The two-particle scattering in the many body Hilbert space considered in this manuscript can be mapped to this setting by taking out the conservation of total momentum and only looking at the relative wave function, which is encoded in the coefficients $c^j(n)$. For the remainder of this section, we assume to have a Hilbert space spanned by states $\{\ket{x,j}\}$ where $x$ is a spatial coordinate that can be discrete ($x\in\mathbb{Z}$) or continuous ($x\in\mathbb{R}$) and $j=1,\ldots,L$ labels different internal levels at every position (corresponding to different particle types). We assume we have some Hamiltonian $\ham$, which can be written as the sum of a free part $\ham_0$ and a potential $\hat{V}$. The free Hamiltonian is translation invariant ($\braket{x',j'|\hat{H_0}|x,j}= (\mathsf{A}_{x-x'})_{j',j}$ with $\mathsf{A}_x = (\mathsf{A}_{-x})^\dagger$) and also assumed to be short-ranged ($\mathsf{A}_{x-x'}=0$ for $\lvert x-x'\rvert >  M$). The potential is centered around $x=0$ and goes to zero quickly, e.g. $\braket{x',j'|\hat{V}|x,j}=0$ for $\lvert x\rvert > M+N$ or $\lvert x'\rvert > M+N$. The free Hamiltionian is diagonalized in momentum space and describes the free propagation of a number of types of particles $\alpha=1,\dots N$ with eigenvalues (dispersion relations) $E_{\alpha}(p)$. Indeed, by using the momentum states $\ket{p,j}=\int\d x e^{ipx} \ket{x,j}$ (an integral over $x$ should be read as a sum for the discrete case), the free Hamiltonian $\ham_0$ is brought into block-diagonal form:
\begin{equation}
\braket{p,j|\ham_0|p',j'} = 2\pi \delta(p-p') (\mathsf{A}(p))_{j,j'}
\end{equation}
where the $L \times L$ Hermitian matrix $\mathsf{A}(p)=\int\d x e^{i p x} \mathsf{A}_x$ is an analytic function of $p$ (since $\mathsf{A}_x$ vanishes for $\lvert x \rvert > M+N$). Its eigenvalues $E_\alpha(p)$ and corresponding eigenvectors $\vect{v}_{\alpha}(p)$ define the spectrum of $\ham_0$. Also note for further reference the relation
\begin{equation}\label{eq:Aderivative}
\vect{v}_{\beta}(p)^\dagger \frac{\d \mathsf{A}}{\d p}(p) \vect{v}_{\alpha}(p) = \frac{\d E_{\alpha}}{\d p}(p) \delta_{\alpha,\beta}.
\end{equation}
We will henceforth denote the eigenvalues of the free Hamiltonian $\ham_0$ as $E(p_\alpha)$ and the corresponding eigenstates as $\ket{p_\alpha}$ with coordinate representation $\braket{x,j|p_\alpha} = v^j_{\alpha}(p) e^{i p x}$. By choosing $\vect{v}_{\alpha}(p)^\dagger \vect{v}_{\beta}(p) = \delta_{\alpha,\beta}$, the eigenstates $\ket{p_\alpha}$ of $\ham_0$ are normalized as $\braket{p'_\beta|p_\alpha}=2\pi\delta(p'-p)\delta_{\alpha,\beta}$ and span the whole Hilbert space
\begin{equation*}
\one = \sum_\alpha \int \frac{\d p}{2\pi} \ket{p_\alpha}\bra{p_\alpha}.
\end{equation*}
The range of $p$ determines whether we are dealing with a discrete or continuous system, and will not be specified. In order to describe scattering experiments, one should build wave packets from these momentum eigenstates
\begin{equation*}
\ket{\phi_\alpha} = \int\frac{\d p}{2\pi} \; \phi(p) \ket{p_\alpha}.
\end{equation*}
Typically, we will be interested in wave packets $\phi(p)$ that are strongly centered around some momentum $p_0$, so that it makes sense to express scattering amplitudes (S matrix elements) in the basis of momentum eigenstates.
\par Let $U(t)$ and $U_0(t)$ denote the unitary evolution associated to respectively $\ham$ and $\ham_0$. We now want to describe some orbit $U(t)\ket{\psi}$, which has an in-aymptote and an out-asymptote in the following sense
\begin{align*}
& U(t)\ket{\psi} \rightarrow U^0(t)\ket{\psi_\text{in}} && \text{as} \quad t\rightarrow-\infty \\
& U(t)\ket{\psi} \rightarrow U^0(t)\ket{\psi_\text{out}} &&\text{as} \quad t\rightarrow+\infty.
\end{align*}
For given $\ket{\psi_{\text{in}}}$ or $\ket{\psi_{\text{out}}}$, one can try to define
\begin{align*}
& \ket\psi = \lim_{t\rightarrow-\infty} U(t)\dag U^0(t) \ket{\psi_\text{in}} = \Omega_+\ket{\psi_\text{in}} \\
& \ket\psi = \lim_{t\rightarrow+\infty} U(t)\dag U^0(t) \ket{\psi_\text{out}} = \Omega_-\ket{\psi_\text{out}}
\end{align*}
with $\Omega_\pm$ the M{\o}ller operators. The existence of these limits, and thus of the M\"{o}ller operators, can be proven by studying wave packets and linear combinations thereof. For a quadratic dispersion relation, the dispersive behavior of the wave packet is often sufficient to guarantee convergence. Since we are studying general dispersion relations $E_{\alpha}(p)$, a sufficient condition can be obtained by restricting to wave packets centered around momenta $p_0$ with non-zero group velocity $d E_{\alpha} / d p \neq 0$. As the limit of unitary operators, the M\"{o}ller operators $\Omega_{\pm}$ are isometries. Finally, we need the condition of asymptotic completeness (which is often harder to prove) to ensure that the range of $\Omega_{+}$ and $\Omega_{-}$ is the same: they map every state to the space of scattering states and satisfy the intertwining relations
\begin{align*}
H \Omega_{\pm} = \Omega_{\pm} H_0.
\end{align*}
\par The scattering operator or S matrix can then be defined as the operator mapping the in-asymptote to the out-asymptote
\begin{align*}
& \ket{\psi_\text{out}} = \Omega_-\dag\Omega_+ \ket{\psi_\text{in}} = S \ket{\psi_\text{in}} \rightarrow S=\Omega_-\dag\Omega_+.
\end{align*}
One can easily show that the free Hamiltonian commutes with $S$ so it makes sense to represent the S matrix in the basis of free momentum states
\begin{equation*}
\bra{q_\beta} S \ket{p_\alpha} = 2\pi\delta(E(q_\beta)-E(p_\alpha)) \times S_{q_\beta,p_\alpha}.
\end{equation*}
If asymptotic completeness is obeyed the $S$ matrix is unitary, which can be expressed in the momentum basis as
\begin{equation} \label{unitary}
\bra{q_\beta} S\dag S \ket{p_\alpha} = 2\pi\delta(p_\alpha-q_\beta) \delta_{\alpha\beta}.
\end{equation}
We can translate this condition to the matrix elements $S_{q_\beta,p_\alpha}$ as
\begin{equation*} \begin{split}
\bra{q_\beta} S\dag S \ket{p_\alpha} &= \sum_\gamma \int\frac{\d r}{2\pi} \bra{q_\beta} S\dag \ket{r_\gamma} \bra{r_\gamma} S \ket{p_\alpha} \\
&= \sum_\gamma \int\frac{\d r}{2\pi} 4\pi^2 \delta(E(p_\alpha)-E(r_\gamma)) \delta(E(q_\beta)-E(r_\gamma)) \ol{S}_{r_\gamma,q_\beta} S_{r_\gamma,p_\alpha} \\
&= \sum_\gamma \int\frac{\d r}{2\pi} \; \ol{S}_{r_\gamma,q_\beta} S_{r_\gamma,p_\alpha} \left( \sum_{p_{\alpha'}\in A(p_\alpha)} \left|\frac{\d E}{\d p}(p_{\alpha'})\right|^{-1} 2\pi \delta(p_{\alpha'}-r_\gamma) \right) \; 2\pi\delta(E(q_\beta)-E(r_\gamma)) \\
&= \left( \sum_{r_\gamma\in A(p_\alpha)} \ol{S}_{r_\gamma,q_\beta} S_{r_\gamma,p_\alpha} \left|\frac{\d E}{\d p}(r_\gamma)\right|^{-1} \right) \times 2\pi\delta(E(q_\beta)-E(p_\alpha)) \\
&= (\tilde{S}\dag\tilde{S})_{q_\beta p_\alpha} \times \left|\frac{\d E}{\d p}(q_\beta)\right|^{1/2}  2\pi\delta(E(q_\beta)-E(p_\alpha)) \left|\frac{\d E}{\d p}(p_\alpha)\right|^{1/2}
\end{split} \end{equation*}
where $A(p_\alpha)$ is the set of momenta $\{q_\beta\}$ such that $E(q_\beta)=E(p_\alpha)$, and we have defined the matrix elements of $\tilde{S}$ as
\begin{equation} \label{Stilde}
\tilde{S}_{q_\beta,p_\alpha} = \left|\frac{\d E}{\d p}(q_\beta)\right|^{-1/2} S_{q_\beta,p_\alpha} \left|\frac{\d E}{\d p}(p_\alpha)\right|^{-1/2}.
\end{equation}
Unitariness of the S matrix, Eq.~\eqref{unitary}, implies that $\tilde{S}_{q_\beta,p_\alpha}$ should be a unitary matrix.
\par There are different ways to calculate these $S$ matrix elements; one way is to construct the stationary scattering states, i.e. the eigenstates of the full Hamiltonian $\hat{H}=\hat{H}_0+\hat{V}$. One first introduces the Green's operators as
\begin{align*}
 G^0(z) &= (z-H^0)^{-1} \\
 G(z) &= (z-H)^{-1},
\end{align*}
which are related through the relation
\begin{align*}
 G(z) &= G^0(z) + G^0(z) V G(z) \\
 &= G^0(z) + G(z) V G^0(z).
\end{align*}
The $T$ operator is defined as
\begin{align*}
 T(z) = V + V G(z) V
\end{align*}
for which we can easily derive the Lippman-Schwinger equation \cite{Lippmann1950}
\begin{align*}
 T(z) = V + V G^0(z) T(z),
\end{align*}
and the equations
\begin{align*}
 & G^0(z) T(z) = (G^0(z) + G^0(z) V G(z)) V = G(z) V \\
 & T(z) G^0(z) = V ( G^0(z) + G(z) V G^0(z) ) = V G(z).
\end{align*}
The Lippman-Schwinger equation can be rewritten as an integral equation for the matrix elements of $T(z)$
\begin{equation*}
 \bra{q_\beta}T(z)\ket{p_\alpha} = \bra{q_\beta}V\ket{p_\alpha} + \sum_\gamma\int\frac{\d r_\gamma}{2\pi} \frac{\bra{q_\beta}V\ket{r_\gamma}}{z-E(r_\gamma)} \bra{r_\gamma} T(z) \ket{p_\alpha}.
\end{equation*}
One can derive a related equation for the M{\o}ller operators
\begin{align*}
 \Omega_+\ket\phi &= \lim_{t\rightarrow-\infty} U(t)\dag U^0(t) \ket\phi  \\
 &= \ket\phi - i \int_{-\infty}^0 \d\tau U(\tau)\dag V U^0(\tau) \ket\phi  \\
 &= \ket\phi - i \int_{-\infty}^0 \d\tau \e^{\epsilon\tau}U(\tau)\dag V U^0(\tau) \ket\phi  \\
 &= \ket\phi - i \sum_\alpha \int\frac{\d p}{2\pi} \int_{-\infty}^0 \d\tau \e^{\epsilon\tau}U(\tau)\dag V U^0(\tau) \ket{p_\alpha} \braket{p_\alpha|\phi}  \\
 &= \ket\phi + \sum_\alpha \int\frac{\d p}{2\pi} \; G(E(p_\alpha)+i0) \, V \ket{p_\alpha} \braket{p_\alpha|\phi},
\end{align*}
where we have introduced the time-dependent damping factor to the potential $V\rightarrow V\e^{-\epsilon t}$, which is allowed for $\epsilon\rightarrow0$ according to the adiabatic theorem. The S matrix
\begin{align*}
\bra{q_\beta} S \ket{p_\alpha} &= \bra{q_\beta} \Omega_-\dag \Omega_+ \ket{p_\alpha} = \lim_{t\rightarrow\infty} \bra{q_\beta} \e^{iH_0t} \e^{-2iHt} \e^{iH_0t} \ket{p_\alpha}
\end{align*}
can be worked out by writing it as the integral of its derivative
\begin{align*}
\bra{q_\beta} S \ket{p_\alpha} &= \braket{q_\beta|p_\alpha} - i \int_0^\infty \d t \bra{q_\beta} \left( \e^{iH_0t} V \e^{-2iHt} \e^{iH_0t} + \e^{iH_0t} \e^{-2iHt} V \e^{iH_0t} \right) \ket{p_\alpha} \\
&= \braket{q_\beta|p_\alpha} - i \lim_{\epsilon\rightarrow0} \int_0^\infty \d t \bra{q_\beta} \left( V \e^{i(E(q_\beta)+E(p_\alpha)+i\epsilon - 2H)t} + \e^{i(E(q_\beta)+E(p_\alpha)+i\epsilon -2H)t} V \right) \ket{p_\alpha} \\
&= \braket{q_\beta|p_\alpha}+ \frac{1}{2} \lim_{\epsilon\rightarrow0} \bra{q_\beta} \left( V  G\left(\frac{1}{2}(E(p_\alpha)+E(q_\beta))+i\epsilon\right)  + G\left(\frac{1}{2}(E(p_\alpha)+E(q_\beta))+i\epsilon\right) V \right) \ket{p_\alpha} \\
&= \braket{q_\beta|p_\alpha} + \lim_{\epsilon\rightarrow0} \left( \frac{1}{E(q_\beta)-E(p_\alpha) + i\epsilon}+\frac{1}{E(p_\alpha)-E(q_\beta) + i\epsilon} \right) \bra{q_\beta} T\left( \frac{1}{2}(E(p_\alpha)+E(q_\beta))+i\epsilon\right) \ket{p_\alpha} \\
&= 2\pi\delta(q_\beta-p_\alpha)\delta_{\beta\alpha} - 2 \pi \delta(E(q_\beta)-E(p_\alpha)) \; i \bra{q_\beta}T(E(p_\alpha)+i0)\ket{p_\alpha}.
\end{align*}
The off-diagonal elements of the S matrix are given by the on-shell T-matrix elements. We define the amplitudes $f$
\begin{equation*}
f(q_\beta\leftarrow p_\alpha) = - i  \left|\frac{\d E}{\d p}(p_\alpha)\right|^{-1/2} \bra{q_\beta}T(E(p_\alpha)+i0)\ket{p_\alpha} \left|\frac{\d E}{\d p}(q_\beta)\right|^{-1/2}
\end{equation*}
which are the off-diagonal elements of $\tilde{S}$ as defined in the unitary matrix \eqref{Stilde}.
\par We can now define the scattering states
\begin{align*}
 \ket{p_\alpha\pm} = \Omega_\pm \ket{p_\alpha}, \qquad H\ket{p_\alpha\pm} = E(p_\alpha) \ket{p_\alpha\pm}
\end{align*}
which, through the Lippmann-Schwinger equation for the M{\o}ller operators, obey the relation
\begin{equation*}
 \ket{p_\alpha\pm} = \ket{p_\alpha} + G(E(p_\alpha)\pm i0) V \ket{p_\alpha} = \ket{p_\alpha} + G^0(E(p_\alpha)\pm i0) V \ket{p_\alpha\pm}.
\end{equation*}
Another important relation is
\begin{equation} \label{Trelation} \begin{split}
 \bra{q_\beta} T(E(p_\alpha) \pm i0) \ket{p_\alpha}  &= \bra{q_\beta} ( V + V G(E(p_\alpha) \pm i0) V ) \ket{p_\alpha} = \bra{q_\beta} V \ket{p_\alpha\pm}.
\end{split} \end{equation}
An explicit expression for the asymptotic wave functions of the scattering states can thus be obtained:
\begin{equation} \label{pPlus} \begin{split}
 \braket{x,j|p_\alpha+} &= \braket{x,j|p_\alpha} + \sum_{j'}\int\d x' \braket{x,j|G^0(E(p_\alpha)+i0)|x',j'}\braket{x',j'|V|p_{\alpha}+}\\
  &= \e^{ip_\alpha x} v^j_{\alpha}(p) + \int\d x' \sum_{j'} \braket{x,j|\frac{1}{E(p_\alpha)-\ham_0+i0}|x',j'} \braket{x',j'|V|p_{\alpha}+}.
\end{split} \end{equation}
Since we know the exact eigenvalues and eigenvectors of $H_0$, we will now first introduce a resolution of the identity
\begin{equation*}
\sum_{j} \int\frac{\d q}{2 \pi} \ket{q,j}\bra{q,j}
\end{equation*}
which brings the Green's function in block diagonal form
\begin{align*}
 \braket{x,j|p_\alpha+} &= \e^{ip x} v^j_{\alpha}(p) + \int\d x' \sum_{j'}\int \frac{\d q}{2\pi} \left(\frac{1}{E(p_\alpha)-\mathsf{A}(q)+i0}\right)_{j,j'} e^{iq(x-x')} \braket{x',j'|V|p_{\alpha}+}
\end{align*}
with the matrix $\mathsf{A}(q)$ an analytic function of $q$, as defined at the beginning of this section. The integral over $q$ can be calculated with the residue theorem. For continuous systems, where $q$ ranges over the real axis, we will have to close the contour in the upper or lower half plane depending on the whether $x-x'>0$ or $x-x'<0$. A first set of poles will be close to the real axis and can be obtained from the eigenvalue decomposition of $\mathsf{A}(q)$. Together with the analytic dependence on $q$ and Eq.~\eqref{eq:Aderivative}, we obtain
\begin{equation}
\mathsf{A}(q\pm i0) = \sum_{\beta} \left(E(q_\beta) \pm i0 \frac{\d E}{\d p}(q_\beta)\right) \vect{v}_{\beta}(q) \vect{v}_\beta(q)^\dagger.
\end{equation}
We should therefore separate the set $A(p_\alpha)$ of all solutions $q_\beta$ for which $E(q_\beta)=E(p_\alpha)$ into two parts $A^{\pm}(p_\alpha)$ corresponding to solutions for which the energy derivative $\frac{\d E}{\d p}(q_\beta)$ is positive ($+$) or negative ($-$). We then find a first set of poles of $\left(\frac{1}{E(p_\alpha)-\mathsf{A}(q)+i0}\right)_{j,j'}$ which are of the form $q_\beta+i0$ for $q_\beta \in A^{+}(p_\alpha)$ and of the form $q_\beta-i0$ for $q_\beta \in A^{-}(p_\alpha)$. The corresponding residues are given by
\begin{align*}
\lim_{\substack{q\to q_{\beta}\pm i0\\q_\beta \in A^{\pm}(p_\alpha)}} (q-(q_\beta\pm i0)) \left(\frac{1}{E(p_\alpha)-\mathsf{A}(q)+i0}\right)_{j,j'} e^{iq(x-x')} = - \left(\frac{\d E}{\d p}(q_\beta)\right)^{-1} v_\beta^j(p) \overline{v}_\beta^{j'}(p) e^{i q_\beta(x-x')}.
\end{align*}
Aside from those solutions, there could be other solutions $q= i \lambda_\gamma$ further away from the real axis ($\Re \lambda \neq 0$). These correspond to values of $\lambda$ where the analytically continued (but non-hermitian) matrix $\mathsf{A}(i\lambda)$ has a real eigenvalue $E_\gamma(i\lambda)=E(i\lambda_\gamma)$ that equals $E(p_\alpha)$; we denote the corresponding left and right eigenvectors as $\tilde{\vect{w}}_\gamma(\lambda)^\dagger$ and $\vect{w}_\gamma(\lambda)$ (wich will in general not be related by hermitian conjugation). The corresponding residue is then given by $-\frac{\d E}{\d p}(\lambda_\gamma) w_\gamma^j \overline{\tilde{w}}_\gamma^{j'}e^{-\lambda_\gamma (x-x')}$ or more generally $-\frac{\d E}{\d p}(\lambda_\gamma) \mathsf{P}_{j,j'}(i\lambda_\gamma)e^{-\lambda_\gamma (x-x')}$ with $\mathsf{P}(i\lambda_\gamma)$ the corresponding eigenspace projector.
\par Let us now return to the evaluation of the integral over $q$. Depending on the sign of $x-x'$, we will close the contour in the upper or lower half plane and pick up the contributions of the poles in those respective domains. Since we also have an integral over $x'$, it seems we will need to split this into the two regions $x<x'$ and $x>x'$. However, we can make use of the locality of the potential to conclude that $\braket{x',j'|V|p_{\alpha}+}$ is only nonzero for $\lvert x'\rvert \leq M+N$. Thus, if $\lvert x \rvert > M+N$, then $x-x'$ will have a fixed sign throughout the integral over $x'$. For e.g. $x-x'$, we will need to sum up the contributions of all the poles in the upper half plane, corresponding to $q_\beta + i0$ for $p_\beta\in A^{+}(p_\alpha) $ and all $i \lambda_\gamma$ with $\Re \lambda_\gamma > 0$. The latter contributions will actually vanish if we now take the limit $x\to\infty$. We can then write the asymptotic wave function as
\begin{equation*}
 \braket{x,j|p_\alpha+} \approx v_\alpha^j(p)\e^{ip_\alpha x}- i \left\{
 \begin{array}{ll} \sum_{q_\beta\in A^-(p_\alpha)} v_\beta^j(q) \e^{i q_\beta x} \left| \frac{\d E}{\d p}(q_\beta) \right|^{-1} \braket{q_\beta| \hat{V}|p\pm} \qquad & x \rightarrow-\infty \\
  \sum_{q_\beta\in A^+(p_\alpha)} v_\beta^j(q) \e^{i q_\beta x} \left| \frac{\d E}{\d p}(q_\beta) \right|^{-1} \braket{q_\beta|\hat{V}|p\pm} \qquad & x \rightarrow+\infty \end{array} \right.
\end{equation*}
and with Eq.~\eqref{Trelation}
\begin{equation*}
  \braket{x,j|p_\alpha+} \approx v_\alpha^j(p)\e^{ip_\alpha x}  -i  \left\{ \begin{array}{ll}  \sum_{q_\beta\in A^-(p_\alpha)}v_\beta^j(q) \e^{i q_\beta x} \left| \frac{\d E}{\d p}(q_\beta) \right|^{-1} \bra{q_\beta} T(E(p_\alpha)+i0) \ket{p_\alpha} \qquad & x \rightarrow-\infty \\
  \sum_{q_\beta\in A^+(p_\alpha)}v_\beta^j(q) \e^{i q_\beta x} \left| \frac{\d E}{\d p}(q_\beta) \right|^{-1} \bra{q_\beta} T(E(p_\alpha)+i0) \ket{p_\alpha} \qquad & x \rightarrow+\infty \end{array} \right. .
\end{equation*}
The coefficients that appear are the amplitudes that were defined earlier, so we have the nice final result
\begin{equation} \label{asWave}
\braket{x,j|p_\alpha+}  = v_\alpha^j(p) \e^{ip_\alpha x} +  \sum_{q_\beta\in A^\pm(p_\alpha)} \left|\frac{\d E}{\d p}(p_\alpha)\right|^{1/2} f(q_\beta\leftarrow p_\alpha) \left|\frac{\d E}{\d p}(q_\beta)\right|^{-1/2} \; \e^{iq_\beta x} v_\beta^j(q), \qquad x \rightarrow\pm \infty.
\end{equation}
For discrete systems, we can proceed in a similar way. Momentum integrals now range from $0$ to $2\pi$ and we automatically obtain a contour around the unit circle in the complex plane by going to the complex variable $\mu = e^{iq}$ (for $x-x'>0$) or $\mu=e^{-iq}$ (for $x-x'<0$). The derivation then follows analogously.

\end{document}